\documentclass[12pt,a4paper]{article}
\usepackage{jheppub}
\usepackage[utf8]{inputenc}

\usepackage{yfonts}
\usepackage{hyperref}
\usepackage{epsfig}
\usepackage{graphicx}
\usepackage{amsfonts}
\usepackage{dsfont}
\usepackage{amsmath} 
\usepackage{color} 
\usepackage{pstool}
\usepackage[all]{xy}

\usepackage{tikz}
\usetikzlibrary{decorations.pathmorphing}
\usetikzlibrary{decorations.markings}
\usepgflibrary{shapes.geometric}

\def\beq{\begin{equation}} 
\def\eeq{\end{equation}} 
\def\beqrr{\begin{array}} 
\def\eeqrr{\end{array}} 
\def\beqa{\begin{eqnarray*}} 
\def\eeqa{\end{eqnarray*}}

\def\Or[#1]{{\text{O}}\left({#1}\right)}
\def\dotl[#1,#2]{\left\langle #1, #2 \right\rangle}
\def\dotlb[#1,#2]{[ #1, #2 ]}
\def\dotp[#1,#2]{(#1) \cdot (#2)}
\def\aff[#1,#2]{\hat{#1}(#2)}
\def\n4sym{{\cal N}=4 SYM}
\def\>{\rangle}
\def\<{\langle}

\def\weight[#1,#2,#3]{\{(#1),#2,#3\}}
\def\ads[#1]{$\text{AdS}_{#1}$}

\newcommand{\ba}{\begin{eqnarray}}
\newcommand{\ea}{\end{eqnarray}}

\newcommand{\be}{\begin{equation}}
\newcommand{\ee}{\end{equation}}  
\newcommand{\bi}{\begin{itemize}}
\newcommand{\ei}{\end{itemize}}

\newcommand{\Ocal}{{\mathcal O}}

\newcommand{\aslash}[1]{\,\,{\raise.15ex\hbox{/}\mkern-12mu #1}}
\newcommand{\bslash}[1]{\,\,{\raise.15ex\hbox{/}\mkern-9mu #1}}

\renewcommand{\bar}{\overline}
\renewcommand{\tilde}{\widetilde}
\renewcommand{\hat}{\widehat}
\renewcommand{\Im}{{\rm Im\,}}
\renewcommand{\Re}{{\rm Re\,}}

\newcommand\lrpar{\raise .8ex\hbox{$^\leftrightarrow$} \hspace{-9pt}
\partial}
\newcommand\lpar{\raise .8ex\hbox{$^\leftarrow$} \hspace{-9pt}
\partial}
\newcommand\rpar{\raise .8ex\hbox{$^\rightarrow$} \hspace{-9pt}
\partial}
\newcommand\lrd{\raise .8ex\hbox{$^\leftrightarrow$} \hspace{-9pt}
\nabla}

\newcommand{\gsim}{\lower.7ex\hbox{$\;\stackrel{\textstyle>}{\sim}\;$}}
\newcommand{\lsim}{\lower.7ex\hbox{$\;\stackrel{\textstyle<}{\sim}\;$}}

\renewcommand{\Im}{\text{Im}}
\renewcommand{\Re}{\text{Re}}

  \let\g=\gamma \let\d=\delta 
    
  \let\n=\nu

  \let\D=\Delta

\renewcommand{\ba}{\begin{eqnarray}}
\renewcommand{\ea}{\end{eqnarray}}
\newcommand{\bea}{\begin{eqnarray}}
\newcommand{\eea}{\end{eqnarray}}

\newcommand{\ii}{\mathrm{i}}

\makeatletter
\newsavebox{\@brx}
\newcommand{\llangle}[1][]{\savebox{\@brx}{\(\m@th{#1\langle}\)}%
  \mathopen{\copy\@brx\kern-0.5\wd\@brx\usebox{\@brx}}}
\newcommand{\rrangle}[1][]{\savebox{\@brx}{\(\m@th{#1\rangle}\)}%
  \mathclose{\copy\@brx\kern-0.5\wd\@brx\usebox{\@brx}}}
\makeatother

\newcommand{\Hypergeometric}[4]{\ensuremath{\, _2F_1\left(
	\begin{array}{c}
		#1 , #2 \\[1mm] #3
	\end{array}
	; #4 \right)
}}
\newcommand{\GeneralizedHypergeometric}[6]{\ensuremath{\, _3F_2\left(
	\begin{array}{c}
		#1 , #2, #3 \\[1mm] #4, #5
	\end{array}
	; #6 \right)
}}

\begin{document}

\begin{titlepage}

\begin{center}
\vspace{1.5cm}

{\Large \bf  A Scattering Amplitude \\[0.3cm]
 in   Conformal Field Theory}

\vspace{0.8cm}

{\bf Marc Gillioz,$^{\star\diamond}$ Marco Meineri,$^\star$ Jo\~ao Penedones$^\star$}

\vspace{.5cm}

{$^\star$ \it  EPFL, Institute of Physics, Lausanne, Switzerland}
\\
\vspace{.3cm}
{$^\diamond$ \it SISSA, via Bonomea 265, 34136 Trieste, Italy}

\end{center}

\vspace{1cm}

\begin{abstract}

We define form factors and scattering amplitudes  in Conformal Field Theory as
 the coefficient of the singularity of the Fourier transform of time--ordered correlation functions, as $p^2 \to 0$.
 In particular, we study a form factor $F(s,t,u)$ obtained from a four-point function of identical scalar primary operators.
 We show that $F$ is crossing symmetric, analytic and it has a partial wave expansion.
 We illustrate our findings in the $3d$ Ising model, perturbative fixed points and holographic CFTs. \phantom{\cite{Lehmann:1954rq}}
\end{abstract}

\bigskip

\end{titlepage}


\tableofcontents
\vfill\eject

\section{Introduction and summary of results}
\label{sec:intro}

Scattering amplitudes are among the most important observables in the realm of quantum field theory (QFT). Their distinguished role does not just descend from their experimental relevance: they also are a valuable theoretical tool. They are crossing symmetric, and at the same time admit a partial wave decomposition, two features which together allow to explore the space of QFTs using a bootstrap approach \cite{Paulos:2017fhb}. The existence of a scattering amplitude -- \emph{i.e.} an overlap between asymptotic non-interacting eigenstates of the momentum generator -- is tied to the presence of single-particle states, separated by an energy gap from the continuous part of the spectrum.\footnote{Scattering amplitudes can also be defined in gapless theories if the interactions decay sufficiently fast at low energies. Goldstone bosons and photons are examples of this type.}
This condition is not fulfilled in a conformal field theory (CFT), where the spectral density has a simple power-law form. However, the Lehmann-Symanzik-Zimmermann (LSZ) theorem \cite{Lehmann:1954rq} offers a different perspective: scattering amplitudes appear as the residue of singularities in the Fourier transform of time-ordered correlators in Lorentzian signature. Crossing symmetry is inherited from the correlator, while the existence of a partial wave decomposition follows from the fact that only a specific partial ordering of the operators contributes to the singularity. In this work, we explore the structure of singularities of the Fourier transform of  the four-point function of scalar primary operators in a CFT in dimension $d>2$. We define a form factor and an amplitude, which are crossing symmetric and obey a conformal partial wave decomposition.

Of course, in a CFT the correlation function in position space already offers a crossing symmetric observable which also admits an OPE decomposition, and the bootstrap approach has prospered in $2d$ \cite{Belavin:1984vu}, and in higher $d$ \cite{Rattazzi:2008pe, Poland:2018epd}. Still, conformal field theories in momentum space have recently drawn increasing interest \cite{Coriano:2013jba, Bzowski:2013sza, Bzowski:2019kwd, Maglio:2019grh, Isono:2018rrb, Isono:2019wex, Bautista:2019qxj, Gillioz:2019lgs, Arkani-Hamed:2018kmz, Baumann:2019oyu, Sleight:2019mgd, Sleight:2019hfp, Albayrak:2020isk}.
The feature which mostly interests us in this work is the technical simplicity of CFT in momentum space. Eigenstates of momenta are orthogonal linear combinations of all the states in a conformal family. Therefore, each partial wave is obtained by inserting in the form factor (or the amplitude) a projector onto a single eigenstate, as opposed to the conformal blocks in position space which require summing over all descendants \cite{Gillioz:2016jnn, Gillioz:2018kwh, Karateev:2018oml, Erramilli:2019njx}. A practical consequence is that we are able to obtain conformal partial waves in closed form in any spacetime dimension. This may be a useful starting point for analytic explorations of the crossing equation. Furthermore, perturbation theory is simpler in momentum space, and it is convenient to extract CFT data from a four-point function in momentum space, rather than resort to the far more complicated position space formulas \cite{Goncalves:2018nlv}.

Let us introduce the main player, and summarize the results of this work. Consider the Fourier transform of the   Euclidean correlation function of scalar primary operators in CFT in $d>2$ spacetime dimensions:\footnote{We focus on four-point functions but this can be  generalized to higher point functions. }
\be
(2\pi)^d \delta^d\big( \sum p_j \big) G(p_1,\dots,p_4)
=\int \left( \prod_{j=1}^4 d^dx_j e^{i p_j\cdot x_j} \right)
\langle \phi_1(x_1) \dots \phi_4(x_4) \rangle~.
\label{eq:EuclideanFourierTransform}
\ee
We will show that when taken as a function of the six independent Lorentz-invariant scalar products $p_i \cdot p_j$,
$G$ diverges in the limit $p_j^2 \to 0$ for $j=1,2,3$ when $\Delta_j <\frac{d}{2}$. Here the $\Delta_j$ are the scaling dimensions of the operators in eq.~\eqref{eq:EuclideanFourierTransform} and we consider generic momenta so that the disconnected terms vanish.
The leading divergence can be used to define a \emph{form factor}  as follows 
\be
  F(s,t,u) \equiv  \left(  
 \prod_{j=1}^3 \lim_{p_j^2 \to 0_+} (p_j^2 )^{\frac{d}{2}-\Delta_j } \right) G(p_1,\dots,p_4)~,
 \label{def:FormFactor}
 \ee
where $s=-(p_1+p_2)^2 $, $t=-(p_1+p_3)^2 $, $u=-(p_2+p_3)^2 $ are the usual Mandelstam invariants obeying  $s+t+u=-p_4^2 $.
As advertized, this limiting procedure is similar to the spirit of the usual LSZ reduction in massive QFT \cite{Lehmann:1954rq}.

We shall focus on the case of identical scalar operators of dimension $\Delta_\phi$.
In this case, the form factor is also crossing symmetric
\be
F(s,t,u)=F(t,s,u)=F(s,u,t)~.
\label{crossing}
\ee
This property also applies to $G$ if we set $p_1^2=p_2^2=p_3^2$. The main advantage of the limit $p_j^2  \to 0$, for $j=1,2,3$, is that only in this case we can write a  partial wave expansion.
Introducing the variables
\be
q^2\equiv p_4^2=-s-t-u\,,\qquad\qquad
w\equiv -\frac{s}{q^2}\,,\qquad\qquad
 \cos \theta \equiv \frac{u-t}{u+t}~,
\ee
we will show that
\begin{equation}
	F(s, t,u) = (q^2)^{-\Delta_\phi}
	\sum_{\mathcal{O}} \lambda_{\phi\phi\mathcal{O}}^2
	F_{\Delta,\ell}\left( w, \cos\theta \right)~,
	\label{CBexpansionF}
\end{equation}
where the sum runs over all primary operators $\mathcal{O}$ of dimension $\Delta$ and spin $\ell$, $\lambda_{\phi\phi\mathcal{O}}$ are OPE coefficients 
and the functions $F_{\Delta, \ell}(w,\cos \theta) $ are kinematical and given in \eqref{eq:F:conformalblock}. They are polynomials of degree $\ell$ in  $\cos \theta$ and are analytic in the variable  $w$  except for branch points at $w=0$ and $w=1$ (they are real for $0<w<1$).
In the small $w$ limit, they are controlled by the twist $\tau = \Delta - \ell$ of the exchanged operator,
\begin{equation}
	F_{\Delta, \ell}(w, \cos\theta) \propto  w^{(\tau - 2 \Delta_\phi)/2}
	\mathcal{C}_\ell^{(\tau - 1)/2}(\cos\theta)~,
	\label{CBleadingtwist}
\end{equation}
where $\mathcal{C}_\ell$ is a Gegenbauer polynomial. Unlike the familiar limits of ordinary conformal blocks, the parameter of this Gegenbauer polynomial is also related to the twist of the exchanged operator, and not to the dimension of spacetime.
 
It is also interesting to take the limit $p_4^2 \to 0$ and define the amplitude
\be
 A(s,t,u)  \equiv  
 \left( \prod_{j=1}^4 \lim_{p_j^2 \to 0_+} (p_j^2  )^{\frac{d}{2}-\Delta_j }  \right) G(p_1,\dots,p_4) ~,
 \label{def:amp}
 \ee 
where $s+t+u=0$. 
In this case, there is no Euclidean regime where the amplitude is real.  We shall focus on the physical Lorentzian regime $s>0$ and $t,u<0$.
We shall see that
\begin{equation}
	A(s,t,u) = s^{d/2 - 2\Delta_\phi} \sum_{\mathcal{O}} \lambda^2_{\phi\phi\mathcal{O}}
	A_{\Delta, \ell}(\cos\theta)~,
\end{equation}
where $A_{\Delta, \ell}(\cos\theta)$ are the kinematical polynomials defined in eqs.~(\ref{eq:A},\ref{eq:g}), which we obtain from the limit $w \to -\infty$ of $F_{\Delta, \ell}(w, \cos\theta)$.

\begin{figure}
	\centering
	\includegraphics[width=80mm]{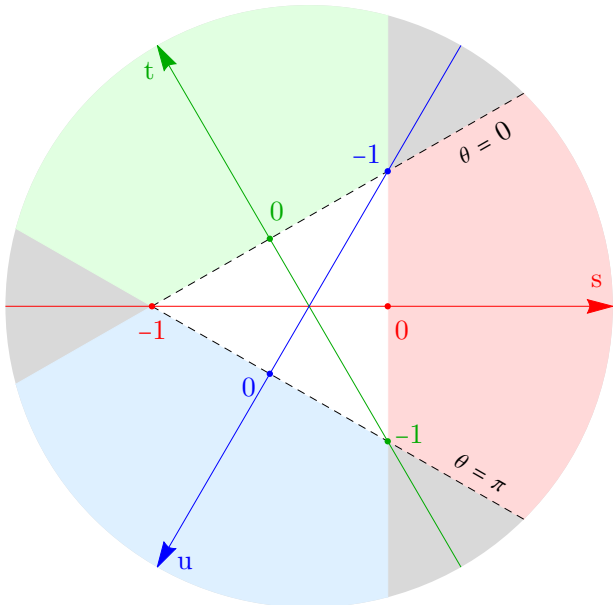}
	\caption{Regions of convergence of the partial wave expansion for the form factor \eqref{def:FormFactor}
	in terms of the Mandelstam invariants $s$, $t$ and $u$,
	in units where $s + t + u = -1$.
	The $s$-channel partial wave expansion~\eqref{CBexpansionF} is defined 
	for $s > 0$ and $t, u < 0$, corresponding to the red region,
	and it can be analytically continued to the white triangle with $s \leq 0$,
	as long as $t$ and $u$ remain negative.
	In other words, the expansion converges as long as the scattering angle is physical,
	$\cos\theta \in (-1,1)$, which corresponds to the wedge delimited by the dashed lines.
	The $t$- and $u$-channel expansions are respectively defined in the green and blue regions
	that are disjoint from the red region.
	However, they can also be analytically continued to the white triangle,
	where all 3 channels converge and the form factor is real.}
	\label{fig:triangle}
\end{figure}

We shall give two derivations of the results presented above. The first approach, which is contained in section~\ref{sec:lsz}, consists in taking the limits $p^2_j\to 0$ in the Lorentzian analogue of expression \eqref{eq:EuclideanFourierTransform}.
This method directly yields the partial wave decomposition of the form factor, eq.~\eqref{CBexpansionF}. In section~\ref{sec:mellin}, on the other hand, we perform the LSZ reduction starting from the Mellin representation of the Euclidean correlation function. This procedure yields the following relation between the form factor and the Mellin amplitude:
\be
F(s,t,u) = C \int [d\gamma] M(\gamma_{ij}) 
\frac{\Gamma(\g_{12})}{(-s)^{\gamma_{12}}}
\frac{\Gamma(\g_{13})}{(-t)^{\gamma_{13}}}
\frac{\Gamma(\g_{23})}{(-u)^{\gamma_{23}}}~,
\ee
where $C$ is a constant defined in eq.~\eqref{Cdef}.
This shows that the form factor is real analytic:
\be
F(s,t,u)^* = F(s^*,t^*, u^*)~.
\label{realanalytic}
\ee
More precisely, it is real for $s,t,u<0$,\footnote{This follows from the observation that $G$ is real for spacelike momenta $p_j^2>0$: In Euclidean space and for real operators $\phi_j$, it is clear that $G(p_j)^*=G(-p_j)$. Moreover, $G(p_j)=G(-p_j)$ if the position-space correlator is invariant under $x_j \to -x_j$. This is is just a rotation for even spacetime dimension $d$ and it is parity for odd $d$. The four-point function of scalar primary operators is always parity-symmetric in CFT$_d$ for $d\ge 3$.}
and it is analytic in, say, the $s$  complex plane minus the branch cut along the positive real axis, for fixed $t$ and $u$ negative.
We will also argue that the $s$-channel partial wave expansion \eqref{CBexpansionF} is convergent in the same region.
The intersection of this region with its images under crossing is
the Euclidean region $s, t, u < 0$, as illustrated in figure~\ref{fig:triangle}.

Moreover, the second derivation allows to relate the amplitude \eqref{def:amp} to the same scaling limit of the Mellin transform which controls the Landau singularity relevant to the bulk-point limit \cite{Penedones:2010ue,Gary:2009ae, Maldacena:2015iua}. Precisely, the amplitude is proportional to the coefficient $\mathcal{L}(z)$ of the Landau singularity: 
\beq
A(s,t,u)
 =\kappa\,
s^{\frac{d}{2}-2\Delta_\phi}  \,
(z-1)^{2-d/2}z^{2-d}\mathcal{L}(z)~, \qquad
 z=-\frac{s}{t}=\frac{2}{1-\cos\theta}~.
\eeq
where $\kappa$ is a constant defined in eq.~\eqref{eq:kappa}, and the coefficient $\mathcal{L}(z)$ of the Landau singularity is defined by
eq.~\eqref{GaroundLandau}.

Finally, section~\ref{sec:examples} is dedicated to some applications of our results. We compute the form factor at the Wilson-Fisher fixed points in $\phi^4$ and $\phi^3$ theories, and we show that eq.~\eqref{CBexpansionF} can be efficiently used to extract CFT data in perturbation theory. We also consider two non-perturbative examples: the $3d$ Ising model, and holographic theories. In particular, we use the available information on the spectrum of the $3d$ Ising model \cite{Simmons-Duffin:2016wlq} to approximate the form factor in this theory, and we find the approximation to be crossing symmetric with remarkable accuracy. 
 
An important question concerns the non-perturbative existence of the Fourier transform \eqref{eq:EuclideanFourierTransform}.
The Euclidean Fourier transform is well defined for $\Delta_i <\frac{d}{2}$, because this makes the leading OPE singularity integrable. Notice that this is the same condition which makes the $p_i^2\to0$ limit divergent, and allows us to extract the form factor and the amplitude. In appendix ~\ref{app:wick}, we discuss the convergence of the Euclidean Fourier transform in detail. Furthermore, we expect that $G(p_1,\dots p_4)$ can be analytically continued to Lorentzian momenta. If the path corresponds to the usual Wick rotation, the result is the Fourier transform of the time-ordered correlator. This expectation lacks a non-perturbative proof, but in appendix~\ref{app:wick} we offer a strategy to explicitly construct the analytic continuation of eq.~\eqref{eq:EuclideanFourierTransform} to Lorentzian signature. A complete analysis is left to future work.


\section{LSZ reduction}
\label{sec:lsz}

We begin with a study of Lorentzian correlators in the spirit of the LSZ reduction formula in QFT.
For simplicity of the argument, we will focus here on the case of 4 identical scalar operators with scaling dimension $\Delta_\phi$. The more general case involving distinct scalar operators is treated in Appendix~\ref{sec:lszdistinct}. 

\subsection{Momentum eigenstates and completeness relation}
\label{sec:states}

The cornerstone of this LSZ reduction procedure is the use of momentum eigenstates
\begin{equation}
	| \mathcal{O}^\alpha(p) \rangle \equiv
	(-p^2)^{d/2 - \Delta}
	\int d^dx \, e^{i p \cdot x} \mathcal{O}^\alpha(x^0 + i \epsilon, x^i) | 0 \rangle~,
	\label{eq:state}
\end{equation}
satisfying $P^\mu | \mathcal{O}(p) \rangle = p^\mu | \mathcal{O}(p) \rangle$. Here $\mathcal{O}$ is a conformal primary operator, and the index $\alpha$ denotes collectively all its spin indices.
The small imaginary time component $i \epsilon$ ($\epsilon > 0$) ensures the finiteness of the norm
\begin{equation}
	\langle \mathcal{O}^{\alpha'}(p') | \mathcal{O}^\alpha(p) \rangle
	= (2\pi)^d \delta^d(p' - p) \Theta(p^0) \Theta(-p^2) (-p^2)^{d/2 - \Delta} \Pi^{\alpha'\alpha}(p)~,
	\label{eq:norm}
\end{equation}
where $\Delta$ is the scaling dimension of the operator and $\Pi^{\alpha'\alpha}(p)$ a positive-definite tensor structure.
The Heaviside $\Theta$-functions ensure that the momentum $p$ lies in the forward lightcone, and thus so does $p'$. The state \eqref{eq:state} is null otherwise.

For a scalar operator canonically normalized such that
$\langle 0 | \mathcal{O}(x) \mathcal{O}(0) | 0 \rangle = (x^2)^{-\Delta}$, then $\Pi(p)$ is just a constant given in terms of the scaling dimension $\Delta$ of the operator,
\begin{equation}
	\Pi(p) = \frac{(4\pi)^{d/2 + 1} }{2^{2\Delta + 1}
	\Gamma\left( \Delta \right) \Gamma\left( \Delta - \frac{d}{2} + 1 \right)}~,
\end{equation}
which is indeed positive when the unitarity bound $\Delta \geq \frac{d-2}{2}$ is satisfied.
For traceless symmetric tensors with spin $\ell$, the tensor $\Pi^{\mu_1 \ldots \mu_\ell, \nu_1 \ldots \nu_\ell}$ is given for instance in Ref.~\cite{Gillioz:2018mto}. For all our purposes it is sufficient to know that such a tensor exists and that its $p$-independent part is given by  
\begin{equation}
	\Pi^{\mu_1 \ldots \mu_\ell, \nu_1 \ldots \nu_\ell}(p)
	= \frac{\pi^{d/2 + 1}}{2^{2\Delta - d - 1} ( \Delta + \ell - 1) \Gamma( \Delta - 1 )
	\Gamma\left( \Delta - \frac{d}{2} + 1 \right)}
	\eta^{\mu_1(\nu_1} \cdots \eta^{\nu_\ell) \mu_\ell } +\dots 
\label{eq:Pi:0}
\end{equation}

Since the tensors $\Pi^{\alpha'\alpha}(p)$ are positive-definite they can always be inverted and one can write the completeness relation
\begin{equation}
	\mathds{1} = | 0 \rangle \langle 0 | + \sum_{\mathcal{O}}
	\int\limits_{k \in \bigvee} \frac{d^dk}{(2\pi)^d} 
	(-k^2)^{\Delta - d/2} \Pi_{\alpha\alpha'}^{-1}(k) 
	| \mathcal{O}^\alpha(k) \rangle \langle \mathcal{O}^{\alpha'}(k) |~,
\label{eq:completenessrelation}
\end{equation}
where the sum is over primary operators only and the integral is over the forward light cone $ \bigvee$, i.e. $k^0 > 0$, $k^2 < 0$.
This relation will be instrumental in the derivation of the LSZ formula.

The definition \eqref{eq:state} contains the conventional factor $(-p^2)^{d/2 - \Delta}$ that is chosen so that 
\begin{equation}
	\langle 0| \mathcal{O}^{\alpha'}(x) | \mathcal{O}^\alpha(p) \rangle
	= e^{i p \cdot x} \Pi^{\alpha'\alpha}(p)~,
	\label{eq:matrixelement}
\end{equation}
in analogy with scattering states in quantum field theory. The key property that we will use in deriving our LSZ reduction formula is that this matrix element remains finite even when $p^2 \to 0$, provided that the limit is taken from inside the forward lightcone (the state is null otherwise).
In fact, it will be convenient to introduce the notation
\begin{equation}
	| \mathcal{O}^\alpha(\vec{p}) \rangle \equiv \lim_{p^0 \to |\vec{p}|_+} | \mathcal{O}^\alpha(p) \rangle
	\label{eq:state:limit}
\end{equation}
to indicate the state that is obtained in this limit. 
More generally, using the OPE recursively, it can be shown that the correlation function $\langle 0 | \mathcal{O}(x_1) \cdots \mathcal{O}(x_n) | \mathcal{O}(\vec{p}) \rangle$ is finite as well.\footnote{More precisely, this matrix element is finite as a distribution: the OPE can be used recursively to argue that $\mathcal{O}(x_1) \cdots \mathcal{O}(x_n) \sim f\left( x_1 - x_n, \ldots, x_{n-1} - x_n \right) \mathcal{O}'(x_n)$, where the equivalence is understood after integrating against test functions~\cite{Mack:1976pa, Kravchuk:2020scc}.}

Note that we will focus on scalar operators with $\Delta < \frac{d}{2}$, in which case the norm of the state \eqref{eq:state} vanishes as $p^2 \to 0$ (it diverges in the opposite case $\Delta > \frac{d}{2}$).
This is not actually an issue since correlation functions such as the norm \eqref{eq:norm} should be understood as distributions in the momenta: the actual states are obtained after integrating against test functions, such as wavepackets, and their norm is always finite.

\subsection{LSZ for the first operator}
\label{sec:lsz:first}

Consider the time-ordered 4-point function of a scalar operator $\phi$ with $\Delta_\phi < d/2$. For now we take the Fourier transform with respect to the first point only,
\begin{equation}
	G(p_1; x_2, x_3, x_4) \equiv
	i \int d^dx_1 \, e^{i p_1 \cdot x_1} \langle 0 | T\{ \phi(x_1) \phi(x_2) \phi(x_3) \phi(x_4) \} | 0 \rangle~.
\end{equation}
Note that this is the Fourier transform of the Lorentzian correlator, as opposed to the Euclidean Fourier transform~\eqref{eq:EuclideanFourierTransform}, hence the conventional factor of $i$ that follows from Wick rotation of the integration measure.
The time integral can be split into 3 regions, writing $G = G_+ + G_0 + G_-$, where
\begin{align}
	G_+ &= i \int_{t_+}^\infty dx_1^0 \int d^{d-1}\vec{x}_1 \, e^{i p_1 \cdot x_1}
	\langle 0 | \phi(x_1) T\{ \phi(x_2) \phi(x_3) \phi(x_4) \} | 0 \rangle,
	\nonumber \\
	G_0 &=  i \int_{t_-}^{t_+} dx_1^0 \int d^{d-1}\vec{x}_1 \, e^{i p_1 \cdot x_1}
	\langle 0 | T\{ \phi(x_1) \phi(x_2) \phi(x_3) \phi(x_4) \} | 0 \rangle,
	\\
	G_- &= i \int_{-\infty}^{t_-} dx_1^0 \int d^{d-1}\vec{x}_1 \, e^{i p_1 \cdot x_1}
	\langle 0 | T\{ \phi(x_2) \phi(x_3) \phi(x_4) \} \phi(x_1) | 0 \rangle~,
	\nonumber
\end{align}
$t_+$ and $t_-$ are chosen such that $t_- < x_2^0, x_3^0, x_4^0 < t_+$, and therefore the operator $\phi(x_1)$ can be taken out of the time-ordered product in $G_+$ and $G_-$, respectively to the left and to the right.
Since the integral $G_0$ is bounded in time and $p$ appears analytically in the integrand, its contribution to $G$ is finite when $p^2 \to 0$.
This is not necessarily the case for $G_+$ and $G_-$.
Focusing on $G_-$, we can insert the complete set of momentum eigenstates \eqref{eq:completenessrelation} in between $\phi(x_1)$ and the time-ordered product, after which we obtain
\begin{equation}
	G_- = \frac{1}{2\pi} \int\limits_{|\vec{p}_1|}^\infty dk^0 \, e^{i (k^0 - p_1^0) t_-}
	\frac{(-k^2)^{\Delta_\phi - d/2}}{k^0 - p_1^0 - i \epsilon}
	\langle 0 | T\{ \phi(x_2) \phi(x_3) \phi(x_4) \} | \phi(k) \rangle
	\Big|_{\vec{k} = \vec{p}_1}~,
\label{eq:k0integral:p1}
\end{equation}
\begin{figure}
	\includegraphics[width=0.48\linewidth]{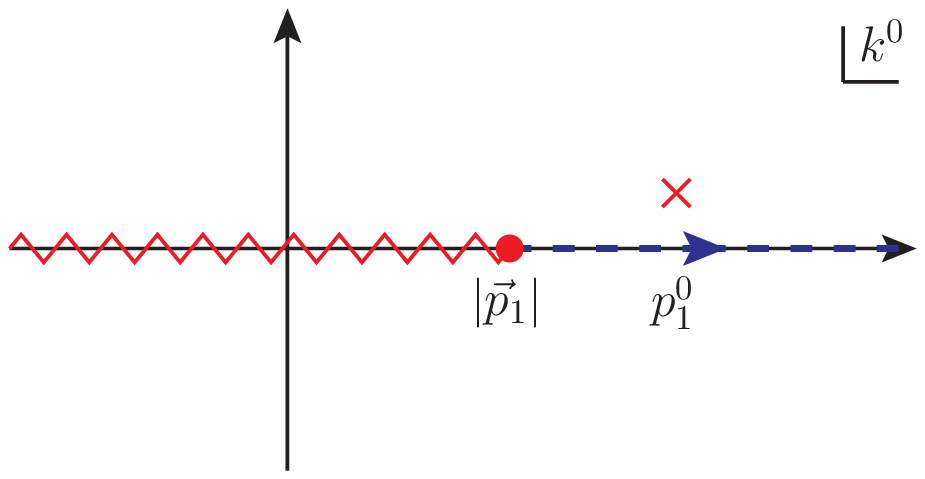}
	\hfill
	\includegraphics[width=0.48\linewidth]{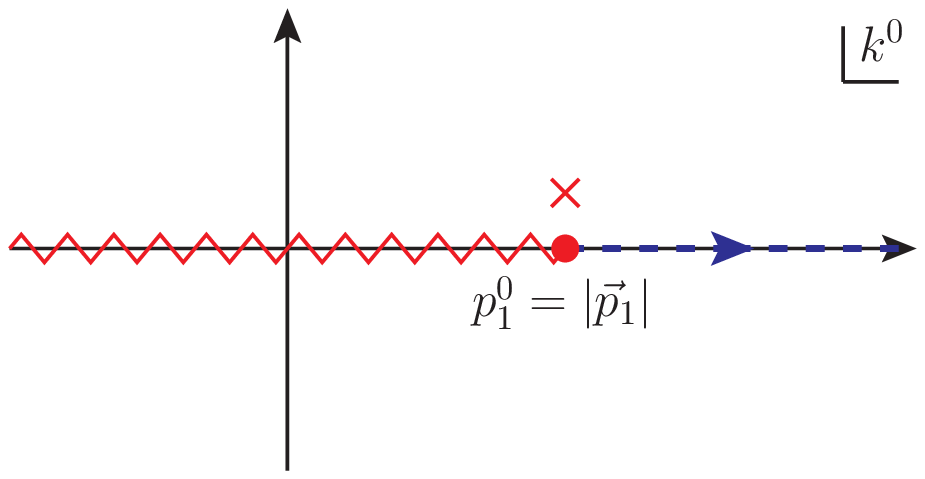}
	\caption{Integration contour of eq.~\eqref{eq:k0integral:p1} in the complex $k^0$ plane.
	The integral is finite when $p_1$ is timelike ($p_1^0 > |\vec{p}_1|$, left-hand side)
	since the contour can be deformed away from the pole,
	and similarly finite when $p_1$ is spacelike.
	There is a singularity in the limit $p_1^0 \to |\vec{p}_1|$ (right-hand side) 
	since the pole approaches the end point of the integration contour.}
\label{fig:complexplane:p1}
\end{figure}%
Since the exponent $\Delta_\phi - \frac{d}{2} > -1$ by the unitarity bound, the term $(-k^2)^{\Delta_\phi - d/2}$ is integrable at the branch point $k^0 = |\vec{p}_1|$. Similarly, the integration contour can be deformed away from the pole at $k^0 = p_1^0 + i \epsilon$, so the denominator is integrable as well. Since the matrix element is finite, the integral is therefore convergent in general. The only exception is when the pole coincides with the branch point, $p_1^0 \approx |\vec{p}_1|$, that is when the momentum $p_1$ is lightlike, as illustrated in figure~\ref{fig:complexplane:p1}.
In this case we have%
\footnote{The simplest way to derive this result is to consider $p_1$ spacelike and change integration variable using $k^0=|\vec{p}_1| +(|\vec{p}_1|-p_1^0)q$.
In the limit $p_1^0 \to |\vec{p}_1|$ from below, we obtain the integral 
\be
\int_0^\infty dq \frac{q^{\Delta_\phi-\frac{d}{2}}}{1+q} = \frac{\pi}{ \sin\left[ \pi \left( \frac{d}{2} - \Delta_\phi \right) \right] }~.
\ee
}
\begin{equation}
	G_- \approx
	\frac{(p_1^2 - i \epsilon)^{\Delta_\phi - d/2}}
	{2 \sin\left[ \pi \left( \frac{d}{2} - \Delta_\phi \right) \right]}
	\langle 0 | T\{ \phi(x_2) \phi(x_3) \phi(x_4) \} | \phi(p_1) \rangle~.
\end{equation}
The integral $G_+$ has a similar singularity when $p_1^0 = -|\vec{p}_1|$, but it remains finite otherwise.
Therefore, one obtains the limit
\begin{equation}
\begin{aligned}
	\lim_{p_1^0 \to |\vec{p}_1|} (p_1^2 - i \epsilon)^{d/2 - \Delta_\phi}
	& i \int d^dx_1 \, e^{i p_1 \cdot x_1}
	\langle 0 | T\{ \phi(x_1) \phi(x_2) \phi(x_3) \phi(x_4) \} | 0 \rangle
	\\
	&= \frac{1}{2 \sin\left[ \pi \left( \frac{d}{2} - \Delta_\phi \right) \right]}
	\langle 0 | T\{ \phi(x_2) \phi(x_3) \phi(x_4) \} | \phi( \vec{p}_1) \rangle~,
\end{aligned}
\label{eq:LSZ:first}
\end{equation}
where the momentum eigenstate on the right-hand side is the state defined in eq.~\eqref{eq:state:limit}.
Similarly, in the opposite limit, one obtains
\begin{equation}
\begin{aligned}
	\!
	\lim_{p_1^0 \to -|\vec{p}_1|} (p_1^2 - i \epsilon)^{d/2 - \Delta_\phi}
	& i \int d^dx_1 \, e^{i p_1 \cdot x_1}
	\langle 0 | T\{ \phi(x_1) \phi(x_2) \phi(x_3) \phi(x_4) \} | 0 \rangle
	\\
	&= \frac{1}{2 \sin\left[ \pi \left( \frac{d}{2} - \Delta_\phi \right) \right]}
	\langle \phi(-\vec{p}_1) | T\{ \phi(x_2) \phi(x_3) \phi(x_4) \} | 0 \rangle~.
\end{aligned}
\label{eq:LSZ:first:opposite}
\end{equation}
These two equations are the equivalent of the LSZ reduction formula for conformal correlators: they state that the time-ordered correlator diverges in the ``on-shell'' limit $p_1^2 \to 0$, and that the coefficient of the divergence can be computed as a matrix element involving the ``scattering state'' \eqref{eq:state}.
Note that our derivation does not require $p_1$ to be timelike: it is possible to approach the limit $p_1^2 \to 0$ from the spacelike region.
The next step consists in iterating the procedure for a second operator, where we will see a more interesting structure emerge.

\subsection{LSZ for the second operator}
\label{sec:lsz:second}

\begin{figure}
	\includegraphics[width=0.48\linewidth]{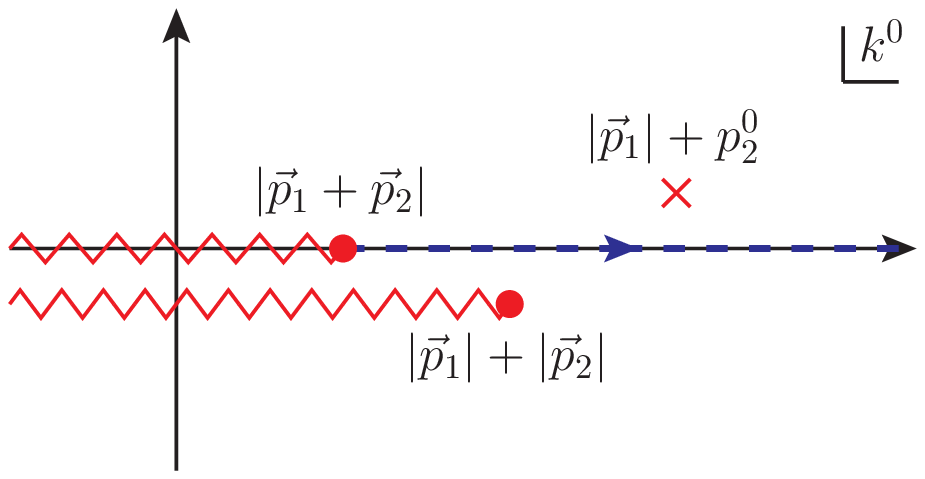}
	\hfill
	\includegraphics[width=0.48\linewidth]{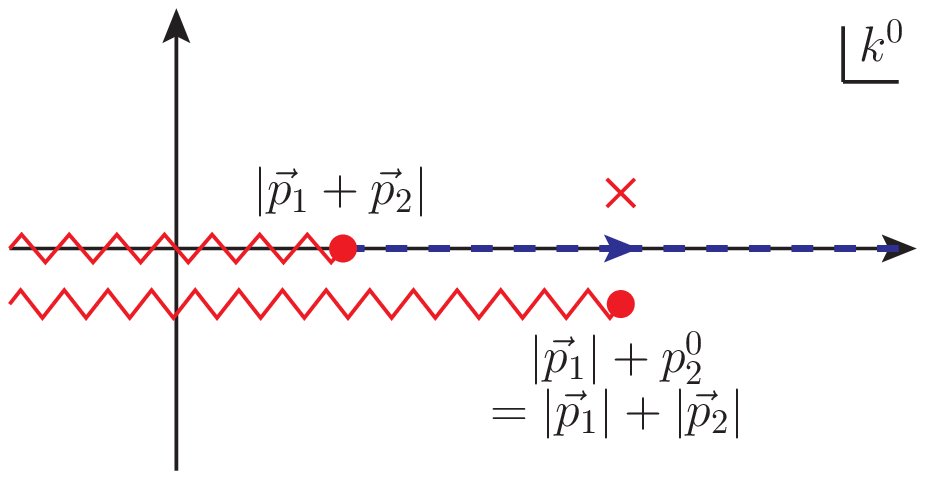}
	\caption{Integration contour of eq.~\eqref{eq:k0integral:p2} in the complex $k^0$ plane.
	The integral is finite when $p_2$ is timelike ($p_2^0 > |\vec{p}_2|$, left-hand side)
	since the contour can be deformed away from the pole.
	When $p_2$ is spacelike ($p_2^0 < |\vec{p}_2|$) the contour can also be deformed;
	it enters the branch cut but the integral remains finite.
	The singularity happens in the limit $p_2^0 \to |\vec{p}_2|$ (right-hand side),
	where the contour is pinched between the branch point and the pole.}
\label{fig:complexplane:p2}
\end{figure}

We now start from the result~\eqref{eq:LSZ:first} and Fourier transform another point of the correlation function, defining
\begin{equation}
	G(\vec{p}_1; p_2; x_3, x_4) = i \int d^dx_2 \, e^{i p_2 \cdot x_2} 
	\langle 0 | T\{ \phi(x_2) \phi(x_3) \phi(x_4) \} | \phi(\vec{p}_1) \rangle~.
\end{equation}
The idea is again to split the time integral into 3 regions and study their singularities. The region of intermediate time is regular as before. In the late time region $x_2^0 > x_3^0, x_4^0$ there is a unique singularity at $p_2^0 = -|\vec{p}_2|$, precisely as in the previous case.
But we are interested in studying the singularity around $p_2^0 = |\vec{p}_2|$, which must come from the early time region $x_2^0 < x_3^0, x_4^0$.
Defining
\begin{equation}
	G_-(\vec{p}_1; p_2; x_3, x_4) = i \int_{-\infty}^{t_-} dx_2^0 \int d^{d-1}\vec{x}_2 \, e^{i p_2 \cdot x_2}
	\langle 0 | T\{ \phi(x_3) \phi(x_4) \} \phi(x_2) | \phi(\vec{p}_1) \rangle~,
\end{equation}
where $t_- < x_3^0, x_4^0$, the use of the completeness relation \eqref{eq:completenessrelation} now gives rise to a genuine operator product expansion,
\begin{equation}
\begin{aligned}
	G_-(\vec{p}_1; p_2; x_3, x_4)
	&= \frac{1}{2\pi} \sum_{\mathcal{O}} \int\limits_{|\vec{p}_1 + \vec{p}_2|}^\infty \!\! dk^0 \,
	e^{i(k^0 - |\vec{p}_1| - p_2^0) t_-}
	\frac{(-k^2)^{\Delta - d/2}}{k^0 - |\vec{p}_1| - p_2^0 - i \epsilon}
	\Pi_{\alpha\alpha'}^{-1}(k) 
	\\
	& \quad \times
	\langle 0 | T\{ \phi(x_3) \phi(x_4) \} | \mathcal{O}^{\alpha'}(k) \rangle
	\langle \mathcal{O}^\alpha(k) | \phi(0) | \phi(\vec{p}_1) \rangle
	\Big|_{\vec{k} = \vec{p}_1 + \vec{p}_2}~.
\end{aligned}
\label{eq:k0integral:p2}
\end{equation}
where $\Delta$ now refers to the scaling dimensions of the intermediate operator $\mathcal{O}$.
The singularities of this integral depend on the matrix element $\langle \mathcal{O}^\alpha | \phi | \phi \rangle$, which in general is a complicated function of $k^0$~\cite{Bautista:2019qxj, Gillioz:2019lgs}.
Its computation is detailed in Appendix~\ref{sec:wightman3pt}. 
Importantly for us, it is a regular function of $k^0$, except for a singularity at $(k - p_1)^2 = 0$, around which
\begin{equation}
\begin{aligned}
	\langle \mathcal{O}^{\mu_1 \ldots \mu_\ell}(k) | \phi(0) | \phi(\vec{p}_1) \rangle
	&\approx
	\lambda_{\phi\phi\mathcal{O}}
	\frac{(2\pi)^{d+1}
	\Gamma\left( \frac{d}{2} - \Delta_\phi \right)
	(\Delta - 1)_\ell}
	{2^{2\Delta_\phi + \Delta + \ell - d}
	\Gamma\left( \Delta_\phi - \frac{d}{2} + 1 \right)
	\Gamma\left( \frac{\Delta + \ell}{2}  \right)^2}
	\\
	& \quad \times
	\left[ e^{i \pi (2\Delta_\phi - \Delta + \ell - 1)/2}
	\left( (k-p_1)^2 - i \epsilon \right)^{\Delta_\phi - d/2}
	+ \text{c.c.} \right]
	\\
	& \quad \times
	(-k^2)^{(d - 2\Delta_\phi - \Delta)/2}
	\mathcal{H}^{\mu_1 \ldots \mu_\ell}(k-p_1, p_1)~.
\end{aligned}
\label{eq:Wightman3pt:divergence}
\end{equation}
$\lambda_{\phi\phi\mathcal{O}}$ is the OPE coefficient%
\footnote{We use the convention of Eq.~\eqref{eq:positionspace3pt} for the 3-point function.}
and $\mathcal{H}$ is a dimensionless tensor structure defined in Eq.~\eqref{eq:Htensor}.
Note that only traceless symmetric tensors appear in the OPE of two scalars, so the generic spin index $\alpha$ has been replaced by $\ell$ Lorentz indices.
This matrix element gives rise to two branch point singularities at $k^0 = |\vec{p}_1| \pm |\vec{p}_2| \mp  i \epsilon$.
Assuming for simplicity of the argument that $\vec{p}_1$ and $\vec{p}_2$ are not collinear, the complex structure of the integrand is as in figure~\ref{fig:complexplane:p2}:
besides the integrable end-point singularity at $k^0 = |\vec{p}_1 + \vec{p}_2|$, there is a pole at $k^0 = |\vec{p}_1| + p_2^0 + i \epsilon$ and a branch cut at $k^0 = |\vec{p}_1| + |\vec{p}_2| - i \epsilon$ (the other branch cut at $k^0 = |\vec{p}_1| - |\vec{p}_2| + i \epsilon$ is not shown as it is to the left of the integration contour).
Taken separately, the pole and the branch cut are integrable, since the contour can be deformed around them.
However, when $p_2^0 \to |\vec{p}_2|$, there is a divergence given by
\begin{equation}
\begin{aligned}
	G_-(\vec{p}_1; p_2; x_3, x_4)
	&\approx \sum_{\mathcal{O}} \lambda_{\phi\phi\mathcal{O}}
	\frac{(2\pi)^{d+1}
	\Gamma\left( \frac{d}{2} - \Delta_\phi \right)
	(\Delta - 1)_\ell}
	{2^{2\Delta_\phi + \Delta + \ell - d}
	\Gamma\left( \Delta_\phi - \frac{d}{2} + 1 \right)
	\Gamma\left( \frac{\Delta + \ell}{2}  \right)^2}
	\\
	& \quad \times
	\frac{e^{i \pi (2\Delta_\phi - \Delta + \ell)/2}
	\left( p_2^2 - i \epsilon \right)^{\Delta_\phi - d/2}}
	{[-(p_1 + p_2)^2 ]^{\Delta_\phi - \Delta/2}}
	\Pi_{\mu_1 \ldots \mu_\ell, \nu_1 \ldots \nu_\ell}^{-1}(p_1 + p_2) 
	\\
	& \quad \times
	\mathcal{H}^{\nu_1 \ldots \nu_\ell}(p_2, p_1)
	\langle 0 | T\{ \phi(x_3) \phi(x_4) \} | \mathcal{O}^{\mu_1 \ldots \mu_\ell}(p_1 + p_2) \rangle~.
\end{aligned}
\end{equation}
Combining this with the result \eqref{eq:LSZ:first} of the previous section, one arrives at
\begin{equation}
\begin{aligned}
	& \left( \prod_{i=1}^2\lim_{p_i^0 \to |\vec{p}_i|}
	(p_i^2 - i \epsilon)^{d/2 - \Delta_\phi}
	i \int d^dx_i \, e^{i p_i \cdot x_i} \right)
	\langle 0 | T\{ \phi(x_1) \phi(x_2) \phi(x_3) \phi(x_4) \} | 0 \rangle
	\\
	& \qquad = \sum_{\mathcal{O}} \lambda_{\phi\phi\mathcal{O}}
	\frac{(2\pi)^{d/2-1}
	\Gamma\left( \frac{d}{2} - \Delta_\phi \right)^2
	\Gamma(\Delta + \ell)
	\Gamma\left( \Delta - \frac{d}{2} + 1 \right)}
	{2^{2\Delta_\phi - \Delta + \ell - d/2}
	\Gamma\left( \frac{\Delta + \ell}{2}  \right)^2}
	\frac{e^{i \pi (2\Delta_\phi - \Delta + \ell)/2}}
	{[-(p_1 + p_2)^2 ]^{\Delta_\phi - \Delta/2}}
	\\
	& \quad\qquad\qquad \times
	\tilde{\mathcal{H}}_{\mu_1 \ldots \mu_\ell}(p_2, p_1)
	\langle 0 | T\{ \phi(x_3) \phi(x_4) \} | \mathcal{O}^{\mu_1 \ldots \mu_\ell}(p_1 + p_2) \rangle~.
\end{aligned}
\label{eq:LSZ:second}
\end{equation}
Note that we have used the definition \eqref{eq:H:shadow} of the shadow transform $\tilde{\mathcal{H}}$ of the tensor $\mathcal{H}$ to simplify the notation.
This equation is symmetric under the exchange of $p_1$ and $p_2$, even though we have derived it in two steps where the symmetry was not at all obvious.
An  interesting feature of this result is the phase that vanishes when the operator $\mathcal{O}$ has the scaling dimension of a double-trace operator $\Delta = 2 \Delta_\phi + l + 2n$.

At this stage the right-hand side of Eq.~\eqref{eq:LSZ:second} is a sum of 3-point functions that can in principle be computed directly. However, it is even more convenient to use once more the tool that we have now used twice.

\subsection{LSZ for the third operator and form factor}
\label{sec:lsz:third}

We proceed and take the Fourier transform of Eq.~\eqref{eq:LSZ:second} with respect to $x_3$. When we send $p_3^0 \to -|\vec{p}_3|$, there is a divergence arising in the late-time region.%
\footnote{Note that the opposite limit $p_3^0 \to |\vec{p}_3|$ is also interesting, but it is also more complicated as it involves studying the limit $p_3^2 \to 0$ of the Wightman 3-point function $\langle \phi(k) | \phi(p_3) | \mathcal{O}(p_1 + p_2) \rangle$ in which neither of the momenta $k$ or $p_1 + p_2$ is lightlike.
 }
In this case, eq.~\eqref{eq:LSZ:first:opposite} can be used to write
\begin{equation}
\begin{aligned}
	\prod_{i=1}^3 & \left( \lim_{p_i^2 \to 0}
	(p_i^2 - i \epsilon)^{d/2 - \Delta_\phi}
	i \int d^dx_i \, e^{i p_i \cdot x_i} \right)
	\langle 0 | T\{ \phi(x_1) \phi(x_2) \phi(x_3) \phi(x_4) \} | 0 \rangle
	\\
	&= \sum_{\mathcal{O}} \lambda_{\phi\phi\mathcal{O}}
	\frac{(2\pi)^{d/2-2}
	\Gamma\left( \frac{d}{2} - \Delta_\phi \right)^3
	\Gamma\left( \Delta_\phi - \frac{d}{2} + 1 \right)
	\Gamma(\Delta + \ell)
	\Gamma\left( \Delta - \frac{d}{2} + 1 \right)}
	{2^{2\Delta_\phi - \Delta + \ell - d/2}
	\Gamma\left( \frac{\Delta + \ell}{2}  \right)^2}
	\\
	& \quad\qquad \times
	\frac{e^{i \pi (2\Delta_\phi - \Delta + \ell)/2}}
	{[-(p_1 + p_2)^2 ]^{\Delta_\phi - \Delta/2}}
	\tilde{\mathcal{H}}_{\mu_1 \ldots \mu_\ell}(p_2, p_1)
	\langle \phi(\vec{p}_3) | \phi(x_4) | \mathcal{O}^{\mu_1 \ldots \mu_\ell}(p_1 + p_2) \rangle~.
\end{aligned}
\label{eq:LSZ:third}
\end{equation}
The dependence on $x_4$ is now trivial: if we Fourier transform the last point, we simply recover the $\delta$-function imposing momentum conservation. 
Alternatively, if we use translation invariance to set the point $x_4 = 0$, this equation defines an object that depends on three null-momenta $p_1$, $p_2$ or $p_3$, or equivalently on the Mandelstam invariants $s$, $t$ and $u$ that we define as
\begin{equation}
	s = -(p_1 + p_2)^2,
	\qquad
	t = -(p_1 + p_3)^2,
	\qquad
	u = -(p_2 + p_3)^2~.
	\label{eq:stu}
\end{equation}
This is precisely the Lorentzian analogue of the form factor $F(s,t,u)$ defined in \eqref{def:FormFactor},
\begin{equation}
	F(s,t,u) \equiv \prod_{i=1}^3 \left( \lim_{p_i^2 \to 0}
	(p_i^2 - i \epsilon)^{d/2 - \Delta_\phi}
	i \int d^dx_i \, e^{i p_i \cdot x_i} \right)
	\langle 0 | T\{ \phi(x_1) \phi(x_2) \phi(x_3) \phi(0) \} | 0 \rangle~.
	\label{eq:F:Lorentzian}
\end{equation}
The two definitions are related by a Wick rotation, and they only differ in the range of their arguments:
while in Euclidean signature $s, t, u < 0$, the Lorentzian LSZ reduction requires $s > 0$.
We shall see that the latter definition leads to an expression that is analytic in $s$, and we will therefore be able to identify the two definitions.

It is convenient to introduce the notation
\begin{equation}
	q^2 = - s - t - u~,
	\label{eq:q2}
\end{equation}
which correspond to the squared momentum of the last operator, as well as the scattering angle
\begin{equation}
	\cos\theta = \frac{u - t}{u + t}~,
	\label{eq:costheta}
\end{equation}
and the dimensionless quantity
\begin{equation}
	w = -\frac{s}{q^2}~.
	\label{eq:w}
\end{equation}
The form factor can then be viewed a function of $q^2$, $w$ and $\cos\theta$. Since only $q^2$ is dimensionful and the overall scaling dimension of the form factor is fixed, it is really a function of the two dimensionless variables $w$ and $\cos\theta$.
In other words, eq.~\eqref{eq:LSZ:third} states that the form factor \eqref{eq:F:Lorentzian} admits a partial wave expansion of the form
\begin{equation}
	F(s, t, u) = (q^2)^{-\Delta_\phi}
	\sum_{\mathcal{O}} \lambda_{\phi\phi\mathcal{O}}^2
	F_{\Delta,\ell}\left( w, \cos\theta \right)~.
	\label{formFactorDecomp}
\end{equation}
The conformal partial waves $F_{\Delta, \ell}$ are defined by the right-hand side of eq.~\eqref{eq:LSZ:third}.
They are expressed in terms of a 3-point function for which we have an explicit expression given in Appendix~\ref{sec:wightman3pt}. However, it is quite impractical to contract the tensor indices by hand.
Instead, one can make use of the constraints imposed by conformal symmetry:
using a method based on the quadratic conformal Casimir operator, we show in Appendix~\ref{sec:casimir} that the partial waves are fixed by symmetry up to an overall multiplicative factor. This factor can then be obtained from eq.~\eqref{eq:LSZ:third} taking a convenient limit: we send $s/q^2 \to 0$ and keep only the leading term in $\cos\theta$, that is $(\cos\theta)^\ell$ in the block of spin $\ell$. In this limit the two-point tensor $\Pi^{\mu_1 \ldots \mu_\ell, \nu_1 \ldots \nu_\ell}$ is directly given by eq.~\eqref{eq:Pi:0}. It is proportional to the identity and can be trivially inverted.
Trace terms in $\mathcal{H}^{\nu_1 \ldots \nu_\ell}$ are subdominant in the limit $s \to 0$, and the Wightman 3-point function is dominated by in a single term at leading order in $(\cos\theta)^{-1}$, given by eq.~\eqref{eq:Wightman3pt:result}:
\begin{equation}
\begin{aligned}
	& \langle \phi(\vec{p}_3) | \phi(0) | \mathcal{O}^{\mu_1 \ldots \mu_\ell} (p_1 + p_2) \rangle
	\\
	& \quad 
	\approx i^\ell \,
	\frac{(2\pi)^{d+2} 2^{d - 2\Delta_\phi - \Delta}
	\left( \Delta - 1 \right)_\ell}
	{\Gamma\left( \Delta_\phi - \frac{d}{2} + 1 \right)
	\Gamma\left( \Delta - \frac{d}{2} + 1 \right)
	\Gamma\left( \Delta_\phi - \frac{\Delta - \ell}{2} \right)
	\Gamma\left( \frac{\Delta + \ell}{2} \right)}
	\frac{p_3^{\mu_1} \cdots p_3^{\mu_\ell}}{(q^2)^{(\Delta + \ell)/2}}~.
\end{aligned}
\end{equation}
Putting the pieces together, we arrive at
\begin{equation}
\begin{aligned}
	F_{\Delta,\ell}( w, \cos\theta) &=
	\frac{(4\pi)^{3d/2}
	\Gamma\left( \frac{d}{2} - \Delta_\phi \right)^3
	\Gamma( \Delta + \ell)
	\left( \Delta - 1 \right)_\ell}
	{2^{4\Delta_\phi + 2\ell}
	\Gamma\left( \frac{\Delta + \ell}{2}  \right)^3
	\Gamma\left( \Delta_\phi - \frac{\Delta - \ell}{2} \right)}
	\\
	& \quad \times
	(w - i \epsilon)^{(\Delta - \ell - 2\Delta_\phi)/2} 
	(\cos \theta)^\ell 
	\left[ 1 + \mathcal{O}( w, \cos\theta^{-1} ) \right]~.
\end{aligned}
\label{eq:F:zcosthetalimit}
\end{equation}
Note that we have absorbed the phase into the power of $w$ with the addition of an infinitesimal $i \epsilon$, since $w$ is negative: in taking the limit $s \to 0_+$ we force $q$ to be spacelike, i.e.~$q^2 > 0$, and therefore $-s/q^2 < 0$.

When combining this limit with the computations of Appendix~\ref{sec:casimir}, we obtain the final result for the partial wave expansion of the form factor,
\begin{equation}
\begin{aligned}
	F_{\Delta, \ell}(w, \cos\theta) &= 
	\frac{(4\pi)^{3d/2} \Gamma\left( \frac{d}{2} - \Delta_\phi \right)^3
	\Gamma( \Delta + \ell)
	\left( \Delta - 1 \right)_\ell}
	{2^{4\Delta_\phi + 2\ell}
	\Gamma\left( \frac{\Delta + \ell}{2}  \right)^3
	\Gamma\left( \Delta_\phi - \frac{\Delta - \ell}{2} \right)}
	\\
	& \quad \times
	(w - i \epsilon)^{(\Delta - \ell - 2 \Delta_\phi)/2} (1 - w)^{1 - \Delta_\phi}
	\frac{\ell!}{2^\ell \left( \frac{d - 2}{2} \right)_\ell}
	\sum_{n = 0}^{\ell/2}
	\frac{\left( \frac{d - 2}{2} \right)_n \left( \frac{d - \Delta + \ell - 1}{2} \right)_n}
	{n! \left( \frac{3 - \Delta - \ell}{2} \right)_n}
	\\
	& \quad\qquad \times
	\GeneralizedHypergeometric{\frac{\Delta - \ell - d + 3 + 2n}{2}}
	{\frac{\Delta - \ell - d + 2}{2}}{\frac{\Delta - \ell - 2\Delta_\phi + 2}{2}}
	{\frac{\Delta - \ell - d + 3 - 2n}{2}}{\Delta - \frac{d}{2} + 1}{w}
	\mathcal{C}_{\ell - 2n}^{d/2 - 1 + n}(\cos\theta)~,
\end{aligned}
\label{eq:F:conformalblock}
\end{equation}
given in terms of a generalized hypergeometric function ${}_3F_2$ in $w$ and of a Gegenbauer polynomial $\mathcal{C}_j^\alpha$ in $\cos\theta$.
Each individual conformal partial wave has zeros when $\Delta - \ell = \Delta_\phi + 2n$ with $n \in \mathbb{N}$ due to the presence of $\Gamma\left( \Delta_\phi - \frac{\Delta - \ell}{2} \right)$ in the denominator of eq.~\eqref{eq:F:conformalblock}. This is the twist of double-trace operators, which means in particular that the form factor is identically vanishing in (generalized) free field theory.
The form of the conformal partial wave suggests moreover that $F$ can be analytically continued to the unphysical regime of $s < 0$ by taking $w > 0$,
and that the form factor is real in this case. This is consistent with our expectations discussed in the introduction and the Mellin representation of section~\ref{sec:mellin}.
On the other hand, the regime of timelike $q$ ($q^2 > 0$) is not accessible from this expression, because it involves continuing $w$ past infinity. The limit $w \to -\infty$ does exist, as we will see next, but there is additional non-analyticity there.

\subsection{LSZ for the fourth operator and amplitude}
\label{sec:lsz:fourth}

To complete the LSZ reduction procedure and obtain the amplitude defined in eq.~\eqref{def:amp}, it only remains to consider the limit $p_4^2 \to 0$ in eq.~\eqref{eq:LSZ:third}, after Fourier transforming the point $x_4$.
However, the result depends on whether this limit is taken with $p_4$ timelike or spacelike:
when taken as a function of the complex variable $p_4^2$, the Wightman 3-point function that appears in eq.~\eqref{eq:LSZ:third} is non-analytic at $p_4^2 = 0$. This can also be understood from the form factor, which does not only have a branch cut along the Lorentzian regime $w \in (-\infty, 0]$, but also along $w \in [1, \infty)$: the two branch cuts meet at $|w| \to \infty$, where the form factor is non-analytic in $w$.

We define the amplitude $A(s,t,u)$ as the limit $p_4^2 \to 0_+$ of the form factor, i.e.~with space-like $p_4$, or equivalently $w \to -\infty$. In this way, two important properties of a scattering amplitude are inherited from the form factor:  these are the triviality of the amplitude in free field theory, and the positivity of its imaginary part.
From the Fourier transform of eq.~\eqref{eq:LSZ:third} and the results of appendix~\ref{sec:wightman3pt}, one obtains the partial wave expansion
\begin{equation}
	A(s,t,u) = s^{d/2 - 2\Delta_\phi} \sum_{\mathcal{O}} \lambda^2_{\phi\phi\mathcal{O}}
	A_{\Delta, \ell}(\cos\theta)~,
	\label{eq:A:expansion}
\end{equation}
where
\begin{equation}
\begin{aligned}
	A_{\Delta, \ell}(\cos\theta) &= i \,
	\frac{(4\pi)^{3d/2 - 1}
	\left[ \Gamma\left( \tfrac{d}{2} - \Delta_\phi \right) \right]^4
	\Gamma(\Delta + \ell)
	\Gamma\left( \Delta - \frac{d}{2} + 1 \right) (\Delta - 1)_\ell}
	{2^{4\Delta_\phi + 2\ell - 1}
	\left[ \Gamma\left( \frac{\Delta + \ell}{2}  \right) \right]^4}\\
	& \quad \times	
	\left[ 1 - e^{i \pi (2\Delta_\phi  - \Delta + \ell)} \right]	
	g_{\Delta, \ell}(\cos\theta)~.
\end{aligned}
\label{eq:A}
\end{equation}
and we have denoted the contraction of the two tensors
\begin{equation}
	g_{\Delta, \ell}(\cos\theta) \equiv \mathcal{H}^{\mu_1 \ldots \mu_\ell *}(p_3, p_4)
	\tilde{\mathcal{H}}_{\mu_1 \ldots \mu_\ell}(p_2, p_1)~,
\end{equation}
which is understood in the limit of all $p_i^2 \to 0$.
Since $\mathcal{H}$ and $\tilde{\mathcal{H}}$ are dimensionless tensors, $g_{\Delta, \ell}$ is a function of the scattering angle $\cos\theta$ only and not of the center-of-mass energy $s$, which we have factored out of the expansion~\eqref{eq:A:expansion}.
The $g_{\Delta, \ell}$ are polynomials of degree $\ell$ in $\cos\theta$, whose explicit form is found to be
\begin{equation}
	g_{\Delta, \ell}(\cos\theta) = \sum_{n=0}^{\ell/2}
	\frac{(-1)^n \ell!\left[ \left( \frac{\Delta - \ell - d + 2}{2} \right)_n \right]^2
	\mathcal{C}_{\ell - 2n}^{(\Delta - \ell - 1)/2 + n}(\cos\theta)}
	{2^\ell n! \left( 2 - \frac{d}{2} - \ell \right)_n
	\left( \frac{3 - \Delta - \ell}{2} \right)_n
	\left( \frac{3 - d + \Delta - \ell}{2} \right)_n
	\left( \frac{\Delta - \ell + 2n - 1}{2} \right)_{\ell - 2n}}~.
\label{eq:g}
\end{equation}
It can be verified that this result matches with the limit $w \to -\infty$ of the $F_{\Delta,\ell}(w, \cos\theta)$ (see appendix~\ref{sec:casimir:amplitude}).
In fact, the polynomials $g_{\Delta,\ell}$ coincide with the polynomials of Ref.~\cite{Gillioz:2018mto}, although the representation \eqref{eq:g} is new. They were found to have several remarkable properties:
they interpolate between $\text{SO}(d-1)$ partial waves proportional to $\mathcal{C}_\ell^{(d - 3)/2}(\cos\theta)$ at the unitarity bound $\Delta - \ell = d - 2$, and $\text{SO}(d)$ partial waves proportional to $\mathcal{C}_\ell^{(d - 2)/2}(\cos\theta)$ in the limit of large $\Delta$.
Moreover, they are positive in the forward and backward scattering limits $\cos\theta = \pm 1$,
which implies that the imaginary part of $A_{\Delta, \ell}$ is itself positive in the same limit.
This is consistent with the interpretation of $A$ as a scattering amplitude in CFT.
As with the form factor, the amplitude also vanishes in generalized free field theory, or when all exchanged operators have $\Delta - \ell = 2\Delta_\phi + 2n$.

\subsection{Convergence of the partial wave expansion}
\label{sec:LSZ:convergence}

The partial wave expansions of the scattering amplitude and the form factor rely on the use of the momentum-space completeness relation \eqref{eq:completenessrelation}.
We would like now to examine the convergence properties of these expansions without making references to a particular theory. Several specific examples are discussed instead in section~\ref{sec:examples}.

It is in general difficult to make precise statements about the spectrum of primary operators in CFT at large scaling dimension and/or spin. One of the best known property of this spectrum is the organizational principle into Regge trajectories \cite{Caron-Huot:2017vep}.  
Our first step is therefore to study the convergence of the expansion for the form factor along a single Regge trajectory. 
It is well known for instance that the OPE $\phi \times \phi$ always contains a tower of operators denoted by $[\phi\partial^\ell\phi]$ with $\ell = 0, 2, 4, \ldots$, whose twist approaches asymptotically the double-trace dimension $2\Delta_\phi$ as~\cite{Fitzpatrick:2012yx, Komargodski:2012ek}
\begin{equation}
	\tau \equiv \Delta - \ell \stackrel{\ell \to \infty}{\approx} 2 \Delta_\phi + \frac{c}{\ell^{\tau_\text{min}}}~,
\end{equation}
where $\tau_\text{min}$ is the lowest twist in that OPE. Their OPE coefficients tend to the generalized free field theory value satisfying \cite{Fitzpatrick:2011dm}
\begin{equation}
	\lambda_{\phi\phi[\phi\partial^\ell\phi]}^2 \propto \frac{\ell^{2\Delta_\phi -3/2}}{2^\ell}~.
\end{equation}
For the form factor, the asymptotic behavior of the conformal blocks $F_{\Delta, \ell}$ at fixed twist and large spin is discussed in Appendix~\ref{sec:convergence}. Away from the forward and backwards limits $\cos\theta = \pm 1$, we have
\begin{equation}
	F_{\Delta, \ell}(w, \cos\theta)
	\propto 
	\frac{(-1)^{\ell/2}2^\ell \ell^{1-\tau/2}}{\Gamma\left( \Delta_\phi - \frac{\tau}{2} \right)}
	\cos\left[ \left( \theta - \tfrac{\pi}{2} \right) \left( \ell + \tfrac{\tau - 1}{2} \right) \right]~,
\label{eq:F:asymptotics}
\end{equation}
and therefore
\begin{equation}
	\lambda_{\phi\phi[\phi\partial^\ell\phi]}^2 F_{\Delta, \ell}(w, \cos\theta) \propto
	\frac{(-1)^{\ell/2} 
	\cos\left[ \left( \theta - \tfrac{\pi}{2} \right) \left( \ell + \frac{\tau - 1}{2} \right) \right]}
	{\ell^{\tau_\text{min}-\Delta_\phi + 1/2}}~.
\label{eq:Reggetrajectory:asymptotics}
\end{equation}
Several lessons can be learned from the series defined by this Regge trajectory:
\begin{enumerate}

\item
First of all, the convergence of the series is at best power-like, and not exponential as in the usual OPE in Euclidean position space.

\item
Moreover, the series defined by this leading Regge trajectory is only absolutely convergent if $\Delta_\phi < \tau_\text{min} - \frac{1}{2}$. This condition can easily be violated in practice: the simplest example is the operator $\sigma$ in the $3d$ Ising model, which has $\Delta_\sigma \cong 0.52$ and the energy-momentum tensor with $\tau_\text{min} = 1$ as lowest-twist operator in its OPE.
However, this lack of absolute convergence does not necessarily mean that the series diverges, as it is oscillatory. We shall see in section~\ref{sec:ising} that the CFT data of the $3d$ Ising model actually gives rise to a convergent expansion.

\item
The partial wave expansion can only be convergent at real scattering angle: if one tries to analytically continue $\theta$ to the complex plane, the cosine in eq.~\eqref{eq:Reggetrajectory:asymptotics} turns into an exponential growing with $\ell$,
\begin{equation}
	F_{\Delta, \ell}(w, \cos\theta) \propto
	e^{|\Im \theta| \ell}~,
\end{equation}
and the series diverges, no matter how small the imaginary part is.
In the limiting case $\cos\theta = \pm 1$ corresponding to forward or backward scattering, the asymptotic behavior \eqref{eq:Reggetrajectory:asymptotics} is replaced by
\begin{equation}
	\lambda_{\phi\phi[\phi\partial^\ell\phi]}^2 F_{\Delta, \ell}(w, \cos\theta = \pm 1) \propto
	\frac{1}{\ell^{\tau_\text{min}-2\Delta_\phi + 1}}~.
\end{equation}
The convergence of this series is generically worse than for $|\cos\theta| < 1$. Since all the terms contribute with the same sign, it is for instance divergent in the $3d$ Ising model.

\end{enumerate}
In summary, the study of the leading Regge trajectory indicates that the conformal partial wave decomposition of the form factor converges at most conditionally, i.e.~not absolutely. The region of convergence is limited to real scattering angles, excluding the forward and backward scattering limits, but it is independent of the kinematic variable $w$. The same conclusions apply therefore to the amplitude, as well as to the analytic continuation of the form factor in the complex $w$ plane.

The other case that can be examined conclusively is the asymptotic behavior of the conformal partial wave expansion at large $\Delta$ and at fixed spin $\ell$. An asymptotic limit for the integrated density of OPE coefficients has been worked out in \cite{Mukhametzhanov:2018zja}.
Defining the (renormalized) density of OPE coefficients
\begin{equation}
	\rho_\ell(\Delta') \equiv
	\sum_{\mathcal{O}} \frac{\lambda_{\phi\phi \mathcal{O}}^2 }{K_{\ell,\Delta}} \,
	\delta(\Delta-\Delta')~,
\end{equation}
where the sum is over all primary operators of a given spin $\ell$, and
\begin{equation}
	K_{\ell, \Delta}=\frac{\Gamma(\Delta-1)\Gamma\left(\frac{\Delta+\ell}{2}\right)^4}
	{2\pi^2 \Gamma \left(\Delta-\frac{d}{2}\right) \Gamma(\Delta+\ell)\Gamma(\Delta+\ell-1)}
	\stackrel{\Delta \to \infty}{\propto} 
	4^{-\Delta} \Delta^{(d-2)/2}~,
\label{KlDelta}
\end{equation}
the integrated density satisfies%
\footnote{Eq.~\eqref{largeDeltaMZ} is in fact only proven for spin $\ell > 1$.}
\begin{equation}
	R_\ell(\Delta)=\int_0^\Delta d\Delta' \rho_\ell(\Delta')
	\stackrel{\Delta \to \infty}{\propto} \Delta^{4\Delta_\phi- 2d + 2}~, \qquad \Delta_\phi > \frac{d-1}{2}~.
\label{largeDeltaMZ}
\end{equation}
The restriction on the scaling dimension $\Delta_\phi$ will be removed momentarily.
The contribution of all the primaries of spin $\ell$ and scaling dimension $\Delta < \Delta_*$ to the form factor is then
\begin{equation}
	\sum_{\substack{\Delta < \Delta_* \\ \ell~\text{fixed}}} \lambda_{\phi\phi\mathcal{O}}^2
	F_{\Delta,\ell}(w,\cos\theta)
	= \int_0^{\Delta_*} d\Delta \, \rho_\ell(\Delta) K_{\ell, \Delta} F_{\Delta, \ell}(w, \cos\theta)~.
\label{eq:largeDelta:partialsums}
\end{equation}

In order to use the asymptotics \eqref{largeDeltaMZ}, we simply rewrite the previous equation in terms of $R_\ell(\Delta)$ by means of an integration by parts:
\begin{multline}
\sum_{\substack{\Delta < \Delta_* \\ \ell~\text{fixed}}} \lambda_{\phi\phi\mathcal{O}}^2
	F_{\Delta,\ell}(w,\cos\theta) \\
	= R_\ell(\Delta_*) K_{\ell, \Delta_*} F_{\Delta_*,\ell}(w,\cos\theta) - \int_0^{\Delta_*}\! d\Delta\,
	 R_\ell(\Delta)
	\frac{d}{d\Delta} \left(K_{\ell, \Delta}F_{\Delta,\ell}(w,\cos\theta)\right)~.
\end{multline}
In appendix~\ref{sec:convergence}, we derive the following asymptotic form for the conformal partial wave:
\begin{equation}
	F_{\Delta,\ell}(w,\cos\theta) \stackrel{\Delta \to \infty}{\propto} 
	4^\Delta \Delta^{3/2-\Delta_\phi} \left(\frac{\sqrt{w}}{1+\sqrt{1-w}}\right)^\Delta
	\sin \left[\frac{\pi}{2}(2\Delta_\phi-\Delta+\ell)\right]~.
\end{equation}
Notice that 
\begin{equation}
	\left|\frac{\sqrt{w}}{1+\sqrt{1-w}}\right|< 1
	\qquad \Leftrightarrow \qquad w \in \mathbb{C} / [1, \infty)~,
	\label{fixedSpinOPE}
\end{equation}
with equality reached when $w\geq 1$. Therefore, the product $K_{\ell,\Delta} F_{\Delta,\ell}(w,\cos\theta)$ decays exponentially unless $w \in [1,\infty)$. Since $R_\ell(\Delta)$ is polynomially bounded at large $\Delta$ -- see eq.~\eqref{largeDeltaMZ} -- we conclude that the conformal partial wave expansion at fixed spin converges at an exponential rate for $w$ away from $[1,\infty)$. It should be noticed that the estimate in eq.~\eqref{largeDeltaMZ} is valid only when $\Delta_\phi>(d-1)/2$, \emph{i.e.} when the exponent on the r.h.s.~is positive. In the opposite case, a similar estimate holds for the appropriate moments of the OPE density \cite{Mukhametzhanov:2018zja}:
\begin{subequations}
\begin{align}
	R^{(1)}_\ell(\Delta) &=\int_0^\Delta d\Delta' \Delta' \rho_\ell(\Delta')
	\stackrel{\Delta \to \infty}{\propto} \Delta^{4\Delta_\phi- 2d + 3}~,  & \frac{d}{2}-\frac{3}{4}<\Delta_\phi < \frac{d}{2}- \frac{1}{2}~, \\
	R^{(2)}_\ell(\Delta) &=\int_0^\Delta d\Delta' (\Delta')^2 \rho_\ell(\Delta')
	\stackrel{\Delta \to \infty}{\propto} \Delta^{4\Delta_\phi- 2d + 4}~,  &\frac{d}{2}-1<\Delta_\phi < \frac{d}{2}-\frac{3}{4}~.
\end{align}
\label{largeDeltaMZ_cases}
\end{subequations}
It is then trivial to see that the exponential convergence at fixed spin is valid for any unitary value of $\Delta_\phi$. It is sufficient to multiply and divide by $\Delta$ or $\Delta^2$ the integrand on the r.h.s.~of eq.~\eqref{eq:largeDelta:partialsums}, so that the estimates \eqref{largeDeltaMZ_cases} can be employed after an integration by parts.

Even if one can establish conditional convergence along a single Regge trajectory or for all primaries of a given spin, the overall convergence of the conformal partial wave expansion is not guaranteed. To establish such a convergence, one should appeal to 
bounds on the density of states at large $\Delta$ and arbitrary spin as in Refs.~\cite{Pappadopulo:2012jk, Rychkov:2015lca}. 
It turns out however that these bounds are too weak to obtain conclusive results for our form factor and amplitude.  
Nevertheless, there exists a physical argument for the distributional convergence of the conformal block expansion.
This argument is based on the observation that Wightman functions admit an OPE whose (distributional) convergence properties are independent of the Lorentzian ordering of the operators~\cite{Mack:1976pa, Kravchuk:2020scc}.
In our case, the amplitude and form factor can be expressed in terms of the state
\begin{equation}
	|\phi(\vec{p}_1) \phi(\vec{p}_2) \rangle \equiv
	\left( \prod_{j=1}^2
	\lim_{p_j^2 \to 0}
	(p_j^2)^{\frac{d}{2} - \Delta_\phi}
	\int_{-\infty}^{t_-} dx_j^0 \int d^{d-1}\vec{x}_j \, e^{i p_j \cdot x_j} \right)
	\phi(x_1) \phi(x_2)  | 0 \rangle~.
\label{eq:2particlestate}
\end{equation}
The LSZ discussion above shows that the matrix elements 
\begin{equation}
	\langle 0 | \Ocal(x_1)\dots \Ocal(x_n) |\phi(\vec{p}_1) \phi(\vec{p}_2) \rangle
\end{equation}
are finite, Lorentz covariant, independent of $t_-$ and symmetric under permutations of the momenta $\vec{p}_1$ and $\vec{p}_2$.
The form factor is given by
\begin{equation}
	F(s,t,u) \propto \langle \phi(\vec{p}_3) | \phi(x=0) |\phi(\vec{p}_1) \phi(\vec{p}_2) \rangle
\end{equation}
and the amplitude by
\begin{equation}
	A(s,t,u) \propto \langle \phi(\vec{p}_3) \phi(\vec{p}_4) |\phi(\vec{p}_1) \phi(\vec{p}_2) \rangle~.
\end{equation}
The conformal partial wave decomposition corresponds simply to using the Hilbert space completeness relation \eqref{eq:completenessrelation} in these Wightman functions.
The integrals in the definition of the state~\eqref{eq:2particlestate} do not spoil the distributional convergence of the OPE.%
\footnote{The sharp integration boundary at $x^0 = t_-$ could be replaced by a smooth cutoff without affecting the results.}
Therefore, one expects that the expansions for both $F$ and $A$ converge at least in the sense of distributions.%
\footnote{In fact, it was shown in $d = 2$ dimensions that there exists a kinematic range in which the momentum-space OPE does converge point-wise, and not only as a distribution~\cite{Gillioz:2019iye}. These results have not yet been generalized to $d > 2$ dimensions.}


\section{Mellin representation}
\label{sec:mellin}

This section is dedicated to an alternative derivation of the form factor and of the amplitude, whose starting point is the Mellin representation of the four-point function. This derivation does not require the insertion of a complete set of states, hence it establishes the existence of the form factor, independently of the convergence of its conformal block decomposition.   It further allows to establish a direct relation between the  amplitude and the Landau singularity depicted in figure~\ref{fig:landau}, namely eq.~\eqref{ALandau}. Through this relation, it is possible to match the partial waves \eqref{eq:A} to a certain limit of the conformal blocks in position space.  

\subsection{The form factor}

Consider the Mellin representation of the correlation function \cite{Mack:2009mi, Penedones:2010ue}  
\be
\langle \phi_1(x_1) \dots \phi_4(x_4) \rangle =
\int \frac{d\g_{12} d\g_{13}}{(2\pi i)^2} M(\gamma_{ij}) \prod_{i<j}^4 \frac{\Gamma(\g_{ij})}{(x_{ij}^2)^{\g_{ij}}}~,
\label{eq:Mellindef}
\ee
where the Mellin variables satisfy the constraints
\be
\sum_{j=1 \atop j\neq i}^4 \gamma_{ij}=\Delta_i~,
\ee
and the integration contours (over the two independent Mellin variables) run parallel to the imaginary axis.
More precisely, as explained in detail in \cite{Penedones:2019tng}, the integration contours are usually deformed to pass on the appropriate side of poles of the integrand.%
\footnote{For identical external operators $\phi_i =\phi$, there are also extra terms corresponding to disconnected contributions and possibly the sum of collinear blocks associated to the exchange of $\phi$ in the three OPE channels.  We expect  the latter to produce  similar extra terms  in \eqref{FFMellin}.}

In appendix~\ref{app:Mellin}, we show that Fourier transforming and taking the limit $p_j^2 \to 0$ for $j=1,2,3$ leads to the following Mellin representation of the form factor \eqref{def:FormFactor}, 
\be
F(s,t,u) = C \int [d\gamma] M(\gamma_{ij}) 
\frac{\Gamma(\g_{12})}{(-s)^{\gamma_{12}}}
\frac{\Gamma(\g_{13})}{(-t)^{\gamma_{13}}}
\frac{\Gamma(\g_{23})}{(-u)^{\gamma_{23}}}~,
\label{FFMellin}
\ee
where $s+t+u=-p_4^2$ and 
\be
C=\pi^{\frac{3d}{2}} 2^{3 d - \sum_{i=1}^4 \Delta_i}  
 \prod_{j=1}^3   \Gamma\left(\frac{d}{2}- \Delta_j \right)~. 
 \label{Cdef}
\ee
 
In the case of four identical operators $\phi$, we have $\gamma_{23}=\gamma_{14}$ and $\gamma_{12}+\gamma_{13}+\gamma_{14}=\Delta_\phi$. Moreover, the Mellin amplitude $M(\gamma_{12},\gamma_{13},\gamma_{14})$ is invariant under permutations
of its variables.
This representation makes crossing symmetry and analyticity manifest.
To see this we can choose the following straight contours in \eqref{FFMellin}:%
\footnote{As explained in \cite{Penedones:2019tng}, choosing a straight contour often leads to extra contributions from picking up poles as we deform the original contour. Luckily these are explicitly real for negative $s,t,u$.}
\beq
\gamma_{12}= \frac{\Delta_\phi}{3}+i x~,\qquad
\gamma_{13}= \frac{\Delta_\phi}{3}+i y~,\qquad
\gamma_{14}= \frac{\Delta_\phi}{3}-i x-iy~.
\eeq
With this choice and using $M(\gamma_{ij})^*= M ( \gamma_{ij}^*  )$, it is clear that
\beq
F(s,t,u)^*=F(s^*,t^*,u^*)~.
\eeq
Moreover, $F(s,t,u)$ is real for negative $s,t,u$ and it has branch cuts along the positive real axis of each Mandelstam variable.

One can also recover the conformal block expansion from the Mellin representation above.
Recall that the Mellin amplitude has poles associated to every exchanged operator 
\be
M \approx \lambda_{\phi\phi\mathcal{O}}^2 \frac{ \mathcal{Q}_{\ell,m}(-2\gamma_{13})}{
2\Delta_\phi -\Delta+\ell-2m-2\gamma_{12}}~,\qquad\qquad
m=0,1,2,\dots
\ee 
where $\Delta$ and $\ell$ are the scaling dimension and spin of the primary operator $\mathcal{O}$. The residue is proportional to the degree-$\ell$ Mack polynomial $\mathcal{Q}_{\ell,m}(-2\gamma_{13})$  defined in \cite{Costa:2012cb}. 
Performing the integral over $\gamma_{12}$ by picking up the OPE poles we obtain the conformal block expansion \eqref{CBexpansionF} with
\begin{align}
F_{\Delta,\ell} (w,\cos \theta) = 
 -C w^{ -\Delta_\phi}
 \sum_{m=0}^\infty &  \left(\frac{2w}{1-w}\right)^{\frac{\Delta-\ell}{2}+m}
 \Gamma\left(\Delta_\phi- \tfrac{\Delta-\ell}{2}-m \right) \times \\
 &\times
\int \frac{d\gamma_{13}}{4\pi i}    
\mathcal{Q}_{\ell,m}(-2\gamma_{13}) 
\frac{\Gamma(\g_{13})}{(1-\cos\theta)^{\gamma_{13}}}
\frac{\Gamma(\g_{23})}{(1+\cos\theta)^{\gamma_{23}}}~,
\nonumber
\end{align}
where $\gamma_{23} = \frac{\Delta-\ell}{2}+m -\gamma_{13}$.
We have explicitly checked that this agrees with  \eqref{eq:F:conformalblock}.
Notice that the integral in the second line is indeed a polynomial of degree $\ell$ in $\cos \theta$. This follows from the identity \eqref{eq:FeynmanId}.

\subsection{The amplitude from a position-space Landau singularity}

The amplitude \eqref{def:amp} can also be obtained from the Mellin representation of the correlation function.
It is sufficient to study the limit $p_4^2 \to 0$ of the form factor \eqref{FFMellin}.
This
is  subtle because we first need to analytically continue $s$ to positive values. We expect the function to have a branch point at the threshold $s=0$. We analytically continue to positive $s$ just above the branch cut, \emph{i.e.}~we take  $(-s)^{-\gamma_{12}} \to s^{-\gamma_{12}} e^{i\pi \gamma_{12}}$.
This implies that the Mellin integral in \eqref{FFMellin} is no longer exponentially suppressed for large and negative $\Im{\g_{12}}$. In fact, the Mellin integral is dominated by large values of $\g_{ij}$ when $p_4^2 \to 0$. To see this, parametrize the Mellin variables by $\zeta$ and $x$ as follows:
\be
\g_{12}= \frac{c}{3} -i \zeta~,\qquad
\g_{13}= \frac{c}{3} +i \zeta x~,\qquad
\g_{23}= \frac{c}{3} +i \zeta(1-x)~, 
\ee
where $c=\frac{1}{2}(\Delta_1+\D_2+\D_3-\D_4)$.
Let us compute the contribution from large values of $\zeta$.
We only need to know the Mellin amplitude in this scaling limit \cite{Penedones:2019tng}:
\be
M(\g_{ij}) \approx (\g_{12})^{\frac{d}{2}-\frac{1}{2}\sum\Delta_j} \tilde{M}\left(-\frac{\g_{13}}{\g_{12}}\right)~.
\label{scalingMellin}
\ee
For large $\zeta$ the integral over $x$ is dominated by a saddle point at $x=-t/s$. Performing this integral,  the integral over $\zeta$ simplifies and we find
\be
F(s,t,u)\approx C  e^{-i\pi(\frac{d}{2}-\frac{1}{2}\sum\Delta_j)}
s^{\frac{d}{2}-\frac{1}{2}\sum\Delta_j}
\tilde{M}\left(-\frac{t}{s}\right) (p_4^2)^{\Delta_4 -\frac{d}{2}} \Gamma\left(\frac{d}{2}- \Delta_4\right) 
\ee
in the limit $p_4^2 \to 0$.
For $\Delta_4 < \frac{d}{2}$ this contribution diverges. Hence, the  large $\zeta$ region is responsible for the amplitude  \eqref{def:amp}:
\beq
A(s,t,u)= 4^d\pi^{3d/2} e^{-i\pi(\frac{d}{2}-\frac{1}{2}\sum\Delta_j)}
(4s)^{\frac{d}{2}-\frac{1}{2}\sum\Delta_j} \prod_{j=1}^4   \Gamma\left(\frac{d}{2}- \Delta_j \right)\,\tilde{M}\left(-\frac{t}{s}\right)~.
\label{AMtilde}
\eeq

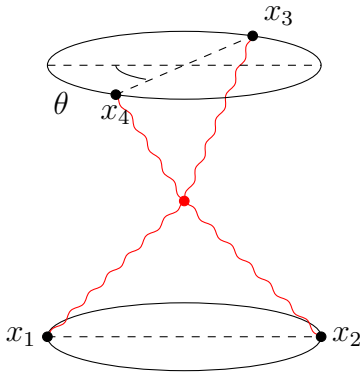
\begin{figure}
\centering
\begin{tikzpicture}[scale=1.8]
\draw (0,-1) ellipse (1 and 0.25);
\draw [red,decorate,decoration={snake,amplitude=0.8pt}] (-1,-1) -- (0,0);
\draw [red,decorate,decoration={snake,amplitude=0.8pt}] (1,-1) -- (0,0);
\draw [red,decorate,decoration={snake,amplitude=0.8pt}] (0,0) -- (0.5,1.2165);
\draw [red,decorate,decoration={snake,amplitude=0.8pt}] (0,0) -- (-0.5,0.7835);
\filldraw[red]  (0,0) circle [radius=1 pt];
\filldraw[black]  (-1,-1) circle [radius=1 pt] node[left] {$x_1$};
\filldraw[black]  (1,-1) circle [radius=1 pt]  node[right] {$x_2$};
\filldraw[black]  (-0.5,0.7835) circle [radius=1 pt] node[anchor=north] {$x_4$};
\filldraw[black]  (0.5,1.2165) circle [radius=1 pt] node[anchor=south west] {$x_3$};
\draw (0,1) ellipse (1 and 0.25);
\draw [dashed] (-0.5,0.7835) --(0.5,1.2165);
\draw [dashed] (-1,-1) -- (1,-1);
\draw [dashed] (-1,1) -- (1,1);
\draw (-0.5,1) arc [start angle=180, end angle=236, x radius=0.5, y radius = 0.125];
\node at (-0.9,0.72) {$\theta$};
\end{tikzpicture}
\caption{The configuration corresponding to the Landau singularity at $z=\bar{z}$. The red dot is lightlike separated from all the insertions. Notice that we can freely move the $x_i$ along the light-rays without affecting the cross-ratios, which are invariant under independent rescalings $x_i \to \lambda_i x_i$.}
\label{fig:landau}
\end{figure}

It is well known that the limit of large Mellin variables corresponds to a particular Landau singularity of the correlation function in position space \cite{Penedones:2010ue}.%
\footnote{Sometimes this sigularity is called the bulk point limit. However, this is only appropriate for $d=2$ \cite{Gary:2009ae, Maldacena:2015iua}. }
This singularity appears in Minkowski spacetime when the 4 external points are all lightlike-related to another point, with $x_{12}$ and $x_{34}$ spacelike and $x_3$ and $x_4$ in the future of $x_1$ and $x_2$, as shown in figure~\ref{fig:landau}.
From here on, for simplicity, we restrict to the case of four identical external operators with dimension $\Delta_\phi$.
Writing the correlation function in terms of the usual cross ratios
\be
\langle \phi(x_1) \dots \phi(x_4) \rangle = \frac{\mathcal{G}(z,\bar{z})}{(x_{12}^2 x_{34}^2)^{\Delta_\phi}}~, \qquad 
z \bar{z} = \frac{x_{12}^2x_{34}^2}{x_{13}^2x_{24}^2}~, \quad (1-z)(1- \bar{z}) = \frac{x_{14}^2x_{23}^2}{x_{13}^2x_{24}^2}~,
\label{4pointzzb}
\ee
the singular configuration corresponds to
\beq
z=\bar{z}=
\frac{2}{1-\cos\theta}~.
\label{zbulkpoint}
\eeq
Here, $\theta$ is the angle defined in figure~\ref{fig:landau}, and will only later be identified with the scattering angle in the amplitude. The corresponding singularity in the function $\mathcal{G}(z,\bar{z})$ can be reached as follows: Starting from a Euclidean configuration, where the four insertions lie on the same time slice, we move the points $x_3$ and $x_4$ forward in time, being careful to shift the insertions in Euclidean time so that the correlator is time-ordered. The path in cross-ratio space is equivalent to the following: In the initial configuration, $\bar{z}=z^*$. Then $\bar{z}$ is taken once around the origin -- not around 1 -- clockwise, while $z$ is held fixed. Finally $z$ and $\bar{z}$ reach the Landau point $z=\bar{z}$ on complex conjugated paths. The whole path can be chosen to lie within the region of convergence of the $(12)$ OPE, with the exception of the Landau point itself which lies at its boundary. Each block is singular in the $z \to \bar{z}$ limit, and generically the correlation function will share the same singularity \cite{Maldacena:2015iua}:
\be
\mathcal{G}^{\circlearrowright}(z,\bar{z}) \approx \frac{\mathcal{L}(z)}{\left(-(z-\bar{z})^2\right)^{\frac{d-3}{2}}} \qquad (z\to\bar{z})~,
\label{GaroundLandau}
\ee
where the symbol $\circlearrowright$ denotes the path in cross-ratio space just described.
The precise relation of the Landau singularity with the Mellin amplitude can be obtained again by means of the scaling limit \eqref{scalingMellin} \cite{Penedones:2010ue}.
We find\footnote{Notice that the relation between our $\mathcal{G}$ and the $\mathcal{A}$ defined in \cite{Penedones:2010ue} is $\mathcal{G}(z,\bar{z})=(z \bar{z})^{2\Delta_\phi}\mathcal{A}(z,\bar{z})$ \emph{in Euclidean signature}. Therefore the relation between the analytically continued functions contains an important additional phase: $\mathcal{G}^{\circlearrowright}(z,\bar{z})=e^{-2\pi i \Delta_\phi}(z \bar{z})^{2\Delta_\phi}\mathcal{A}^{\circlearrowright}(z,\bar{z})$.}
\beq
\mathcal{L}(z)= i \pi^{3/2} 
e^{-i\pi d/2} 
\Gamma\left(\frac{d-3}{2}\right)
 (2z)^{d-2}(z-1)^{d/2-2} \tilde{M}(1/z)~.
 \label{LMtilde}
\eeq
By comparing eqs. \eqref{AMtilde} and \eqref{LMtilde}, we conclude that the amplitude is completely determined by the position-space Landau singularity:
\begin{equation}
A(s,t,u) =\kappa\,
s^{\frac{d}{2}-2\Delta_\phi}(z-1)^{2-d/2}z^{2-d}\mathcal{L}(z)~, 
 \label{ALandau}
\end{equation}
where
\begin{equation}
\kappa = -ie^{2\pi i \Delta_\phi} 4^{d+1-2\Delta_\phi}  \,
\frac{\pi^{\frac{3}{2}(d-1)}\Gamma\left(\frac{d}{2}-\Delta_\phi\right)^4}{\Gamma\left(\frac{d-3}{2}\right)} \,,\qquad \qquad
	z=-\frac{s}{t}=\frac{2}{1-\cos\theta}~. \label{eq:kappa}
\end{equation}
The relation between the position-space cross-ratio and the Mandelstam variables nicely identifies the angle $\theta$ in figure~\ref{fig:landau} with the scattering angle in momentum space. Eq.~\eqref{ALandau} then provides an intuitive physical picture for the conformal scattering amplitude $A(s,t,u)$: the Landau diagram in figure~\ref{fig:landau} describes propagation of massless particles from the external points $x_1$ and $x_2$ towards the collision point, and from there to the other external points $x_3$ and $x_4$.

\subsection{Partials waves  from conformal blocks}

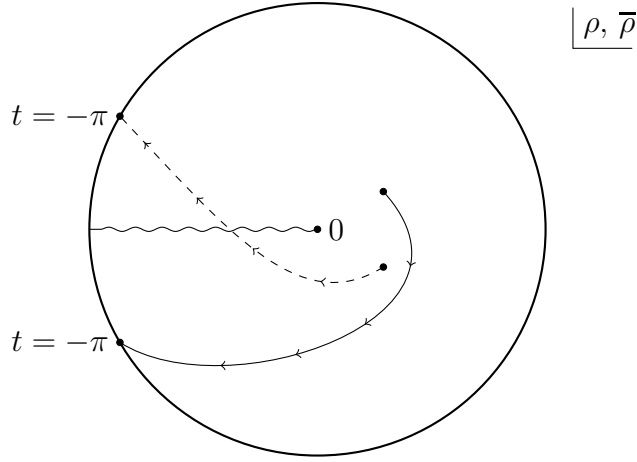
\begin{figure}
\centering
\begin{tikzpicture}[scale=1.2]
\draw [thick] (0,0) circle [radius=2.5];
\draw [decorate,decoration={snake,amplitude=0.8pt}] (0,0) -- (-2.5,0);
\filldraw [black] (0,0) circle [radius=1pt] node[right] {$0$};
\filldraw [black] (-2.165,-1.25) circle [radius=1pt] node[left] {$t=-\pi$};
\filldraw [black] (-2.165,1.25) circle [radius=1pt] node[left] {$t=-\pi$};
\filldraw [black] (0.722,0.417) circle [radius=1pt];
\filldraw [black] (0.722,-0.417) circle [radius=1pt];
\draw [dashed,postaction = {decoration={markings, mark=between positions 0.2 and 0.9 step 1cm with {\arrow{>};}},decorate}] (0.722,-0.417) .. controls (-0.2,-1) and (-1,0) .. (-2.165,1.25);
\draw [postaction = {decoration={markings, mark=between positions 0.2 and 0.9 step 1cm with {\arrow{>};}},decorate}](0.722,0.417) .. controls (2,-1) and (-1,-2) .. (-2.165,-1.25);
\draw [black] (3.45,2)-- (2.8,2) -- (2.8,2.45);
\node at (3.2,2.26) {$\rho,\, \bar{\rho}$};
\end{tikzpicture}
\caption{The paths in the $\rho$ and $\bar{\rho}$ planes needed to reach the Landau singularity starting from a Euclidean configuration. 
The path of $\rho = e^{i(t+ \theta)}$  is the continuous line, while the path of $\bar{\rho} =  e^{i(t-\theta)}$ is dashed. The time coordinate $t$ varies from $t=i \pi$ in the Euclidean regime to $t=-\pi$ in the Lorentzian regime.
The wavy line marks the position of the cut in each conformal block.}
\label{fig:rho_path}
\end{figure}

The relation \eqref{ALandau} between the amplitude and the position space Landau singularity allows us to obtain the partial wave expansion directly from the conformal block expansion of the position-space correlator. As we already emphasized, we can approach the Landau singularity from within the convergence region of the $s$-channel conformal block expansion. In this subsection, we would like to show that the expansion, evaluated at the Landau configuration, precisely reduces to the partial wave expansion \eqref{eq:A:expansion} of the amplitude. Our strategy will be to show that this correspondence is valid block by block. We shall find it convenient to use $\rho$-coordinates \cite{Hogervorst:2013sma}, whose relation to the $z,\, \bar{z}$ cross-ratios is
\beq
\rho=\frac{1-\sqrt{1-z}}{1+\sqrt{1-z}}~, \qquad z=\frac{4\rho}{(1+\rho)^2}~.
\label{ztorho}
\eeq
and similarly for $\bar{\rho}$. If we pose $\rho = e^{i(t+ \theta)}$ and $\bar{\rho} =  e^{i(t-\theta)}$, the bulk point limit is reached as $t\to -\pi $ for instance along the path $t=i\pi e^{i \alpha}$, $\alpha:\ 0 \to \pi/2$ -- see figure~\ref{fig:rho_path}. In the final configuration $\bar{\rho}=\rho^*$ and $|\rho|=1$, but along the path $\bar{\rho}$ is sent clockwise around the origin, before joining a complex conjugate path to $\rho$. Now, as we send $\bar{\rho}\to e^{-2\pi i}\bar{\rho}$, the conformal blocks for the four-point function of identical operators only change by an overall phase, to which we will come back at the end. Therefore, the limit we are interested in can be computed in Euclidean kinematics, where $\rho=\bar{\rho}^*=r e^{i(\theta+\pi)}$, $0<r<1$. We then introduce the usual shorthand notation $\eta=-\cos\theta$, and define the $s$-channel OPE expansion of the correlator in eq.~\eqref{4pointzzb} as follows:
\beq
\mathcal{G}(z,\bar{z})=\sum_{\mathcal{O}} \lambda^2_{\phi\phi \mathcal{O}} h_{\D,\ell}(r,\eta)~.
\eeq
We make the following ansatz for the expansion around $r=1$:
\beq
h_{\D,\ell}(r,\eta) = (1-r^2)^\alpha \sum_{n=0}^\infty h_n(\eta) 
(1-r^2)^n~,
\label{r1blockAnsatz}
\eeq
where we suppressed the dependence of $h_n$ on $\Delta$ and $\ell$. This ansatz is justified by the explicitly known cases $d=4$ and $d=6$ and by the large $\Delta$ limit.  By plugging the ansatz in the quadratic Casimir equation we find  two possible values
\beq
\alpha = 0~, \qquad \alpha=3-d~.
\eeq
The ansatz \eqref{r1blockAnsatz} needs therefore to be modified by allowing for two towers of terms, which will mix if we set $d$ to an integer. Since we are interested in the leading singularity as $r\to 1$, we set $\alpha=3-d$. The quadratic Casimir defines a recurrence relation for the $h_n$'s, and does not provide an equation for $h_0$. Instead, by combining the quadratic and quartic Casimir, and by expanding both up to fourth order around $r=1$, we do eventually find an equation for $h_0(\eta)$ which we report here: 
\begin{equation}
\begin{aligned}
	\left(\mathcal{C}_2 \left(\mathcal{C}_2+\frac{d^2}{2}-\frac{5}{2}d+2\right)+\frac{2 (d-2) (\mathcal{C}_2+d-3)}{1-\eta ^2}-\mathcal{C}_4-\frac{2 (d-3)^2}{\left(1-\eta ^2\right)^2}\right) h_0(\eta ) & \\
   -2\left( (d+1) (\mathcal{C}_2-2d)+\frac{2  \left(d^2-3 d+3\right)}{1-\eta ^2}\right)
   \eta\, h_0'(\eta ) & \\
   \left( 2(\mathcal{C}_2-(d+6) (d+1))(1-\eta^2)+2d (d+6)\right) h_0''(\eta ) & \\
   +4 (d+3)\eta  \left(\eta ^2-1\right)  h_0^{(3)}(\eta ) 
   +2 \left(1-\eta ^2\right)^2
   h_0^{(4)}(\eta ) &= 0~.
   \label{Casg0}
\end{aligned}
\end{equation}
$\mathcal{C}_2$ and $\mathcal{C}_4$ are the eigenvalues of the quadratic and quartic Casimir:
\begin{align}
\mathcal{C}_2 &= \Delta  (\Delta -d)+\ell  (d+\ell -2)~, \\
\mathcal{C}_4 &= \Delta ^2 (\Delta -d)^2 +  \frac{d (d-1)}{2} \Delta(\Delta -d)
\nonumber \\
& \quad +\ell ^2 (d+\ell -2)^2+\frac{1}{2} (d-4) (d-1)\, \ell  (d+\ell -2)~.
\end{align}
One can easily check that eq.~\eqref{Casg0} is solved by
\beq
h_0(\eta) \propto \frac{1}{(1-\eta^2)^{1/2}} g_{\D,\ell}(\eta)~,  
\label{g0sol}
\eeq
where $g_{\D,\ell}(\eta)$ are the polynomials defined in eq.~\eqref{eq:g}. More precisely, the prefactor in eq.~\eqref{g0sol} is motivated by the relation between the amplitude and the coefficient of the Landau singularity, eq.~\eqref{ALandau}. After extracting it, we expect the correct solution to be a polynomial in $\eta^2$, and eq.~\eqref{g0sol} is the unique solution with this feature. 
Furthermore, one can explicitly check that the 4d and 6d position-space blocks, which are known exactly, reduce to eq.~\eqref{g0sol} in the $r\to 1$ limit.

This nicely confirms the relationship between the amplitude and the Landau singularity. In principle, we would like to extract the proportionality coefficient in eq.~\eqref{g0sol} from eq.~\eqref{ALandau}.
However, we must be careful that the expansion of $\mathcal{L}(z)$ in conformal blocks is ambiguous. Indeed, if $\tilde{\mathcal{G}}(z,\bar{z})$ is a function which admits a conformal block expansion and is less singular than $(z-\bar{z})^{3-d}$ as $z\to \bar{z}$, the equality
\beq
\mathcal{L}(z) = \lim_{z\to \bar{z}}\left(-(z-\bar{z})^2\right)^\frac{d-3}{2} \left(\mathcal{G}^{\circlearrowright}(z,\bar{z})+\tilde{\mathcal{G}}(z,\bar{z})\right)~,\quad z>1~,
\label{LGtilde}
\eeq 
yields a block decomposition of $\mathcal{L}$ which term by term depends on $\tilde{\mathcal{G}}$.   
 Since the momentum space blocks  \eqref{eq:A} vanish for double-trace operators, we can easily fix this ambiguity by choosing $\tilde{\mathcal{G}}=-e^{-2\pi i \Delta_\phi}\mathcal{G}$, i.e.~the Euclidean correlator. Indeed, $\mathcal{G}$ is obviously regular as $z$ and $\bar{z}$ approach each other on complex conjugated paths, away from $z=1$. At the level of the conformal blocks, the singularity in eq.~\eqref{r1blockAnsatz} is suppressed by delicate cancellations.\footnote{Notice that the $\rho$ series does not have positive coefficients when $\eta\neq 1$.} At the same time, double traces do not contribute to the difference $\mathcal{G}^{\circlearrowright}-e^{-2\pi i \Delta_\phi}\mathcal{G}$, since each conformal block picks up a phase $e^{-\pi i (\Delta-\ell)}$ as $\bar{z}$ is brought around the origin. Combining eq.~\eqref{LGtilde} with eq.~\eqref{ALandau}, and comparing with eqs.~(\ref{eq:A},\ref{eq:A:expansion}), we are led to the following equality:
 \begin{multline}
 \lim_{r\to 1} (1-r^2)^{d-3} h_{\Delta,\ell}(r,\eta) =\frac{\mathcal{C}_{\Delta,\ell}}{(1-\eta^2)^{1/2}} \, g_{\Delta,\ell}(\eta)~, \\
 \mathcal{C}_{\Delta,\ell}=2^{d-2-2\ell}\sqrt{\pi}
 \frac{\Gamma\left(\frac{d-3}{2}\right)\Gamma(\Delta+\ell)\Gamma\left(\Delta-\frac{d}{2}+1\right)
 (\Delta-1)_\ell}{\Gamma\left(\frac{\Delta+\ell}{2}\right)^4}~,
 \end{multline}
where $g_{\Delta,\ell}$ are the polynomials defined in \eqref{eq:g}. This formula can be easily checked against the known 4d and 6d blocks.%
\footnote{Recall that our convention for the three-point coefficients descends from eq.~\eqref{eq:positionspace3pt}. This corresponds to the following normalization for the position space blocks:
 \beq
 h_{\Delta,\ell}(z,\bar{z}) \underset{z,\bar{z}\to 0}{\approx} \frac{\ell!}{2^\ell (d/2-1)_\ell}
  (z\bar{z})^{\Delta/2}C_\ell^{d/2-1}\left(\frac{z+\bar{z}}{2\sqrt{z\bar{z}}}\right)~.
 \eeq}
 It would be nice to prove this formula for any $d$. 
This should be possible  using the diagonal limit of conformal blocks studied in \cite{Hogervorst:2013kva}.


\section{Examples}
\label{sec:examples}


In this section, we will consider the form factor $F(s,t,u)$ in several examples. We begin with two perturbative examples that are both deformations of the free boson theory in different spacetime dimensions,
and then move on to non-perturbative examples with the Ising model in $d = 3$, and to a brief discussion of holographic theories.

In perturbation theory, our definition of the form factor $F$ and of the amplitude $A$ matches the usual LSZ prescription that consists in amputating external legs in Feynman diagrams computations, provided that the singularity $(p^2)^{\Delta_\phi - d/2}$ of the two-point function at $p^2 \to 0$ is reproduced at all orders.
The perturbative examples below are both truncated at one-loop order, in which case the singularity is indeed reproduced.

\subsection{$\phi^4$ in $4-\varepsilon$ dimensions}
\label{sec:phifour}

\begin{figure}
	\centering
	\begin{tabular}{c@{\qquad\qquad}c}
		\includegraphics[width=35mm]{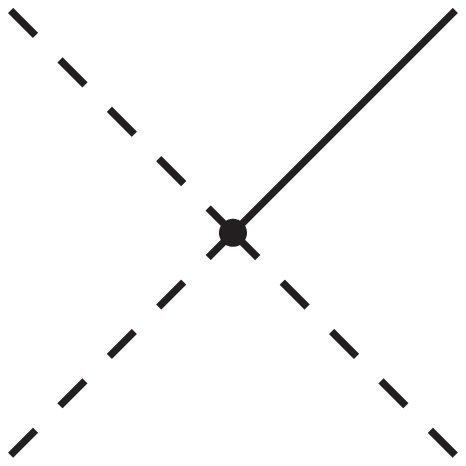} &
		\includegraphics[width=35mm]{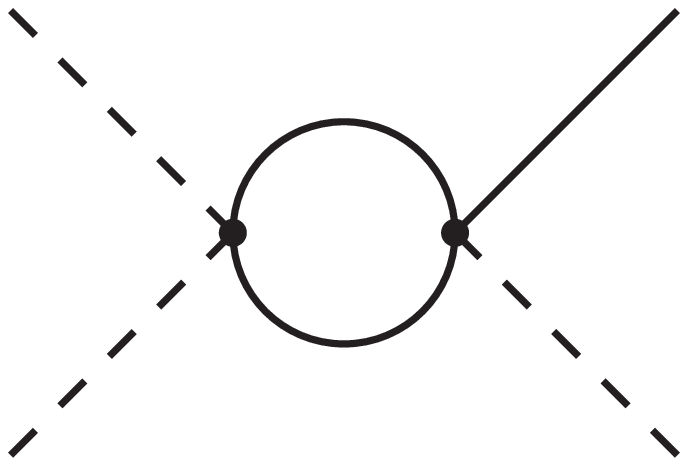}
		\\
		(a) & (b)
	\end{tabular}
\caption{The tree-level (a) and one-loop (b) Feynman diagrams that enter the computation of the form factor \eqref{eq:phifour:formfactor} in the $\phi^4$ theory. Dashed lines indicate legs that are amputated in the LSZ reduction procedure. The full one-loop form factor is obtained summing over the permutations of the three amputated legs.}
\label{fig:phifour:diagrams}
\end{figure}

The first example is $\phi^4$ theory in $d = 4 - \varepsilon$ dimensions, where $0< \varepsilon \ll 1$.
It is given by the action
\begin{equation}
	S = \int d^{4 - \varepsilon}x \left[ - \frac{1}{2} (\partial \phi)^2 - \frac{g}{4!} \, \phi^4 \right]~.
	\label{eq:phifour:action}
\end{equation}
It is well known that this theory has a fixed point at a perturbative value of the coupling $g = g_*$, with 
\begin{equation}
	\frac{g_*}{(4\pi)^2} = \frac{\varepsilon}{3} + \mathcal{O}(\varepsilon^2)~.
	\label{eq:phifour:fixedpoint}
\end{equation}
The computation of the form-factor $F$ can be performed in an expansion in $g$, which is equivalent to a loop expansion. The relevant Feynman diagrams with 3 amputated legs up to order $g^2$ are shown in figure~\ref{fig:phifour:diagrams}.
After renormalization of the coupling $g$, the form factor is found to be%
\footnote{The factor $16 \pi^4$ arises from the fact that in our canonical CFT normalization of the two-point function the free-field propagator in momentum space takes the form
\begin{equation}
	\langle 0 | T\{ \phi(p') \phi(p) \} | 0 \rangle = (2\pi)^d \delta^d(p'+p)
	\left[ \frac{4 \pi^{d/2}}{\Gamma\left( \frac{d-2}{2} \right)} \right]
	\frac{-i}{p^2 - i \varepsilon}~.
\end{equation}
where the factor in squared brackets differs from the propagator that would be derived from the action \eqref{eq:phifour:action}.
\label{footnote:normalization}}
\begin{equation}
	F(s,t,u) = \frac{16 \pi^4}{s + t + u}
	\left[ g + \frac{g^2}{(4\pi)^2} \frac{1}{2} \left( \log(-s) 
	+ \log(-t) + \log(-u) + c \right)
	+ \mathcal{O}(g^3) \right]~.
	\label{eq:phifour:formfactor}
\end{equation}
We are working for simplicity in the unphysical regime $s,t,u < 0$ where the form factor is real, but our derivation holds for $s > 0$ as well.
$c$ is a constant that depends on the choice of renormalization scheme. It can be fixed by requiring that the two-loop renormalization of the wavefunction obeys the conformal normalization of the two-point function, but we choose to keep our computations down to the first loop order and leave $c$ undetermined.

To write the form factor~\eqref{eq:phifour:formfactor} in the form of the conformal partial wave expansion \eqref{CBexpansionF}, we begin with noting that the scalar field $\phi$ has scaling dimension
\begin{equation}
	\Delta_\phi = \frac{d}{2} - 1 + \varepsilon \gamma_\phi + \mathcal{O}(\varepsilon^2)
	= 1 + \varepsilon \left( \gamma_\phi - \tfrac{1}{2} \right) + \mathcal{O}(\varepsilon^2)~,
	\label{eq:phifour:phianomalousdimension}
\end{equation}
where $\gamma_\phi$ is its anomalous dimension.
The dimensionality of $F$ is consistent with $(q^2)^{-\Delta_\phi}$ provided that 
\begin{equation}
	\varepsilon \left( \gamma_\phi - \tfrac{1}{2} \right) = -\frac{3}{2} \frac{g}{(4\pi)^2}
	+ \mathcal{O}(g^2)
\end{equation}
or equivalently that $\gamma_\phi = 0$ at the fixed point. This is again a well-known fact, because the wavefunction of $\phi$ does not get renormalized at the first loop order.
We can therefore write
\begin{equation}
	(q^2)^{\Delta_\phi} F(s,t,u)=
	\sum_\mathcal{O} \lambda_{\phi\phi\mathcal{O}}^2 F_{\Delta, \ell}(w, \cos\theta)
	= \varepsilon f_1(w, \cos\theta) + \varepsilon^2 f_2(w, \cos\theta) + \mathcal{O}(\varepsilon^3)~,
	\label{eq:phifour:Fexpansion}
\end{equation}
where
\begin{align}
	f_1(w, \cos\theta) &= -\frac{2^8 \pi^6}{3},
	\\
	f_2(w, \cos\theta) &= -\frac{2^8 \pi^6}{9}
	\left[ \frac{1}{2} \log(w) 
	+ \log(1-w)
	+ \log\left( \frac{\sin\theta}{2} \right) + \text{constant} \right]~,
\end{align}
expressed in terms of the variables $w = s/(s+t+u)$ and $\cos\theta = (u-t)/(u+t)$.

We can now compare this expression with the conformal partial wave expansion. In the OPE $\phi \times \phi$, there are operators with $\ell$ derivatives and $2n$ powers of $\phi$ (schematically of the form $\partial^\ell \phi^{2n}$). In the $\varepsilon$ expansion, the scaling  dimensions  of the these operators are $2n+\ell +\mathcal{O}(\varepsilon)$. This implies that $F_{\Delta, \ell} \sim \varepsilon$. Moreover, the OPE coefficients start as $\varepsilon^{(n-1)}$.
Therefore, up to order $\varepsilon^2$ we only need to consider the tower of double-trace operators of the form $[ \phi \partial^{\ell} \phi]$, with even spin $\ell$ and scaling dimension
\begin{equation}
	\Delta_{[ \phi \partial^{\ell} \phi]} = 2\Delta_\phi + \ell + \varepsilon \gamma_\ell 
	+ \varepsilon^2 \gamma'_\ell + \mathcal{O}(\varepsilon^3)~.
	\label{eq:phifour:doubletraceanomalousdimensions}
\end{equation}
Because the $[ \phi \partial^{\ell} \phi]$ are close to being double-trace operators, their OPE coefficient is given at leading order in $\varepsilon$ by the free-field theory value~\cite{Fitzpatrick:2011dm},
\begin{equation}
	\lambda_{\phi\phi[\phi\partial^\ell\phi]}^2
	= \left[ 1 + (-1)^\ell \right] \frac{2^\ell (\ell!)^2}{(2\ell)!}
	\left[ 1 + \mathcal{O}(\varepsilon) \right]~.
	\label{eq:phifour:OPEcoefficient}
\end{equation}
The leading contribution to the conformal partial waves is proportional to the anomalous dimension of the operator:
\begin{equation}
	F_{2\Delta_\phi + \ell, \ell}(w,\cos\theta)
	= -2^8 \pi^6 (2\ell + 1) P_\ell(\cos\theta) \,
	\left[ \varepsilon \, \gamma_\ell + \varepsilon^2 \gamma'_\ell
	+ \mathcal{O}(\varepsilon^3) \right]
	\left[ 1 + \mathcal{O}(\varepsilon) \right]~,
	\label{eq:F:doubletrace}
\end{equation}
where $P_\ell$ are the Legendre polynomials, a special case of the Gegenbauer polynomials $\mathcal{C}_\ell^{(d-3)/2}$ in $d = 4$.
Matching these results with eq.~\eqref{eq:phifour:Fexpansion} at linear order in $\varepsilon$ implies immediately
\begin{equation}
	\gamma_0 = \frac{1}{3},
	\qquad\qquad\qquad
	\gamma_{\ell} = 0
	\quad 
	(\ell > 0)~.
	\label{eq:phifour:gamma}
\end{equation}
At quadratic order in $\varepsilon$, it is not possible to obtain information about the next-to-leading order correction $\gamma_0'$ to the anomalous dimension of the spin-zero operator without knowledge of the correction to the OPE coefficient~\eqref{eq:phifour:OPEcoefficient} and of the anomalous dimension of $\phi$ at two-loop order.
However, since the leading anomalous dimension of operators with spin $\ell \neq 0$ vanishes, it is possible to match $\gamma_\ell'$ with the result~\eqref{eq:phifour:Fexpansion}:
using the expansion
\begin{equation}
	\log\left( \frac{\sin\theta}{2} \right) 
	= - 1 - \sum_{\substack{\ell = 2 \\ \text{even}}}^\infty \frac{2\ell + 1}{\ell (\ell + 1)} P_\ell(\cos\theta)~,
\end{equation}
one obtains
\begin{equation}
	\gamma'_{\ell} = -\frac{1}{9 \ell (\ell + 1)}
	\qquad
	(\ell > 0)~.
\end{equation}
These anomalous dimensions, as well as Eq.~\eqref{eq:phifour:gamma}, are well-known results in the $\phi^4$ theory.
The goal of this exercise was to show that they can be obtained very simply from a one-loop computation using Feynman diagrams.
Extending the computation to higher orders would allow to access other operators in the $\phi \times \phi$ OPE.

\subsection{$\phi^3$ in $6+\varepsilon$ dimensions}
\label{sec:phicube}

\begin{figure}
	\centering
	\begin{tabular}{c@{\qquad}c@{\qquad}c}
		\includegraphics[width=40mm]{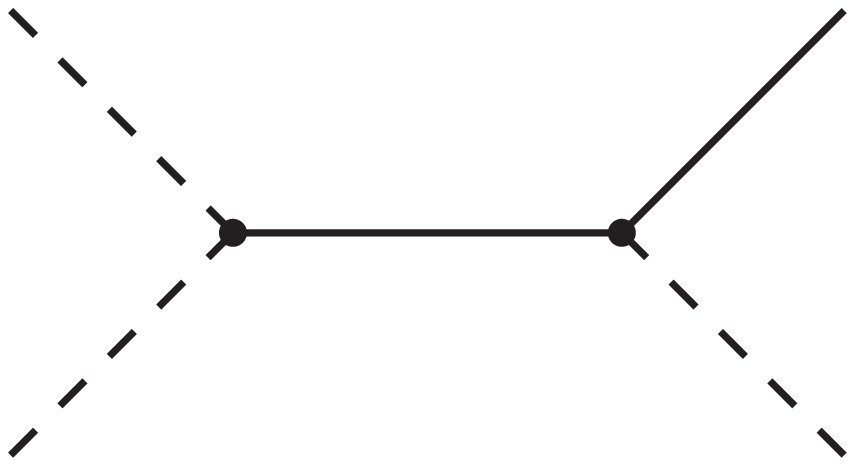} &
		\includegraphics[width=40mm]{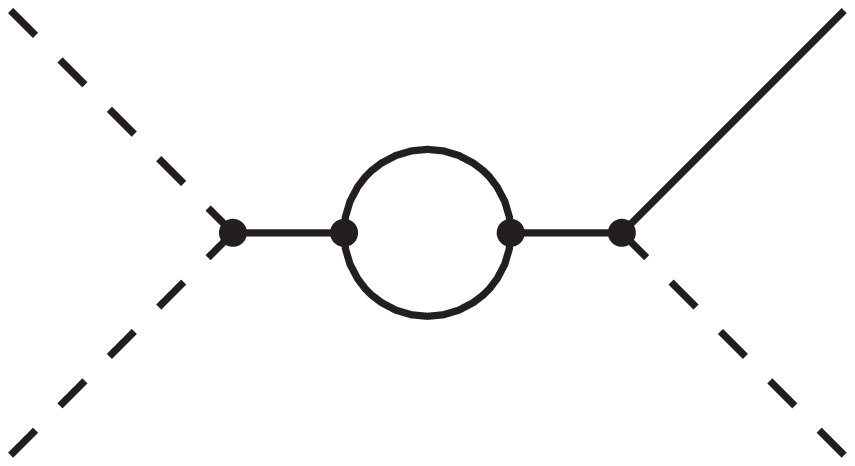} &
		\includegraphics[width=40mm]{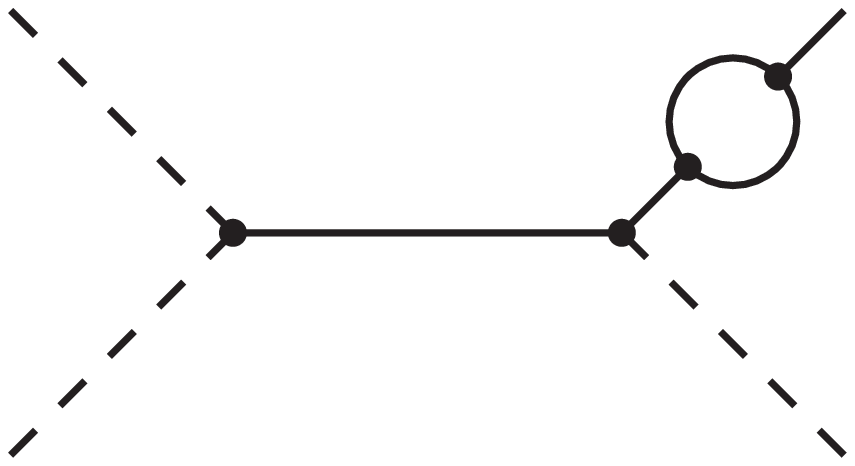}
		\\
		(a) & (b) & (c)
		\\[5mm]
		\includegraphics[width=40mm]{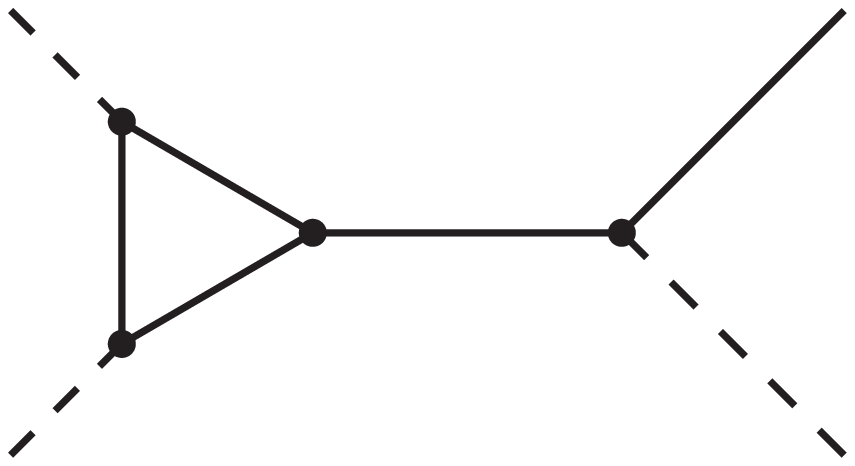} &
		\includegraphics[width=40mm]{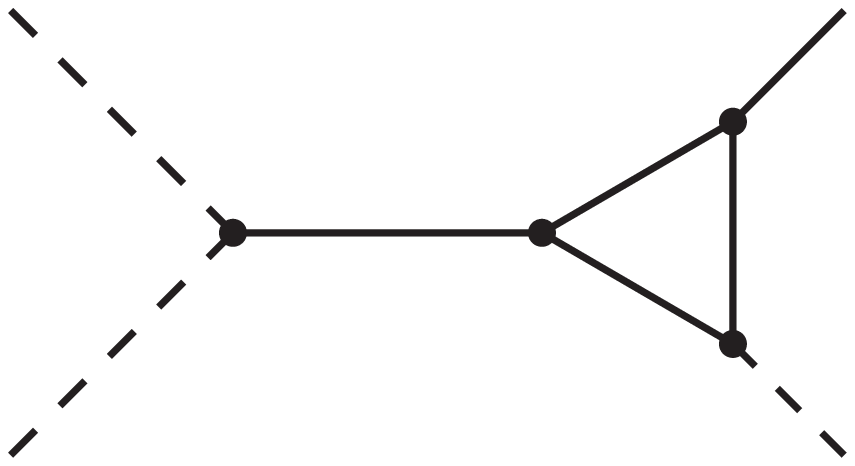} &
		\includegraphics[width=40mm]{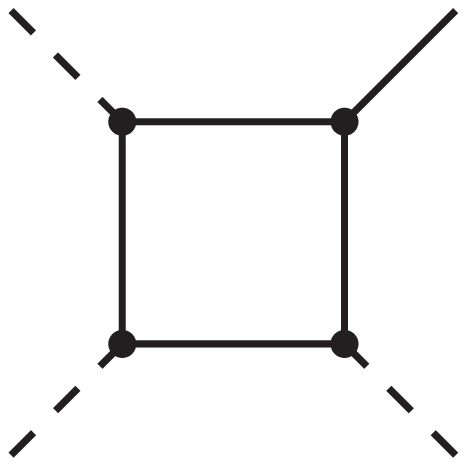}
		\\
		(d) & (e) & (f)
	\end{tabular}
\caption{The tree-level (a) and one-loop (b)--(f) Feynman diagrams that enter the computation of the form factor \eqref{eq:phicube:formfactor} in the $\phi^3$ theory. Dashed lines indicate legs that are amputated in the LSZ reduction procedure. The full one-loop form factor is obtained summing over the permutations of the three amputated legs.}
\label{fig:phicube:diagrams}
\end{figure}

We consider next the theory given by the action
\begin{equation}
	S = \int d^{6 + \varepsilon}x \left[ - \frac{1}{2} (\partial \phi)^2 - \frac{g}{3!} \, \phi^3 \right]~.
	\label{eq:phicube:action}
\end{equation}
This theory has a fixed point at $g = g_*$ with
\begin{equation}
	\frac{g_*^2}{(4\pi)^3} = \frac{2}{3} \, \varepsilon + \mathcal{O}(\varepsilon^2)~.
	\label{eq:phicube:fixedpoint}
\end{equation}
The perturbative computation of the form factor $F$ now involves more Feynman diagrams as shown in figure~\ref{fig:phicube:diagrams}. In this case both the wavefunction and the coupling are renormalized at one-loop order.
We find
\begin{equation}
	F(s,t,u) = 16 \pi^6 g^2 \left[ F_\text{tree}(s,t,u) + \frac{g^2}{(4\pi)^3} F_\text{loop}(s,t,u)
	+ \mathcal{O}(g^3) \right]
	\label{eq:phicube:formfactor}
\end{equation}
where, denoting $q^2 = - s - t - u$,
\begin{align}
	F_\text{tree}(s,t,u) &= -\frac{1}{q^2} \left( \frac{1}{s} + \frac{1}{t} + \frac{1}{u} \right), \\
	F_\text{loop}(s,t,u) &= -\frac{1}{q^2 s} \bigg[
	-\frac{5}{6} \log(q^2)
	+ \frac{1}{2} \left( \frac{s}{t + u} - \frac{5}{6} \right) \log\left( -\frac{s}{q^2} \right)
	\nonumber \\
	& \quad\qquad\quad
	+ \log\left( \frac{s + t}{t} \right) \log\left( -\frac{q^2}{u} \right)
	+ \log\left( \frac{s + u}{u} \right) \log\left( -\frac{q^2}{t} \right)
	\nonumber \\
	& \quad\qquad\quad
	+ \text{Li}_2\left( \frac{q^2 s}{t u} \right)
	- \text{Li}_2\left( -\frac{s}{t} \right) - \text{Li}_2\left( -\frac{s}{u} \right) 
	+ \text{constant} \bigg]
	\nonumber \\
	& \qquad
	+ \left( s \leftrightarrow t \right)
	+ \left( s \leftrightarrow u \right)~.
	\label{eq:phicube:Floop}
\end{align}
As before, we can determine the anomalous dimension of the field $\phi$ by matching the overall dimension of this form factor with $(q^2)^{-\Delta_\phi}$: using the convention
\begin{equation}
	\Delta_\phi = \frac{d}{2} - 1 + \varepsilon \gamma_\phi + \mathcal{O}(\varepsilon^2)
	= 2 + \varepsilon \left( \gamma_\phi + \tfrac{1}{2} \right) + \mathcal{O}(\varepsilon^2)~,
	\label{eq:phicube:phianomalousdimension}
\end{equation}
we must have
\begin{equation}
	F(s,t,u) \propto \frac{1}{(q^2)^2} \left[ 1 - \varepsilon \left( \gamma_\phi + \tfrac{1}{2} \right) \log(q^2)
	+ \mathcal{O}(\varepsilon^2) \right]~,
\end{equation}
which matches Eq.~\eqref{eq:phicube:formfactor} if $\gamma_\phi + \frac{1}{2} = \frac{5}{9}$, i.e.
\begin{equation}
	\gamma_{\phi} = \frac{1}{18}~.
\end{equation}

We can then proceed and extract the CFT data of the operators in the $\phi \times \phi$ OPE using the conformal partial wave expansion for the form factor.
At leading order in $\varepsilon$, we must have
\begin{equation}
	\sum_\mathcal{O} \lambda_{\phi\phi\mathcal{O}}^2 F_{\Delta, \ell}(w, \cos\theta)
	= \frac{2^{11} \pi^9}{3} \varepsilon
	\left[ \frac{1}{w} + \frac{4}{(1-w) (\sin\theta)^2} \right] + \mathcal{O}(\varepsilon^2)~.
\end{equation}
We learned in particular that the expansion around $w = 0$ is in one-to-one correspondence with the twist expansion in the OPE: 
Eq.~\eqref{CBleadingtwist} shows that the contribution of an operator of twist $\tau = \Delta - \ell$ starts at order $w^{(\tau - 4)/2}$.
The $1/w$ terms on the right-hand side of eq.~\eqref{eq:phifour:Fexpansion} indicate therefore the existence of an operator of twist 2 in the spectrum. This is obviously the scalar $\phi$ itself.
Its conformal partial wave obeys
\begin{equation}
	F_{\Delta_\phi, 0}(z, \cos\theta) = 2^{10} \pi^9
	\frac{1 + 5w}{w(1-w)} + \mathcal{O}(\varepsilon)~.
\end{equation}
From this we can determine the value of the OPE coefficient at leading order in $\varepsilon$,
\begin{equation}
	\lambda_{\phi\phi\phi}^2 = \frac{2}{3} \, \varepsilon + \mathcal{O}(\varepsilon^2)
	= \frac{g^2}{(4\pi)^3} + \mathcal{O}(g^4)~.
\end{equation}
The remainder of the form factor at order $\varepsilon$ gives then
\begin{equation}
\begin{aligned}
	\sum_{\mathcal{O} \neq \phi} \lambda_{\phi\phi\mathcal{O}}^2 F_{\Delta, \ell}(w, \cos\theta)
	&= \frac{2^{11} \pi^9}{3} \varepsilon \,
	\frac{1}{1-w}
	\left[ \frac{4}{(\sin\theta)^2} - 6 \right] + \mathcal{O}(\varepsilon^2)
	\\
	&= \frac{2^{11} \pi^9}{3} \varepsilon \,
	\frac{1}{1-w} \sum_{\substack{\ell = 2 \\ \text{even}}}^\infty
	\frac{4 (2\ell + 3)}{(\ell + 1) (\ell + 2)} \mathcal{C}_\ell^{3/2}(\cos\theta)
	+ \mathcal{O}(\varepsilon^2)~.
\end{aligned}
\end{equation}
Note that the expansion in the second line is not absolutely convergent: at large $\ell$, the Gegenbauer polynomials grow like $\mathcal{C}_\ell^{3/2}(\cos\theta) \propto \sqrt{\ell}$ when $\cos\theta \neq \pm 1$.
Nevertheless, it is conditionally convergent as discussed in section~\ref{sec:LSZ:convergence}.
From the leading power $w^0$, we deduce that the next operators in the spectrum have twist close to 4.
The second equality also shows that it admits a partial wave expansion of the form of eq.~\eqref{CBleadingtwist}, that is an expansion in $\text{SO}(5)$ partial waves.
Of course, such operators exist in the OPE: they are the double-trace operators $[\phi \partial^\ell \phi]$, with scaling dimension
\begin{equation}
	\Delta_{[\phi \partial^\ell \phi]}  = 2 \Delta_\phi + \ell + \varepsilon \, \gamma_\ell
	+ \mathcal{O}(\varepsilon^2)
\end{equation}
and OPE coefficient given by~\cite{Fitzpatrick:2011dm},
\begin{equation}
	\lambda_{\phi\phi[\phi\partial^\ell\phi]}^2
	= \left[ 1 + (-1)^\ell \right] \frac{2^\ell (\ell + 1)! (\ell + 2)!}{2 (2 \ell + 1)!}
	\left[ 1 + \mathcal{O}(\varepsilon) \right]~.
	\label{eq:phicube:doubletraceOPEcoefficient}
\end{equation}
Given that the conformal partial waves for these operators obey
\begin{equation}
	F_{\Delta, \ell}(w, \cos\theta)
	= - 2^{10} \pi^9 \varepsilon \, \frac{\gamma_\ell}{1-w}
	\frac{(2\ell + 1)! (2\ell + 3)}{2^{\ell-1} (\ell + 1)! (\ell + 2)!}
	\mathcal{C}_\ell^{3/2}(\cos\theta)
	+ \mathcal{O}(\varepsilon^2)~,
\end{equation}
we obtain immediately
\begin{equation}
	\gamma_\ell = - \frac{4}{3(\ell + 1) (\ell + 2)}~.
\end{equation}
No other operators enter the $\phi \times \phi$ OPE at order $\varepsilon$: together $\phi$ and the double-trace operators $[\phi \partial^\ell \phi]$ with even $\ell \geq 2$ saturate the form factor.%
\footnote{There is no primary operator with $\ell = 0$ in the double trace series since $\phi^2 \propto \square \phi$ is a descendant by the equation of motion.}

The form factor at order $\varepsilon^2$ is much more complicated: it does not only contain powers of $w$, but also terms proportional to $\log(w)$. However, after subtracting the contributions from $\phi$ and from $[\phi \partial^\ell \phi]$, all the terms in $\log(w)$ cancel.
Similarly, it can be verified using the expansion of the conformal partial waves to order $\varepsilon^2$ that all terms of order $w^{-1}$ and $w^0$ cancel.
For instance in the case of double-trace operators, the partial waves are
\begin{align}
	\lambda_{\phi\phi[\phi\partial^\ell\phi]}^2
	F_{\Delta, \ell}(w, \cos\theta) & \nonumber
	\\
	= - 2^{10} \pi^9 \, \frac{\gamma_\ell}{1-w}
	\bigg\{ &
	(2\ell + 3) \left( \varepsilon
	+ \frac{1}{2} \gamma_\ell \log(w) \varepsilon^2 +  c_\ell  \varepsilon^2 \right)
	\mathcal{C}_\ell^{3/2}(\cos\theta)\nonumber
	\\
	& + \frac{3}{2} \bigg[ \left( \frac{\gamma_\ell + 2 \gamma_\phi}{\ell + 2} + 2 \gamma_\phi + 1 \right)
	\mathcal{C}_\ell^{5/2}(\cos\theta) \label{eq:phicube:twist4block}
	\\
	& \qquad + \left( \frac{\gamma_\ell + 2 \gamma_\phi}{\ell + 1} - 2 \gamma_\phi - 1 \right)
	\mathcal{C}_{\ell-2}^{5/2}(\cos\theta) 
	\bigg] w \varepsilon^2 \bigg\}
	+ \mathcal{O}(\varepsilon^3)~,
	\nonumber
\end{align}
where the $c_\ell$ are constants that depend on the correction of order $\varepsilon$ to the OPE coefficient \eqref{eq:phicube:doubletraceOPEcoefficient}; they could in principle be matched with the constant term in eq.~\eqref{eq:phicube:Floop}, but we do not attempt to do this because this requires again fixing the renormalization scheme using a two-loop computation.
As suggested by eq.~\eqref{eq:phicube:twist4block}, the subtraction is better done after expanding terms of order $w^0$ in $\text{SO}(5)$ partial waves, and terms of order $w^1$ in $\text{SO}(7)$ partial waves. At the end of this tedious procedure, we are left with
\begin{equation}
\begin{aligned}
	& \sum_{\mathcal{O} \neq \phi, [\phi \partial^\ell \phi]}
	\hspace{-3mm}
	\lambda_{\phi\phi\mathcal{O}}^2 F_{\Delta, \ell}(w, \cos\theta)
	\\
	&\qquad 
	= -\frac{2^{13} \pi^9}{9} \varepsilon^2 \, w \,
	\bigg[ \frac{7}{6 \sin^2 \theta} 
	+ \frac{1 - \cos\theta}{(1 + \cos\theta)^2} \log\left( \frac{1 - \cos\theta}{2} \right)
	\\
	&\qquad \qquad\qquad\qquad\quad
	+ \frac{1 + \cos\theta}{(1 - \cos\theta)^2} \log\left( \frac{1 + \cos\theta}{2} \right)
	+ \mathcal{O}(w)
	\bigg]
	+ \mathcal{O}(\varepsilon^3)~.
\end{aligned}
\end{equation}
It admits a relatively simple expansion in $\text{SO}(7)$ partial waves,
\begin{align}
	\sum_{\mathcal{O} \neq \phi, [\phi \partial^\ell \phi]}
	\hspace{-3mm}
	\lambda_{\phi\phi\mathcal{O}}^2 & F_{\Delta, \ell}(w, \cos\theta)
	\nonumber \\
	&
	= -\frac{2}{3} w \varepsilon^2 \sum_{\substack{\ell = 0 \\ \text{even}}}^\infty
	\frac{(2 \ell + 5) (11 \ell^4 + 110 \ell^3 + 397 \ell^2 + 610 \ell + 360)}
	{(\ell + 1) (\ell + 2)^3 (\ell + 3)^3 (\ell + 4)}
	\mathcal{C}_{\ell}^{5/2}(\cos\theta)
	\nonumber \\
	& \quad + \mathcal{O}(w^2\varepsilon^2, \varepsilon^3)~.
	\label{eq:phicube:twist6term}
\end{align}
This must agree with the contributions of a series of operators with twist close to 6 that we denote by $[\phi \square \partial^\ell \phi]$, with even spin $\ell$ and scaling dimension
\begin{equation}
	\Delta_{[\phi \square \partial^\ell \phi]} = 2\Delta_\phi + 2 + \ell + \varepsilon \tilde{\gamma}_\ell 
	+ \mathcal{O}(\varepsilon^2)~.
\end{equation}
The conformal partial waves for these operators take the form
\begin{equation}
	F_{\Delta, \ell}(w, \cos\theta) = 2^{11} \pi^9
	\frac{3 \tilde{\gamma}_\ell (2\ell + 5)!}
	{2^\ell (\ell + 1) (\ell + 2) (\ell + 2)! (\ell + 4)!} \,
	w \varepsilon \,
	\mathcal{C}_{\ell}^{5/2}(\cos\theta)
	+ \mathcal{O}(w^2\varepsilon, \varepsilon^2)~.
	\label{eq:phicube:twist6block}
\end{equation}
Note that there is more than one such operator at every spin $\ell > 0$.
What we obtain from the matching of eqs.~\eqref{eq:phicube:twist6term} and \eqref{eq:phicube:twist6block} is a constraint on the sum of OPE coefficients multiplying the anomalous dimension $\tilde{\gamma}_\ell$,
\begin{equation}
	\sum \lambda_{\phi\phi[\phi \square \partial^\ell \phi]}^2 \tilde{\gamma}_\ell
	= - \varepsilon \frac{2^\ell [ (\ell + 1) ! ]^2}{(2 \ell + 3)!}
	\frac{11 \ell^4 + 110 \ell^3 + 397 \ell^2 + 610 \ell + 360}{27 (\ell + 2) (\ell + 3)^2}
	+ \mathcal{O}(\varepsilon^2)~.
\end{equation}
This can be compared with the state-of-the-art results obtained using the skeleton expansion technique~\cite{Goncalves:2018nlv}. In this case a sum of the squared OPE coefficients has been computed to be
\begin{equation}
	\sum \lambda_{\phi\phi[\phi \square \partial^\ell \phi]}^2
	= \varepsilon \frac{2^\ell [ (\ell + 1) ! ]^2}{(2 \ell + 3)!}
	\frac{(\ell + 1) (\ell + 4) (\ell^2 + 5 \ell + 18)}{36 (\ell + 3)}
	+ \mathcal{O}(\varepsilon^2)~,
	\label{eq:phicube:twist6OPEcoefficients}
\end{equation}
so that we can express the average anomalous dimension as
\begin{equation}
\begin{aligned}
	\Delta_{[\phi \square \partial^\ell \phi]} - \ell - 6
	&= ( \tilde{\gamma}_\ell + 2 \gamma_\phi + 1 ) \varepsilon
	+ \mathcal{O}(\varepsilon^2)
	\\
	&= \frac{2 \ell (\ell + 5) (5 \ell^4 + 50 \ell^3 + 199 \ell^2 + 370 \ell + 288)}
	{9 (\ell + 1) (\ell + 2) (\ell + 3) (\ell + 4) (\ell^2 + 5 \ell + 18)} \varepsilon
	+ \mathcal{O}(\varepsilon^2)~.
	\label{eq:phicube:twist6anomalousdimensions}
\end{aligned}
\end{equation}
The first few values for $\ell = 0, 2, 4, 6, 8$ are
\begin{equation}
	0~, \quad
	\frac{28}{45}~, \quad
	\frac{262}{315}~, \quad
	\frac{12353}{13230}~, \quad
	\frac{89648}{90585}~,
\end{equation}
which precisely match the average anomalous dimensions computed in ref.~\cite{Goncalves:2018nlv}.
For operators of spin $\ell > 8$ the result \eqref{eq:phicube:twist6anomalousdimensions} is to the best of our knowledge the first computation of these anomalous dimensions.
Note that we relied on eq.~\eqref{eq:phicube:twist6OPEcoefficients} derived elsewhere to obtain this result. However, going to the next order in $\varepsilon$ would also have allowed to separately determine the value of the OPE coefficient and anomalous  dimensions, however at the price of performing a two-loop Feynman diagram computation.

These perturbative examples provide a simple and yet very non-trivial validation of our conformal partial wave expansion formula.
Moreover, they make evident the simplicity of inverting the OPE in perturbation theory: every order in a Feynman diagram expansion can be easily matched with the contribution of primary operators in the conformal field theory.
From the leading term at a given twist one obtains a constraint on the product of OPE coefficient and anomalous dimension (or the average of this product if there is a degeneracy both in twist and spin), and going to the next order in perturbation theory allows to resolve independently the value of the OPE coefficient and of the anomalous dimension.
We expect that higher-order corrections to the CFT data can similarly be obtained from subleading terms in the perturbative expansion, although we did not perform this analysis in our two examples.
Finally, we would like to emphasize that this OPE inversion method is algorithmic and we hope that it can be used to extract the CFT data from Feynman diagram computations in other interesting cases.


\subsection{The $3d$ Ising model}
\label{sec:ising}

We now turn to a truly non-perturbative example:
the Ising model in $d = 3$ dimensions, for which a large amount of CFT data is known, owing to a combination of numerical studies and large-spin perturbation theory~\cite{Simmons-Duffin:2016wlq}.
The model contains a light scalar field $\sigma$ whose scaling dimension $\Delta_\sigma = 0.5181489(10)$ satisfies the condition $\Delta_\sigma < d/2$.
The scaling dimensions and OPE coefficients of many operators in the OPE $\sigma \times \sigma$ have been computed, among which:
\begin{itemize}

\item
the scalar operator $\epsilon$ with $\Delta_\epsilon = 1.412625(10)$;

\item
operators up to spin $\ell \approx 40$ in the families $[\sigma\sigma]_0$, $[\epsilon\epsilon]_0$, $[\sigma\sigma]_1$, whose twist are respectively approaching $2\Delta_\sigma \approx 1.0$, $2\Delta_\epsilon \approx 2.8$, and $2\Delta_\sigma + 2 \approx 3.0$;

\item
a few operators of low spin and scaling dimensions $\Delta \leq 11$ that do not belong to the previous families.

\end{itemize}

\begin{figure}
	\centering
	\includegraphics[width=11cm]{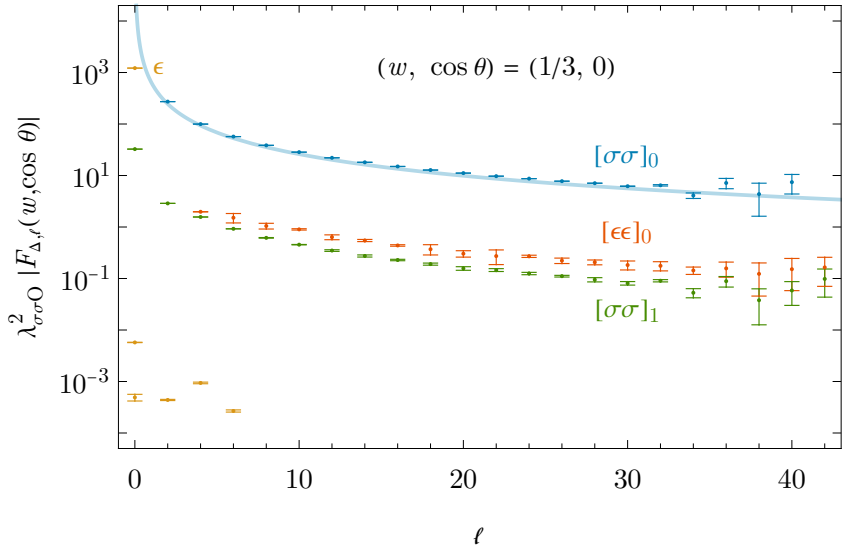}
	\caption{Magnitude of the contribution of the first few operators to the form factor $F$
	at the crossing-symmetric point $(w, \cos\theta) = \left( \frac{1}{3}, 0 \right)$,
	as a function of the spin $\ell$ of the operator,
	and in units in which $q^2 = 1$.
	The $3d$ Ising data is taken from ref.~\cite{Simmons-Duffin:2016wlq}.
	The error bars show the numerical uncertainty on the value of the OPE coefficients only 
	(in comparison the uncertainty on the scaling dimension of the operator is negligible).
	As indicated, the blue, red and green sets of points correspond to operators in the $[\sigma\sigma]_0$,
	$[\epsilon\epsilon]_0$ and $[\sigma\sigma]_1$ families respectively; 
	the yellow points are operators that do not belong to one of these families.
	Note that the largest contribution in absolute value comes from the scalar operator $\epsilon$.
	The solid line shows the asymptotic estimate of eqs.~\eqref{eq:Ising:lambdasigmasigma} and
	\eqref{eq:Ising:Fsigmasigma}.}
	\label{fig:Ising:convergence}
\end{figure}

This data seems to show  that the OPE for the form factor converges rapidly in the Ising model when the scattering angle $\theta \neq 0, \pi$, even though it can be shown formally that the series is not absolutely convergent.
For instance, in the $[\sigma\sigma]_0$ family, the OPE coefficients are approximately given by the generalized free field theory values~\cite{Fitzpatrick:2011dm}, which approach asymptotically
\begin{equation}
	\lambda^2_{\sigma\sigma[\sigma\sigma]_0}
	\approx \frac{\sqrt{\pi} \ell^{2\Delta_\sigma - 3/2}}
	{2^{2\Delta_\sigma + \ell - 3} \left[ \Gamma(\Delta_\sigma) \right]^2}
	\approx \frac{2.36}{2^\ell \ell^{0.46}}~.
\label{eq:Ising:lambdasigmasigma}
\end{equation}
On the other hand, the asymptotic behavior of the conformal partial waves given by \eqref{eq:F:asymptotics}
is  
\begin{equation}
	F_{[\sigma\sigma]_0}(w, \cos\theta) \approx
	336 \, (1 - w)^{1 - \Delta_\sigma}
	\frac{(-1)^{\ell/2} 2^\ell}{\ell^{0.93}}
	\frac{\cos\left[ \left( \theta - \tfrac{\pi}{2} \right)
	\left( \ell + \Delta_\sigma - \tfrac{1}{2} \right) \right]}
	{| 2 \sin\theta |^{\Delta_\sigma - 1/2}}~,
\label{eq:Ising:Fsigmasigma}
\end{equation}
where we have used the fact that the twist of the operators is well approximated by  
\begin{equation}
	\tau_{[\sigma\sigma]_0} \approx 2\Delta_\sigma
	- \frac{2 \lambda_{\sigma\sigma\epsilon}^2
	\Gamma(\Delta_\epsilon) \left[ \Gamma(\Delta_\sigma) \right]^2}
	{\left[ \Gamma\left( \Delta_\sigma - \frac{\Delta_\epsilon}{2} \right)
	\Gamma\left( \frac{\Delta_\epsilon}{2} \right) \right]^2} \,
	\ell^{-\Delta_\epsilon}
	\approx
	1.04 - \frac{0.093}{\ell^{1.4}}~,
\label{eq:Ising:twistasymptotics}
\end{equation}
for a large region in spin.
When combined, eqs.~\eqref{eq:Ising:lambdasigmasigma} and \eqref{eq:Ising:Fsigmasigma} give a series whose coefficients in absolute value decrease like $\ell^{\Delta_\sigma - \Delta_\epsilon - 1/2} \approx \ell^{-1.4}$, and therefore seems absolutely convergent.
However, the fact that the twist of the operators in \eqref{eq:Ising:twistasymptotics} is controlled by $\Delta_\epsilon$ is accidental:
in fact, it is known that their anomalous dimension must eventually be controlled by the lowest-twist operator in the OPE~\cite{Fitzpatrick:2012yx, Komargodski:2012ek}, which in this case is the energy-momentum tensor with $\tau = 1$.%
\footnote{This curiosity of the Ising model is due on the one hand to the OPE coefficient of $\langle \sigma \sigma \epsilon \rangle$ being significantly larger than the one of $\langle \sigma \sigma T \rangle$, and on the other hand to an additional suppression of the influence of $T$ because its twist $\tau = 1$ is  close to the double-trace dimension $2 \Delta_\sigma \approx 1.04$.}
Therefore, at very large spin, the contribution of the family $[\sigma\sigma]_0$ will eventually decay as $\ell^{\Delta_\sigma - 3/2} \approx \ell^{-0.98}$, and the series cannot be absolutely convergent.

\begin{figure}[t]
	\centering
	\includegraphics[width=10cm]{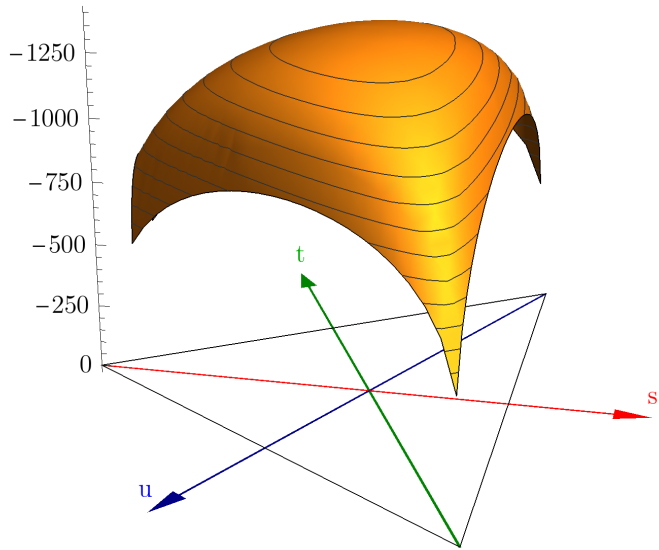}
	\caption{The form factor $F(s,t,u)$ in the $3d$ Ising model, in units where $s + t + u = -1$.
	The horizontal axes correspond to the plane spanned by $s$, $t$ and $u$ as in figure~\ref{fig:triangle},
	and the triangle indicates the region $s, t, u < 0$ in which the form factor is real.
	The form factor is computed using the $s$-channel OPE only (limited to $-0.9\le s\le 0$ for numerical stability),
	but it turns out to be remarkably crossing-symmetric over the full Euclidean triangle. 
	The magnitude of the form factor is maximal at the crossing symmetric point $s = t = u = -\frac{1}{3}$.
	}
	\label{fig:Ising:formfactor}
\end{figure}

Figure~\ref{fig:Ising:convergence} shows the magnitude of the contribution to the form factor of several operators at a specific kinematic point. 
It is obvious that the OPE is dominated by the operators of the lowest-twist family $[\sigma\sigma]_0$, as well as by a handful of low-spin operators, in particular the scalar $\epsilon$.
This situation persists at different kinematic points, and it is therefore possible to obtain a precise estimate of the form factor in the $3d$ Ising model using the data of ref.~\cite{Simmons-Duffin:2016wlq}.
Figure~\ref{fig:Ising:formfactor} shows the result of such an estimate.
The form factor computed in this way is remarkably crossing-symmetric, with discrepancies at the permille level over most of the region $s, t, u < 0$.
For example, using this truncated $s$-channel conformal block decomposition of $F(s,t,u)$, we obtain
\begin{equation}
\begin{aligned}
	F\left( -\tfrac{1}{4},  -\tfrac{1}{2},  -\tfrac{1}{4}\right)  &\approx -1356.3~,
	\\
	F\left( -\tfrac{1}{2},  -\tfrac{1}{4},  -\tfrac{1}{4}\right) &\approx -1353.6~.
\end{aligned}
\end{equation}
in units where $s+t+u=-q^2=-1$.


\subsection{Holographic CFTs}
\label{sec:ads}

\begin{figure}[t]
\centering
\begin{tikzpicture}[scale=1.8]
\draw (-2,0) ellipse (1 and 1);
\draw [black] (-2.7071,-0.7071) --(-1.2929,0.7071);
\draw [black] (-2.7071,0.7071) --(-1.2929,-0.7071);
\filldraw[black]  (-2,0) circle [radius=1 pt];
\draw (2,0) ellipse (1 and 1);
\draw [black] (2.7071,-0.7071) --(2,-0.3);
\draw [black] (2.7071,0.7071) --(2,0.3);
\draw [black] (2,0.3) --(1.2929,0.7071);
\draw [black] (2,-0.3) --(1.2929,-0.7071);
\draw [dashed] (2,-0.3) --(2,0.3);
\filldraw[black]  (2,0.3) circle [radius=1 pt];
\filldraw[black]  (2,-0.3) circle [radius=1 pt];
\node at (-2.8,-0.8) {$1$};
\node at (-2.8,0.8) {$4$};
\node at (-1.2,-0.8) {$2$};
\node at (-1.2,0.8) {$3$};
\node at (2.8,-0.8) {$2$};
\node at (2.8,0.8) {$3$};
\node at (1.2,-0.8) {$1$};
\node at (1.2,0.8) {$4$}; 
\end{tikzpicture}
\caption{\emph{Left:} Scalar quartic contact Witten diagram. \emph{Right:} Tree level $s$-channel exchange Witten diagram.}
\label{fig:Wittendiagrams}
\end{figure}
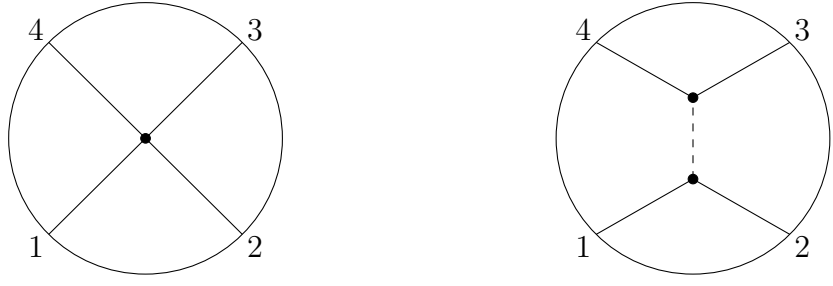

It is instructive to compute the form factor $F(s,t,u)$ associated to some simple Witten diagrams in Anti-de Sitter spacetime (see figure~\ref{fig:Wittendiagrams}).
For simplicity we restrict our attention to the case of identical external scalars.
The simplest case is a contact interaction, which corresponds to a constant Mellin amplitude \cite{Penedones:2010ue}. 
Using \eqref{FFMellin}, we can easily perform the Mellin integrals and obtain
\be
F(s,t,u) = C  M  \,
\Gamma(\Delta_\phi)
(-s-t-u)^{-\Delta_\phi}~.
\label{FcontactAdS}
\ee
Remarkably, the form factor only depends on $q^2=-s-t-u$. This is analogous to a constant scattering amplitude associated to a contact interaction.

We can also consider the limit $q^2 \to 0$ that gives the amplitude \eqref{def:amp}.
We observe that the limit $q^2 \to 0$ of \eqref{FcontactAdS}  is more singular than the general prediction
\eqref{def:amp}. This is to be expected because local bulk interactions give rise to correlators that are more singular in the kinematical limit $z\to \bar{z}$ than a single conformal block. Indeed, this stronger singularity is absent in the non-perturbative bulk theory due to the softness of high-energy scattering in string theory \cite{Okuda:2010ym, Maldacena:2015iua}.
 
Consider now the Witten exchange diagram of a bulk scalar of dimension $\Delta$ in the $s$-channel.
The corresponding Mellin amplitude only depends on $\gamma_{12}$ \cite{Penedones:2010ue}. 
Therefore,  the integral over $\gamma_{13}$ in \eqref{FFMellin} is of the form
\be
  \int \frac{d\gamma_{13}}{2\pi i}  
\frac{\Gamma(\g_{13})}{(-t)^{\gamma_{13}}}
\frac{\Gamma(\Delta_\phi-\g_{12}-\g_{13})}{(-u)^{\Delta_\phi-\g_{12}-\g_{13}}}=\frac{\Gamma(\Delta_\phi-\g_{12})}{(-t-u)^{\Delta_\phi-\g_{12}}}~.
\ee
Thus,
\be
F(s,t,u) = C(q^2)^{-\Delta_\phi} \int \frac{d\gamma_{12}}{2\pi i} M(\gamma_{12}) 
\frac{\Gamma(\g_{12})}{w^{\gamma_{12}}}
\frac{\Gamma(\Delta_\phi-\g_{12})}{(1-w)^{\Delta_\phi-\g_{12}}}~,
\ee
where we used $w=-s/q^2$. Notice that the form factor does not depend on the scattering angle $\theta$ as expected for an $s$-channel scalar exchange.
The analytic structure in the $s$ complex plane is a branch cut from $s=0$ until $s=\infty$ along the positive real axis. This is the expected analytic structure for the scattering amplitude associated with the $s$-channel exchange of a continuum of particles with mass squared $m^2\ge 0$. Notice that there is no branch point at $s=-q^2$ or $w=1$.

Similarly, the $s$-channel exchange of a particle of spin $J$ in AdS leads to a Mellin amplitude which is a polynomial of degree $J$ in the variable $\g_{13}$ 
 \cite{Penedones:2010ue}.
It is convenient to write this polynomial  in a basis of Pochhammer symbols%
\footnote{The Pochhammer symbol is $(x)_k \equiv \frac{\Gamma(x+k)}{\Gamma(x)}=x (x+1) (x+2) \dots (x+k-1)$.}
\be
M(\g_{12},\g_{13})=\sum_{k=0}^J  (\g_{13})_k \,a_k(\g_{12})~,
\ee
 so that we can use the identity
\begin{align}
  \int \frac{d\gamma_{13}}{2\pi i}  (\g_{13})_k
\frac{\Gamma(\g_{13})}{(-t)^{\gamma_{13}}}
\frac{\Gamma(\Delta_\phi-\g_{12}-\g_{13})}{(-u)^{\Delta_\phi-\g_{12}-\g_{13}}}&=\frac{\Gamma(\Delta_\phi-\g_{12}+k)}{(-t-u)^{\Delta_\phi-\g_{12}+k}}
(-t)^k  \label{eq:FeynmanId}\\
&=\frac{\Gamma(\Delta_\phi-\g_{12})}{(-t-u)^{\Delta_\phi-\g_{12}}}
(\Delta_\phi-\g_{12})_k
\left(\frac{1-\cos \theta}{2}\right)^k~.\nonumber
\end{align}
This leads to 
\be
F(s,t,u) = C(q^2)^{-\Delta_\phi} \int \frac{d\gamma_{12}}{2\pi i} H(\gamma_{12},\cos \theta) 
\frac{\Gamma(\g_{12})}{w^{\gamma_{12}}}
\frac{\Gamma(\Delta_\phi-\g_{12})}{(1-w)^{\Delta_\phi-\g_{12}}}~,
\ee
with
\be
H(\gamma_{12},\cos \theta) =
\sum_{k=0}^J    \,a_k(\g_{12})\,(\Delta_\phi-\g_{12})_k
\left(\frac{1-\cos \theta}{2}\right)^k~.
\ee
We conclude that the $s$-channel exchange of a spin $J$ particle in AdS gives a polynomial form factor of degree $J$ in $\cos \theta$. 
This is consistent with the fact that the $s$-channel conformal block decomposition of this diagram only contains blocks with spin $\le J$.
The analytic structure in the $s$ complex plane is the same as for a scalar exchange.


\section{Discussion}
\label{sec:discussion}

In this article we introduced an observable  in CFT that has many similarities with a scattering amplitude. 
The obvious question is, what is being scattered?
To answer   this question, we can 
weakly  couple the CFT to a free massive scalar $\chi$,
\beq
S_{CFT}  - \int d^dx\left[   \frac{1}{2}(\partial \chi)^2 +\frac{1}{2}m^2\chi^2 \right]
+g\, m^{1+\frac{d}{2}-\Delta_\phi} \int d^dx \chi(x) \phi(x)~,
\eeq
where the small coupling $g$ is dimensionless.
The full theory is UV complete if the interaction is relevant, \emph{i.e.} for $\Delta_\phi <\frac{d}{2}+1$.
The propagator of the field $\chi$ is given by
\beq
\frac{1}{p^2+m^2-\Sigma(p^2)}~,
\eeq
where the self energy reads
\beq
\Sigma(p^2) = g^2 m^2 \left( \frac{m^2}{p^2} \right)^{\frac{d}{2}-\Delta_\phi}\frac{(4\pi)^{\frac{d}{2}} \Gamma\left(\frac{d}{2} -\Delta_\phi \right)}
{4^{\Delta_\phi} \Gamma(\Delta_\phi)} + O(g^4)~.
\eeq
We see that for $g \ll 1$ the interaction with the CFT has a very small effect on the propagator near its on-shell pole at 
\beq
p^2=-m^2\left[ 1-
g^2 e^{i\pi(\frac{d}{2}-\Delta_\phi)}\frac{(4\pi)^{\frac{d}{2}} \Gamma\left(\frac{d}{2} -\Delta_\phi \right)}{4^{\Delta_\phi} \Gamma(\Delta_\phi)} + O(g^4)  
\right]~.
\eeq
This means that for $g^2 \ll 1$, $\chi$ is a long-lived quasi-particle with a width $\Gamma \sim g^2 m$.%
\footnote{Notice that there is another pole in the propagator for very small $p^2$ given by $g^2  \left( \frac{m^2}{p^2} \right)^{\frac{d}{2}-\Delta_\phi} \sim 1$.
However this pole is outside the regime of validity of perturbation theory. In fact, in the deep IR the two theories are strongly coupled and generically there will be a mass gap.
}
Consider now 2-to-2  scattering of $\chi$ particles in this theory. The probability of this process happening is controlled by the same coupling $g$, and is in particular of order $g^4$. In other words, this exclusive process is rare: given the initial state of two $\chi$ particles, the non-trivial final states will most of the time include CFT excitations. Nevertheless, the connected amplitude for $\chi\chi \to \chi \chi$ is well defined and given (with all incoming convention) by
\beq
T=g^4 m^{4+2d-4\Delta_\phi} G(p_1,p_2,p_3,p_4) + O(g^6)~,
\eeq
where $p_i^2=-m^2$.
 As any scattering amplitude of identical particles, $T$ is crossing symmetric and  admits a partial wave expansion. However, the full Hilbert space is given by the tensor product of the Hilbert spaces of the CFT and the scalar $\chi$.
Therefore, for finite $m^2$ we cannot write a partial wave expansion for $T$ using only CFT states as intermediate states. 
Remarkably, we have seen that this is possible in the limit $m^2 \to 0$ if
 $\Delta_\phi<\frac{d}{2}$. 
 Using \eqref{def:amp}, we find that\footnote{The imaginary part of $T$ is not positive in the forward limit, since $\Im A>0$. This is not a problem, because the forward limit of the total cross-section is dominated by processes in which one of the two $\chi$ particles decays.}
\beq
T\approx g^4 m^{4-2d+4\Delta_\phi}   e^{-4i\pi \Delta_\phi}   A(s,t,u) ~.
\eeq
Notice that due to scale invariance, the limit $m^2 \to 0$, can also be thought as the limit of high energy scattering at fixed angle, \emph{i.e.}
$|s| \sim |t| \sim |u| \gg m^2$.  
Similarly, the form factor $F$ can be thought of as the high energy limit of the 3-particle form factor of $\phi(x)$ in the asymptotic states of $\chi$ to leading order in the coupling $g$.

From the pragmatic point of view, it would be interesting to understand if the form factor $F$ is amenable to  conformal bootstrap methods.
The expansion in conformal partial waves (in units $s+t+u=-q^2=-1$)
\begin{equation}
	F(s, t,u) =  
	\sum_{\mathcal{O}} \lambda_{\phi\phi\mathcal{O}}^2
	F_{\Delta,\ell}\left( -s , \frac{u-t}{u+t} \right)~,
\end{equation}
and  crossing symmetry impose very non-trivial constraints on the CFT data.
Similarly to the usual conformal bootstrap in position space, the crossing symmetry $t \leftrightarrow u$ just implies that the spin $\ell$ must be even.
On the other hand, imposing $s \leftrightarrow t$ crossing symmetry leads to 
sum rules
\beq
\sum_{\mathcal{O}} \lambda_{\phi\phi\mathcal{O}}^2
	\left[ F_{\Delta,\ell}\left( -s , \frac{u-t}{u+t} \right) 
	-F_{\Delta,\ell}\left( -t , \frac{u-s}{u+s} \right) \right]=0~.
\eeq
We expect these sums to converge for $s,t,u$ real and negative.
Notice that the block $F_{\Delta,\ell}$ vanishes for double-twist operators (i.e. for $\Delta-\ell=2\Delta_\phi+2n$ with $n=0,1,2,\dots$).
 This is similar to the optimal functionals of refs.~\cite{Mazac:2016qev, Mazac:2018mdx, Mazac:2018ycv, Paulos:2019gtx, Mazac:2019shk, Carmi:2019cub, Penedones:2019tng}.
We leave further exploration of this idea for the future.

A related idea is to write a fixed-$t$ dispersion relation that expresses $F$ in terms of its imaginary part (for $s>0$ and $u>0$). Notice that ${\rm Im}\,  F_{\Delta,\ell}$ has double zeros at the position of double twist operators. This is the analogue in momentum space of the double discontinuity in position space \cite{Caron-Huot:2017vep} and of  dispersion relations in Mellin space \cite{Penedones:2019tng}.

The kinematical limit  $0< -s \sim -t \ll 1$ may be amenable to analytic treatment. 
From \eqref{CBleadingtwist}, we conclude that this limit is dominated by low twist and large spin operators. This is similar to the double-lightcone limit in position space that is the cornerstone of several analytic bootstrap methods \cite{Fitzpatrick:2012yx, Komargodski:2012ek, Alday:2015ewa, Alday:2016njk}.  
It would be interesting to generalize these methods for the form factor $F(s,t,u)$. In particular, this may lead to important simplifications in perturbative CFTs.

Another advantage of working in momentum space is that we can benefit from highly developed techniques for perturbative computations. 
It should be possible to adapt some of these techniques to compute the form factor $F$ to higher order than the corresponding position space four-point function. This would make it easier to extract CFT data to higher orders in perturbation theory.

It may be possible to rigorously establish the domain of analyticity of the form factor $F(s,t,u)$ 
using retarded commutators instead of time-ordered products in the LSZ reduction, as reviewed in \cite{Sommer:1970mr} for the case of scattering amplitudes in massive QFTs.
The obvious first task  is to determine the primitive domain of analyticity.

Finally, it would be very interesting to generalize our form factor and amplitude construction to other external operators. 
Firstly, we would like to relax the condition $\Delta_\phi <\frac{d}{2}$ for scalar operators. We suspect that our main formula \eqref{CBexpansionF} is actually valid also for this case.
The reason for this is that if $\Delta_{\phi}>\frac{d}{2}$, then we can still define the form factor as the coefficient of the finite but non-analytic term $(p^2)^{\Delta_\phi-\frac{d}{2}}$ as $p^2 \to 0$ in the Fourier transform of the four-point function.
Secondly, we would like to generalize our construction to non-scalar operators, in particular 
conserved currents and the stress tensor.
This is closely related with previous works that weakly coupled CFTs to gauge fields and gravity \cite{Komargodski:2011vj, Gillioz:2018kwh}.  
The form factor should be specially advantageous because it has a region where it is real and analytic and its conformal partial wave expansion 
gives us access to all CFT data in the current (or stress tensor) four-point function.


\subsection*{Acknowledgements}

We would like to thank Vasco Gon\c{c}alves, Riccardo Rattazzi, Slava Rychkov and Alexander Zhiboedov for useful discussions. MM and JP are supported by the Simons Foundation grant 488649 (Simons collaboration on the Non-perturbative Bootstrap) 
and  by the Swiss National Science Foundation through the project  200021-169132 and through the National Centre of Competence in Research
SwissMAP.


\appendix
\addtocontents{toc}{\protect\setcounter{tocdepth}{1}}

\section{The Wightman 3-point function}
\label{sec:wightman3pt}

This appendix details the computation of the Wightman three-point functions and in particular its singularity \eqref{eq:Wightman3pt:divergence} used in section~\ref{sec:lsz}.
The derivation follows closely the treatment of ref.~\cite{Gillioz:2019lgs}, but in a simplified and (mostly) self-consistent way.

The most general case of interest to us is the three-point function of 2 scalar operators $\phi_1$ and $\phi_2$ with scaling dimensions $\Delta_1$ and $\Delta_2$ and one traceless symmetric tensor $\mathcal{O}^{\mu_1\ldots\mu_\ell}$ with scaling dimension $\Delta$ and spin $\ell$.
In position space, this is given by
\begin{equation}
\begin{aligned}
	& \langle 0 | \mathcal{O}^{\mu_1\ldots\mu_\ell}(x_3) \phi_2(x_2) \phi_1(x_1) | 0 \rangle
	\\
	& \qquad\qquad
	= \lambda_{12\mathcal{O}} \,
	\frac{\left( \frac{x_{31}^{\mu_1}}{x_{31}^2} - \frac{x_{32}^{\mu_1}}{x_{32}^2} \right)
	\cdots \left( \frac{x_{31}^{\mu_\ell}}{x_{31}^2} - \frac{x_{32}^{\mu_\ell}}{x_{32}^2} \right)
	- \text{traces}}
	{(x_{21}^2)^{(\Delta_1 + \Delta_2 - \Delta + \ell)/2}
	(x_{31}^2)^{(\Delta_1 - \Delta_2 +  \Delta - \ell)/2}
	(x_{32}^2)^{(\Delta_2 - \Delta_1 + \Delta - \ell)/2}}~.
\end{aligned}
\label{eq:positionspace3pt}
\end{equation}
where $x_{ij}^2 = -(x_i^0 - x_j^0 - i \epsilon)^2 + (\vec{x}_i - \vec{x}_j)^2$ 
and $\lambda_{12\mathcal{O}}$ is the OPE coefficient.
Performing the Fourier transform of this expression is difficult. It is more efficient to use an approach based on the conformal Ward identities~\cite{Coriano:2013jba}.

To avoid dealing with the $\delta$-function arising from momentum conservation we only consider the Fourier transform of the three-point function with respect to the positions $x_1$ and $x_3$ of the operators $\phi_1$ and $\mathcal{O}$, and we use translation invariance to set $x_2 = 0$. Using the notation of Section~\ref{sec:lsz}, this means that we are interested in the object
\begin{equation}
	\langle \mathcal{O}^{\mu_1\ldots\mu_\ell}(k) | \phi_2(0) | \phi_1(p) \rangle~,
\end{equation}
where it is implicitly assumed that both momenta $p^\mu$ and $k^\mu$ lie inside the forward lightcone, i.e.~$p^2, k^2 < 0$ and $p^0, k^0 > 0$.
When dealing with traceless symmetric tensors it is convenient to introduce a polarization vector $\xi^\mu$ that satisfies the null condition $\xi^2 = 0$, and to contract free indices with $\xi^\mu$ to obtain Lorentz-invariant quantities. The three-point function can then be written in terms of the 5 invariants $k \cdot \xi$, $p \cdot \xi$, $k^2$, $p \cdot k$ and $p^2$. For later convenience we will trade $p \cdot k$ for the quantity $q^2 \equiv (p-k)^2$.
Since we are exclusively interested in the limit $p^2 \to 0$, we make use of the state defined in eq.~\eqref{eq:state:limit} and denote
\begin{equation}
	\xi_{\mu_1} \cdots \xi_{\mu_\ell}
	\langle \mathcal{O}^{\mu_1\ldots\mu_\ell}(k) | \phi_2(0) | \phi_1(\vec{p}) \rangle
	\equiv \lambda_{12\mathcal{O}}
	V( k^2, q^2, k \cdot \xi, p \cdot \xi)~.
	\label{eq:V}
\end{equation}

\subsection{Conformal Ward identities}

In writing Eq.~\eqref{eq:V} we have already made use of Lorentz covariance, but in a conformal field theory correlators are also constrained by the scale and special conformal symmetries.
Scale symmetry simply implies that $V$ must have mass dimension $\Delta_2 - \Delta_1 - \Delta$,
while special conformal symmetry is more complicated. It can be put in the form of a set of differential equations,
\begin{equation}
	\hat{K}^\mu V( k^2, q^2, k \cdot \xi, p \cdot \xi) = 0~,
\end{equation}
where
\begin{equation}
\begin{aligned}
	\hat{K}^\mu &\equiv 2 p_\nu \frac{\partial^2}{\partial p_\mu \partial p_\nu}
	- p^\mu \frac{\partial^2}{\partial p_\nu \partial p^\nu}
	+ 2 \Delta_1 \frac{\partial}{\partial p_\mu}
	- 2 k_\nu \frac{\partial^2}{\partial k_\mu \partial k_\nu}
	+ k^\mu \frac{\partial^2}{\partial k_\nu \partial k^\nu}
	- 2 \Delta \frac{\partial}{\partial k^\mu}
	\\
	& \quad
	+ 2 \left( \frac{\partial}{\partial k^\nu} + ( 2 \Delta - d ) \frac{k_\nu}{k^2} \right)
	\left( \xi^\mu \frac{\partial}{\partial \xi_\nu} - \xi^\nu \frac{\partial}{\partial \xi_\mu} \right)~.
\end{aligned}
\end{equation}
Note that this differential operator does not commute with the limit $p^2 \to 0$ that we want to take. However one might consider the projections 
\begin{equation}
	\hat{K}_p \equiv p_\mu \hat{K}^\mu
	\qquad
	\text{and}
	\qquad
	\hat{K}_\zeta \equiv \zeta_\mu \hat{K}^\mu~,
\end{equation}
where $\zeta$ is a vector taken to be orthogonal to both $p$ and $k$.
They both satisfy the property
\begin{equation}
	\hat{K}_{p, \zeta} \left[ p^2 f(p, k, \xi) \right] \propto p^2~,
\end{equation}
and therefore commute with the limit $p^2 \to 0$.
We will now discuss separately the scalar case $\ell = 0$ from the spinning case $\ell \neq 0$.

\subsection{Scalar operator $\ell = 0$}

When the operator $\mathcal{O}$ is a scalar, there are no polarization vectors and the most general ansatz consistent with scale symmetry can be written
\begin{equation}
	V( k^2, q^2)
	= (q^2)^{(\Delta_2 - \Delta_1 - \Delta)/2}
	F\left( \frac{k^2}{q^2} \right)~,
\end{equation}
where we have assumed for now that $q$ is spacelike, i.e.~$q^2 > 0$.
The action of $\hat{K}_\zeta$ on $V$ is trivial in this case. The constraints from special conformal symmetry follow from
\begin{equation}
\begin{aligned}
	\hat{K}_p V(k^2, q^2) = 2 (q^2)^{(\Delta_2 - \Delta_1 - \Delta)/2} \left( 1 - \frac{k^2}{q^2} \right)
	\bigg[ \frac{k^2}{q^2} \left( 1 - \frac{k^2}{q^2} \right) F''\left( \frac{k^2}{q^2} \right)
	\qquad &
	\\
	+ \left( c - (a + b + 1) \frac{k^2}{q^2}\right)
	F'\left( \frac{k^2}{q^2} \right) - a b F\left( \frac{k^2}{q^2} \right)
	\bigg]~, & 
\end{aligned}
\end{equation}
where we have defined
\begin{equation}
	a = \frac{\Delta_1 - \Delta_2 + \Delta}{2}~,
	\qquad
	b = \frac{\Delta_1 + \Delta_2 + \Delta - d}{2}~,
	\qquad
	c = \Delta - \frac{d}{2} + 1~.
\end{equation}
The Ward identity $\hat{K}_p V(k^2, q^2) = 0$ is therefore a hypergeometric equation whose general solution is of the form
\begin{equation}
	F\left( \frac{k^2}{q^2} \right)
	= A  \Hypergeometric{a}{b}{c}{\frac{k^2}{q^2}}
	+ B \, \left( -\frac{k^2}{q^2} \right)^{d/2 - \Delta} \, \Hypergeometric{a - c + 1}{b - c + 1}{2 - c}{\frac{k^2}{q^2}}~.
\end{equation}
with $A$ and $B$ arbitrary constants.
Of the two terms, only the first one has the expected behavior at small $k^2$: by the arguments of section~\ref{sec:states}, $F$ should remain finite even when $\Delta < d/2$.
We can therefore set $B = 0$. The value of the coefficient $A$ can be determined by direct Fourier integration of eq.~\eqref{eq:positionspace3pt} in the limit $p^2, k^2 \to 0$ with $q^2$ spacelike, which gives
\begin{equation}
\begin{aligned}
	V(k^2,q^2) = \,&
	\frac{(2\pi)^{d+2} 2^{d - \Delta_1 - \Delta_2 - \Delta}}
	{\Gamma\left( \Delta_1 - \frac{d}{2} + 1 \right) \Gamma\left( \Delta - \frac{d}{2} + 1 \right)
	\Gamma\left( \frac{\Delta_1 + \Delta_2 - \Delta}{2} \right)
	\Gamma\left( \frac{\Delta_2 - \Delta_1 + \Delta}{2} \right)}
	\\
	& \times
	(q^2)^{(\Delta_2 - \Delta_1 - \Delta)/2}
	\Hypergeometric{\frac{\Delta_1 - \Delta_2 + \Delta}{2}}{\frac{\Delta_1 + \Delta_2 + \Delta - d}{2}}
	{\Delta - \frac{d}{2} + 1}{\frac{k^2}{q^2}}~.
\end{aligned}
\label{Vspacelikeq}
\end{equation}
$V(k^2,q^2)$ is non-analytic when the exchange momentum crosses the light cone, i.e.~$q^2 = (k-p)^2 = 0$. 
The regime where $k-p$ is future directed timelike ($q^2 < 0$) can be attained 
following the logic of ref.~\cite{Gillioz:2019lgs}:
\begin{equation}
\begin{aligned}
	V(k^2,q^2) = \,&
	\frac{(2\pi)^{d+1}
	2^{d - \Delta_1 - \Delta_2 - \Delta}}
	{\Gamma\left( \Delta_1 - \frac{d}{2} + 1 \right)
	\Gamma\left( \frac{\Delta_2 - \Delta_1 + \Delta}{2} \right)} \,
	(-k^2)^{(\Delta_2 - \Delta_1 - \Delta)/2}
	\\
	& \times \Bigg\{
	\frac{\Gamma\left( \frac{d}{2} - \Delta_2 \right)}
	{\Gamma\left( \frac{\Delta_1 - \Delta_2 + \Delta}{2} \right)}
	\left[ e^{i \pi (\Delta_1 + \Delta_2 - \Delta - 1)/2}
	\left( -\frac{q^2}{k^2} - i \epsilon \right)^{\Delta_2 - d/2}
	+ \text{c.c.} \right]
	\\
	& \qquad\quad \times
	\Hypergeometric{\frac{\Delta_1 + \Delta_2 - \Delta}{2}}{\frac{\Delta_1 + \Delta_2 + \Delta - d}{2}}
	{\Delta_2 - \frac{d}{2} + 1}{\frac{q^2}{k^2}}
	\\
	& \qquad
	+ \frac{\Gamma\left( \Delta_2 - \frac{d}{2} \right)}
	{\Gamma\left( \frac{\Delta_1 + \Delta_2 + \Delta - d}{2} \right)
	\Gamma\left( \frac{\Delta_2 - \Delta_1 + \Delta - d + 2}{2} \right)
	\Gamma\left( \frac{\Delta_1 + \Delta_2 - \Delta}{2} \right)}
	\\
	& \qquad\quad \times
	\Hypergeometric{\frac{\Delta_1 - \Delta_2 + \Delta}{2}}{\frac{\Delta_1 - \Delta_2 - \Delta + d}{2}}
	{\frac{d}{2} - \Delta_2 + 1}{\frac{q^2}{k^2}} \Bigg\}~.
\end{aligned}
\end{equation}
Notice that this last form reduces to \eqref{Vspacelikeq} for $q^2>0$ and 
it makes explicit the behavior of the three-point function around the point $q^2 = 0$.
In particular,   there is a divergence when $\Delta_2 < \frac{d}{2}$. Around that point we have
\begin{equation}
\begin{aligned}
	V(k^2, q^2) \stackrel{q^2 \to 0}{\approx} \, &
	\frac{(2\pi)^{d+1}
	2^{d - \Delta_1 - \Delta_2 - \Delta} \Gamma\left( \frac{d}{2} - \Delta_2 \right)}
	{\Gamma\left( \Delta_1 - \frac{d}{2} + 1 \right)
	\Gamma\left( \frac{\Delta_1 - \Delta_2 + \Delta}{2} \right)
	\Gamma\left( \frac{\Delta_2 - \Delta_1 + \Delta}{2} \right)} \,
	(-k^2)^{(\Delta - \Delta_1 - \Delta_2)/2}
	\\
	& \times
	\left[ e^{i \pi (\Delta_1 + \Delta_2 - \Delta - 1)/2}
	\left( q^2 - i \epsilon \right)^{\Delta_2 - d/2}
	+ \text{c.c.} \right]~.
\end{aligned}
\end{equation}
In the case of identical operators $\Delta_1 = \Delta	_2$ this is precisely Eq.~\eqref{eq:Wightman3pt:divergence} with $\ell = 0$.
Note that besides the branch point singularity at $q^2 = 0$, the three-point function is regular over its full region of support.

\subsection{Spinning operator $\ell > 0$}

In the case of an operator $\mathcal{O}$ that carries spin, one can consider the most general ansatz that is polynomial of degree $\ell$ in the polarization vector $\xi$, for instance
\begin{equation}
	V( k^2, q^2, k \cdot \xi, p \cdot \xi)
	= (q^2)^{(\Delta_2 - \Delta_1 - \Delta + \ell)/2}
	\sum_{n = 0}^\ell \frac{(k \cdot \xi)^n (p \cdot \xi)^{\ell - n}}{(-k^2)^n (q^2)^{\ell - n}}
	F_{\ell, n}\left( \frac{k^2}{q^2} \right)~.
	\label{eq:Vansatz}
\end{equation}
The relative powers of $k^2$ and $q^2$ are arbitrary, since only the overall mass dimension is fixed by scale symmetry, but this particular choice is very convenient: there exist a linear combination of the differential operators $\hat{K}_p$ and $\hat{K}_\xi$ that acts in a simple way on the ansatz, giving
\begin{equation}
\begin{aligned}
	& \left( \hat{K}_p - \frac{p \cdot \xi}{\zeta \cdot \xi} \hat{K}_\zeta \right)
	V(k^2, q^2, k \cdot \xi, p \cdot \xi)
	\\
	& \qquad
	= 2 (q^2)^{(\Delta_2 - \Delta_1 - \Delta + \ell)/2}
	\sum_{n = 0}^\ell \frac{(k \cdot \xi)^n (p \cdot \xi)^{\ell - n}}{(-k^2)^n (q^2)^{\ell - n}}
	\left( 1 - \frac{k^2}{q^2} \right)
	\\
	& \qquad\quad \times
	\left[ y (1-y) F''_{\ell,n}(y)
	+ \left( c_{\ell,n}- (a_{\ell,n} + b_{\ell,n} + 1) y \right)
	F'_{\ell,n}(y) - a_{\ell,n} b_{\ell,n} F_{\ell,n}(y) \right]~,
\end{aligned}
\end{equation}
where we have denoted $y = k^2/q^2$, and now
\begin{equation}
	a_{\ell,n} = \frac{\Delta_1 - \Delta_2 + \Delta + \ell}{2} - n\,,
	\quad
	b_{\ell,n} = \frac{\Delta_1 + \Delta_2 + \Delta - d + \ell}{2} - n\,,
	\quad
	c_{\ell,n} = \Delta - \frac{d}{2} + 1 - n\,.
\end{equation}
Requiring that this quantity vanishes for an arbitrary polarization vector $\xi$ means that each of the unknown functions $F_{\ell,n}$ satisfies a hypergeometric equation.
As in the scalar case, only one of the two solutions to each equation exhibits the right scaling behavior in the limit $k^2 \to 0$, so that one must have
\begin{equation}
	F_{\ell, n}\left( \frac{k^2}{q^2} \right) = A_{\ell,n} 
	\Hypergeometric{\frac{\Delta_1 + \Delta_2 + \Delta - d + \ell}{2} - n}
	{\frac{\Delta_1 - \Delta_2 + \Delta + \ell}{2} - n}
	{\Delta - \frac{d}{2} + 1 - n}{\frac{k^2}{q^2}}~.
\end{equation}
To fix the value of the coefficients $A_{\ell,n}$, one can consider the action of the operator $\hat{K}_\zeta$, which gives
\begin{equation}
\begin{aligned}
	&\hat{K}_\zeta V(k^2, q^2, k \cdot \xi, p \cdot \xi)
	\\
	&= 2 (\zeta \cdot \xi) (q^2)^{(\Delta_2 - \Delta_1 - \Delta + \ell)/2}
	\sum_{n = 0}^{\ell-1} \frac{(k \cdot \xi)^n (p \cdot \xi)^{\ell - n - 1}}{(-k^2)^{n+1} (q^2)^{\ell - n - 1}}
	\\
	& \quad \quad \times
	\bigg[ (n+1) (\Delta - 2 + \ell - n) F_{\ell,n+1}(0)
	+ (\ell - n) \left( \Delta - \tfrac{d}{2} - n \right)  F_{\ell,n}(0)
	+ \mathcal{O}\left( \frac{k^2}{q^2} \right) \bigg].
\end{aligned}
\end{equation}
This gives a recursion relation for the coefficients $A_{\ell,n}$ that is solved by
\begin{equation}
	A_{\ell,n} = \frac{(-1)^{\ell - n} \ell!}{n! (\ell - n)!}
	\frac{\left( \Delta - 1 \right)_{\ell - n}}{\left( \Delta - \ell -  \frac{d}{2} + 1 \right)_{\ell - n}} \,
	A_{\ell,\ell}~.
\end{equation}
As in the scalar case, the constant $A_{\ell,\ell}$ can again be determined using the direct Fourier transform of the three-point function~\eqref{eq:positionspace3pt} in the limit $p^2, k^2 \to 0$, contracting for convenience the indices of the tensor $\mathcal{O}$ with $p^\mu$. From the definition \eqref{eq:positionspace3pt}, it can be shown that
\begin{equation}
\begin{aligned}
	& \frac{\partial^\ell}{\partial x_1^{\mu_1} \cdots \partial x_1^{\mu_\ell}}
	\langle 0 | \mathcal{O}^{\mu_1\ldots\mu_\ell}(x_3) \phi_2(x_2) \phi_1(x_1) | 0 \rangle
	\\
	& \qquad\qquad
	= \lambda_{12\mathcal{O}} \,
	\frac{(-1)^\ell
	\left( \frac{\Delta_1 + \Delta_2 - \Delta + \ell}{2} \right)_\ell
	\left[ 1 + \mathcal{O}\left( \frac{x_{21}^2}{x_{31}^2}, \frac{x_{32}^2}{x_{31}^2} \right) \right]}
	{(x_{21}^2)^{(\Delta_1 + \Delta_2 - \Delta + 3 \ell)/2}
	(x_{31}^2)^{(\Delta_1 - \Delta_2 +  \Delta - 3 \ell)/2}
	(x_{32}^2)^{(\Delta_2 - \Delta_1 + \Delta + \ell)/2}}~,
\end{aligned}
\end{equation}
which implies
\begin{equation}
\begin{aligned}
	& p_{\mu_1} \cdots p_{\mu_\ell} \langle \mathcal{O}^{\mu_1\ldots\mu_\ell}(k) | \phi_2(0) | \phi_1(p) \rangle
	\\
	& \quad
	= \lambda_{12\mathcal{O}} \,
	(-i)^\ell
	\frac{(2\pi)^{d+2} 2^{d - \Delta_1 - \Delta_2 - \Delta - \ell}
	(q^2)^{(\Delta_2 - \Delta_1 - \Delta + 3\ell)/2}
	(-k^2)^{- \ell}
	\left[ 1 + \mathcal{O}(p^2, k^2) \right]}
	{\Gamma\left( \Delta_1 - \frac{d}{2} + 1 \right) \Gamma\left( \Delta - \ell - \frac{d}{2} + 1 \right)
	\Gamma\left( \frac{\Delta_1 + \Delta_2 - \Delta + \ell}{2} \right)
	\Gamma\left( \frac{\Delta_2 - \Delta_1 + \Delta + \ell}{2} \right)}~.
\end{aligned}
\end{equation}
This fixes the coefficient $A_{\ell,\ell}$ and we obtain finally
\begin{equation}
\begin{aligned}
	V( & k^2, q^2, k \cdot \xi, p \cdot \xi) 
	\\
	&= (-i)^\ell
	\frac{(2\pi)^{d+2} 2^{d - \Delta_1 - \Delta_2 - \Delta}
	\left( \Delta - 1 \right)_\ell}
	{\Gamma\left( \Delta_1 - \frac{d}{2} + 1 \right)
	\Gamma\left( \Delta - \frac{d}{2} + 1 \right)
	\Gamma\left( \frac{\Delta_1 + \Delta_2 - \Delta + \ell}{2} \right)
	\Gamma\left( \frac{\Delta_2 - \Delta_1 + \Delta + \ell}{2} \right)}
	\\
	& \quad \times
	\sum_{n=0}^\ell \frac{(-1)^n \ell!}{n! (\ell - n)!}
	\frac{\left( \Delta - \frac{d}{2} + 1 - n \right)_n}
	{\left( \Delta - 1 + \ell - n \right)_n}
	\frac{( k \cdot \xi )^n ( p \cdot \xi )^{\ell - n}}
	{(-k^2)^n (q^2)^{\ell - n}}
	\\
	& \quad\qquad \times
	(q^2)^{(\Delta_2 - \Delta_1 - \Delta + \ell)/2}
	\Hypergeometric{\frac{\Delta_1 - \Delta_2 + \Delta + \ell}{2} - n}
	{\frac{\Delta_1 + \Delta_2 + \Delta - d + \ell}{2} - n}
	{\Delta - \frac{d}{2} + 1 - n}{\frac{k^2}{q^2}}~.
\end{aligned}
\label{eq:Wightman3pt:result}
\end{equation}
Each term in the sum is similar to the scalar case with the replacements $\Delta \to \Delta - n$ and $\Delta_1 \to \Delta_1 + \ell - n$. The singularity of the type $(q^2)^{\Delta_2 - d/2}$ for $\Delta_2 < \frac{d}{2}$ is therefore reproduced term by term when this three-point function is analytically continued around $q^2 = 0$, and we find
\begin{equation}
\begin{aligned}
	V( & k^2, q^2, k \cdot \xi, p \cdot \xi) 
	\\
	& \stackrel{q^2 \to 0}{\approx}
	\frac{(2\pi)^{d+1} 2^{d - \Delta_1 - \Delta_2 - \Delta - \ell}
	\Gamma\left( \frac{d}{2} - \Delta_2 \right) \left( \Delta - 1 \right)_\ell}
	{\Gamma\left( \Delta_1 - \frac{d}{2} + 1 \right)
	\Gamma\left( \frac{\Delta_1 - \Delta_2 + \Delta + \ell}{2}  \right)
	\Gamma\left( \frac{\Delta_2 - \Delta_1 + \Delta + \ell}{2} \right)}
	(-k^2)^{(d - \Delta_1 - \Delta_2 - \Delta)/2}
	\\
	& \quad\qquad \times
	\left[ e^{i \pi (\Delta_1 + \Delta_2 - \Delta + \ell - 1)/2}
	\left( q^2 - i \epsilon \right)^{\Delta_2 - d/2}
	+ \text{c.c.} \right]
	\mathcal{H}^{\mu_1 \ldots \mu_\ell}(k-p, p) \xi_{\mu_1} \cdots \xi_{\mu_\ell},
\end{aligned}
\end{equation}
where the traceless symmetric tensor $\mathcal{H}^{\mu_1 \ldots \mu_\ell}$ is defined by
\begin{equation}
\begin{aligned}
	& \mathcal{H}^{\mu_1 \ldots \mu_\ell}(k-p,p)
	\xi_{\mu_1} \cdots \xi_{\mu_\ell}
	\\
	& \qquad\qquad \equiv (-i)^\ell \, 
	\sum_{n=0}^\ell \frac{2^\ell \ell!}{n! (\ell - n)!}
	\frac{\left( \tfrac{\Delta_1 - \Delta_2 + \Delta + \ell}{2} - n \right)_n}
	{\left( 2 - \Delta - \ell \right)_n}
	\frac{( k \cdot \xi )^n ( p \cdot \xi )^{\ell - n}}
	{(-k^2)^{\ell/2}}~.
\end{aligned}
\end{equation}
An equivalent definition of this tensor is
\begin{equation}
\begin{aligned}
	& \mathcal{H}^{\mu_1 \ldots \mu_\ell}(p_2, p_1)
	\xi_{\mu_1} \cdots \xi_{\mu_\ell}
	\\
	& \quad
	= i^\ell
	\sum_{n=0}^\ell \frac{2^\ell (-1)^n \ell !}{n! (\ell - n)!}
	\frac{\left( \tfrac{\Delta_2 - \Delta_1 + \Delta + \ell}{2} - n \right)_n
	\left( \tfrac{\Delta_1 - \Delta_2 + \Delta - \ell}{2} + n \right)_{\ell - n}}
	{\left( \Delta - 1 \right)_\ell}
	\frac{( p_1 \cdot \xi )^n ( p_2 \cdot \xi )^{\ell - n}}
	{[ -(p_1 + p_2)^2]^{\ell/2}}~,
\end{aligned}
\label{eq:Htensor}
\end{equation}
which makes explicit the symmetry under the simultaneous exchange $p_1 \leftrightarrow p_2$ and $\Delta_1 \leftrightarrow \Delta_2$. Note that this tensor can be combined with the inverse of the two-point function \eqref{eq:Pi:0} to get
\begin{equation}
	\tilde{\mathcal{H}}_{\mu_1 \ldots \mu_\ell}(p_2,p_1)
	\equiv \frac{\pi^{d/2 + 1} \Pi^{-1}_{\mu_1 \ldots \mu_\ell, \nu_1 \ldots \nu_\ell}(p_1 + p_2)
	\mathcal{H}^{\nu_1 \ldots \nu_\ell}(p_2,p_1)}
	{2^{2\Delta - d - 1} ( \Delta + \ell - 1) \Gamma( \Delta - 1 )
	\Gamma\left( \Delta - \frac{d}{2} + 1 \right)}~,
\end{equation}
with the property that 
\begin{equation}
	\tilde{\mathcal{H}}^{\mu_1 \ldots \mu_\ell}(p_2,p_1)
	= \Big[ \mathcal{H}^{\mu_1 \ldots \mu_\ell}(p_2,p_1)
	~\text{with} ~ \Delta \to d - \Delta \Big]~.
\label{eq:H:shadow}
\end{equation}
The two tensors $\tilde{\mathcal{H}}$ and $\mathcal{H}$ are the shadow transform of each other.


\section{The Casimir equation in momentum space}
\label{sec:casimir}

In this appendix we derive a differential equation for the conformal partial waves appearing in the expansion of the four-point function.
The starting point is the Fourier transform of the four-point correlation function
\begin{equation}
	\int \left( \prod_{i=1}^4 d^dx_i \, e^{i p_i \cdot x_i} \right)
	\langle \phi_1(x_1) \phi_2(x_2) \phi_3(x_3) \phi_4(x_4) \rangle
	= (2\pi)^d \delta^d\big(\sum p_i \big) G(p_i)~.
	\label{Gpartialordered}
\end{equation}
In this equation, and in the rest of this appendix, the pair $\phi_1,\,\phi_2$ is on the left of the pair $\phi_3,\,\phi_4$, but the order of $(\phi_1,\,\phi_2)$ and of $(\phi_3,\,\phi_4)$ does not matter.
$G$ is a function of the Lorentz invariant quantities $p_i \cdot p_j$. Given the constraint $p_1 + p_2 + p_3 + p_4 = 0$ there are six such invariants, which we could take to be
\begin{equation}
	m_i^2 \equiv -p_i^2
	\quad (i = 1, \ldots, 4)~,
	\qquad
	t = -(p_1 + p_3)^2~,
	\qquad
	u = -(p_2 + p_3)^2~.
\end{equation}
The third Mandelstam invariant $s = -(p_1 + p_2)^2$ is related to the above by $s = \sum m_i^2 - t - u$.
It will also be useful to use the scattering angle
\begin{equation}
	\cos\theta \equiv \frac{u-t}{u+t}~.
\end{equation}

\subsection{Casimir operator}

The function $G$ admits a conformal partial wave expansion
\begin{equation}
	G(p_i) = \sum_{\mathcal{O}_{\Delta, \ell}} \lambda_{12\mathcal{O}} \lambda_{34\mathcal{O}} G_{\Delta,\ell}(p_i)~,
\end{equation}
where $G_{\Delta,\ell}(p_i)$ represents the contribution of an intermediate primary operator with scaling dimension $\Delta$ and spin $\ell$ and of its descendants. Schematically, we have
\begin{equation}
\begin{aligned}
	\int \left( \prod_{i=1}^4 d^dx_i \, e^{i p_i \cdot x_i} \right)
	\int d^dy &
	\langle \phi_1(x_1) \phi_2(x_2) \mathcal{O}_{\Delta,\ell}(y) \rangle
	\langle \mathcal{O}_{\Delta,\ell}(y) \phi_3(x_3) \phi_4(x_4) \rangle
	\\
	&\sim (2\pi)^d \delta^d(p_1 + p_2 + p_3 + p_4) G_{\Delta,\ell}(p_i)~.
\end{aligned}
\end{equation}
This means that $G_{\Delta,\ell}$ is a solution of the Casimir equation
\begin{equation}
	\hat{C}_2 G(p_i) = \left[ \Delta (\Delta - d) + \ell (\ell + d - 2) \right] G(p_i)~,
\end{equation}
where $\hat{C}_2$ is defined by the equation
\begin{equation}
	\int \left( \prod_{i=1}^4 d^dx_i \, e^{i p_i \cdot x_i} \right)
	\langle \phi_1(x_1) \phi_2(x_2) C_2 \phi_3(x_3) \phi_4(x_4) \rangle
	= (2\pi)^d \delta^d(p_1 + p_2 + p_3 + p_4) \hat{C}_2 G(p_i),
\end{equation}
in terms of the quadratic Casimir invariant $C_2$ of the conformal group $\text{SO}(d, 2)$.
Letting the operator $C_2$ act on the right and working out the conformal algebra gives
\begin{equation}
	\hat{C}_2 = \hat{D}_0 + \left( \hat{D}_1 \cos\theta + \hat{D}'_1 \right) \frac{\partial}{\partial \cos\theta} 
	+ \left( \hat{D}_2 \cos\theta^2 + \hat{D}'_2 \cos\theta + \hat{D}''_2 \right) \frac{\partial^2}{\partial \cos\theta^2}~,
\end{equation}
where for the pieces homogeneous in $\cos\theta$ we have
\begin{align}	
	\hat{D}_0 &= -2 (s - m_3^2 - m_4^2)
	\left( m_3^2 \frac{\partial^2}{\partial (m_3^2)^2}
	+ m_4^2 \frac{\partial^2}{\partial (m_4^2)^2} \right)
	+ 8 m_3^2 m_4^2 \frac{\partial^2}{\partial (m_3^2) \partial (m_4^2)}
	\nonumber \\
	& \quad
	+ 2 \left[ \left( \Delta_3 - \frac{d}{2} - 1 \right) ( s - m_3^2 - m_4^2)
	- 2 \left( \Delta_4 - d \right) m_3^2 \right] \frac{\partial}{\partial (m_3)^2}
	\nonumber \\
	& \quad
	+ 2 \left[ \left( \Delta_4 - \frac{d}{2} - 1 \right) ( s - m_3^2 - m_4^2)
	- 2 \left( \Delta_3 - d \right) m_4^2 \right] \frac{\partial}{\partial (m_4)^2}
	\nonumber \\
	& \quad
	+ (\Delta_3 + \Delta_4 - d) (\Delta_3 + \Delta_4 - 2d)~,
	\\
	\hat{D}_1 &= 2 (d-2) \frac{s - m_3^2 - m_4^2}{s - M^2}
	- 4 \frac{(m_1^2 + m_2^2)(m_3^2 + m_4^2) - 4 m_3^2 m_4^2}{(s - M^2)^2}
	\nonumber \\
	& \quad
	+ 2 \frac{m_1^2 + m_2^2 - 2 m_4^2}{s - M^2} \left( \Delta_3 - d - 2 m_3^2 \frac{\partial}{\partial (m_3^2)} \right)
	\nonumber \\
	& \quad
	+ 2 \frac{m_1^2 + m_2^2 - 2 m_3^2}{s - M^2} \left( \Delta_4 - d - 2 m_4^2 \frac{\partial}{\partial (m_4^2)} \right)~,
	\\
	\hat{D}_2 &= 2 \frac{(s - m_1^2 - m_2^2)^2 - s (m_3^2 + m_4^2) + 4 m_3^2 m_4^2}
	{(s - M^2)^2}~,
\end{align}
with $M^2 = \sum m_i^2$, and for the rest
\begin{align}
	\hat{D}'_1 &= 2 \frac{m_1^2 - m_2^2}{s - M^2}
	\left( \Delta_4 - \Delta_3 + 2 \frac{m_3^2 - m_4^2}{s - M^2} 
	+ 2 m_3^2 \frac{\partial}{\partial (m_3^2)}
	- 2 m_4^2 \frac{\partial}{\partial (m_4^2)} \right)~,
	\\
	\hat{D}'_2 &= 4 \frac{(m_1^2 - m_2^2) (m_3^2 - m_4^2)}{(s - M^2)^2}~,
	\\
	\hat{D}''_2 &= -2 \frac{(s - 2m_1^2 - 2m_2^2)(s - m_3^2 - m_4^2) + (m_1^2 - m_2^2)^2}{(s - M^2)^2}~.
\end{align}
Since $\hat{C}_2$ is a differential operator in $m_3^2$, $m_4^2$ and $\cos\theta$ only, there is no problem in taking the limit $m_1^2, m_2^2 \to 0$ directly. However, we would also like to send $m_3^2 \to 0$, but this latter limit does not commute with the operator $\hat{D}_0$.
One needs more information in order to obtain a differential equation for the four-point function in that limit.

\subsection{Special conformal transformations}

This additional information can be gathered by taking into account the Ward identity for special conformal transformation, which states that
\begin{equation}
	\hat{K}^\mu G(p_1, p_2, p_3) = 0~,
	\label{KonG}
\end{equation}
where we have chosen to set $p_4 = -p_1 - p_2 - p_3$ and
\begin{equation}
	\hat{K}^\mu \equiv \sum_{i = 1}^3 \left[ 2 p_i^\nu \frac{\partial^2}{\partial p_{i\mu} \partial p_i^\nu}
	- p_i^\mu \frac{\partial^2}{\partial p_{i\nu} \partial p_i^\nu}
	- 2 \left( \Delta_i - d \right) \frac{\partial}{\partial p_{i\mu}} \right]~.
\end{equation}
In fact, eq.~\eqref{KonG} is valid for each individual partial wave. Indeed, each addend in the completeness relation \eqref{eq:completenessrelation} is a projector onto the subspace of the Hilbert space spanned by a conformal family, which is left invariant by the action of $\hat{K}^\mu$. Hence, each projector commutes with $\hat{K}^\mu$, and inserting a projector in eq.~\eqref{Gpartialordered} precisely yields the conformal partial wave. Therefore,
\begin{equation}
	\hat{K}^\mu G_{\Delta, \ell}(p_1, p_2, p_3) = 0~.
\end{equation}
The operator $\hat{K}^\mu$ does not commute with the limit $p_1^2, p_2^2 \to 0$ that we want to consider.
Instead, it obeys for instance
\begin{equation}
	\left[ \hat{K}^\mu, p_1^2 \right] =  4  p_1^2 \frac{\partial}{\partial p_1^\mu} 
	- 4 \left( \Delta - \tfrac{d}{2} - 1 \right) p_1^\mu~.
\end{equation}
This means however that we can construct the Lorentz-invariant differential operator
\begin{equation}
	\hat{K} \equiv
	\left[ (p_1 \cdot p_2) p_3^\mu - (p_2 \cdot p_3) p_1^\mu - (p_1 \cdot p_3) p_2^\mu \right] \hat{K}_\mu
\end{equation}
such that 
\begin{equation}
	\left[ \hat{K}, m_1^2 \right] \propto m_1^2
	\qquad
	\text{and}
	\qquad
	\left[ \hat{K}, m_2^2 \right] \propto m_2^2~.
\end{equation}
Now we have got two differential operators $\hat{C}_2$ and $\hat{K}$ for which the limits $m_1^2, m_2^2 \to 0$ can be taken, but which do not individually commute with the limit $m_3^2 \to 0$. We have actually
\begin{align}
	\left[ \hat{C}_2, m_3^2 \right] &= 2 \left( \Delta_3 - \frac{d}{2} - 1 \right) \left( m_1^2 - m_2^2 - t - u \right)
	+ \mathcal{O}(m_3^2)~, \\
	\left[ \hat{K}, m_3^2 \right] &= -4 \left( \Delta_3 - \frac{d}{2} - 1 \right)
	\left( m_1^2 - t \right) \left( m_2^2 - u \right)
	+ \mathcal{O}(m_3^2)~.
\end{align}
This means that the linear combination
\begin{equation}
	\hat{\mathbf{C}}_2 \equiv \hat{C}_2
	+ \frac{\left( m_1^2 - m_2^2 - t - u \right)}{2 \left( m_1^2 - t \right) \left( m_2^2 - u \right)} \hat{K}
\end{equation}
satisfies
\begin{equation}
	\left[ \hat{\mathbf{C}}_2, m_i^2 \right] \propto m_i^2
	\qquad
	(i = 1, 2, 3)
\end{equation}
and the operator $\hat{\mathbf{C}}_2$ now acts directly on the limit $m_1^2, m_2^2, m_3^2 \to 0$.
Still, the action of this operator depends on the leading behavior of the four-point function in that limit. 
In our case, based on the assumptions of section~\ref{sec:lsz}, we make the ansatz
\begin{equation}
	G_{\Delta, \ell}(p_i) = \frac{(m_1^2)^{\Delta_1 - d/2} (m_2^2)^{\Delta_2 - d/2} (m_3^2)^{\Delta_3 - d/2}}
	{(m_4^2)^{(\Delta_1 + \Delta_2 + \Delta_3 - \Delta_4)/2}}
	\left[ f_{\Delta, \ell}\left( w, \cos\theta \right)
	+ \mathcal{O}(m_1^2, m_2^2, m_3^2) \right]~,
\end{equation}
where we have defined $w = s / m_4^2$.
The function $f_{\Delta, \ell}$ will now obey a Casimir equation of the form
\begin{equation}
	\hat{\mathfrak{C}}_2 \, f_{\Delta, \ell}(w, \cos\theta)
	= \left[ \Delta (\Delta - d) + \ell (\ell + d - 2) \right] f_{\Delta, \ell}(w, \cos\theta)~,
	\label{eq:Casimir:f}
\end{equation}
where the right-hand side is the Casimir eigenvalue and the differential operator $\hat{\mathfrak{C}}_2$ is
\begin{equation}
\begin{aligned}
	\hat{\mathfrak{C}}_2 &= 4 w^2 (1-w) \frac{\partial^2}{\partial w^2}
	+ 4 w \left[ \left( \Delta_1 + \Delta_2 - \frac{d}{2} + 1 \right) (1 - w) 
	- \Delta_3 \, w \right] \frac{\partial}{\partial w}
	\\
	& \quad
	+ 4 w \cos\theta \frac{\partial^2}{\partial w \, \partial \cos\theta}
	- 2 \left( 1 - \cos\theta^2 \right) \frac{\partial^2}{\partial \cos\theta^2}
	\\
	& \quad
	+ 2 \left[ \left( \Delta_1 + \Delta_2 - 2 (\Delta_3 - 1) \frac{w}{1-w} \right) \cos\theta
	+ \Delta_1 - \Delta_2 \right] \frac{\partial}{\partial \cos\theta}
	\\
	& \quad
	- \left( \Delta_1 + \Delta_2 + \Delta_3 - \Delta_4 \right)
	\left( \Delta_1 + \Delta_2 + \Delta_3 + \Delta_4 - d \right) w 
	\\
	& \quad
	+ \left( \Delta_1 + \Delta_2 \right) \left( \Delta_1 + \Delta_2 - d \right)~.
\end{aligned}
\label{eq:Casimir}
\end{equation}

\subsection{Solutions}

Since we know that the conformal partial wave can be obtained from the product of three-point functions, we are looking for solutions that are polynomial in $\cos\theta$ with degree equal to the spin $\ell$.
For low spin it is not difficult to construct solutions explicitly. For instance the scalar case only depends on $w$, and in that case the Casimir equation is of hypergeometric type. Choosing the solution that has the correct behavior at small $w$, as in eq.~\eqref{eq:F:zcosthetalimit}, we obtain
\begin{equation}
	f_{\Delta,0}(w) = w^{(\Delta - \Delta_1 - \Delta_2)/2} (1-w)^{1-\Delta_3}
	\Hypergeometric{\frac{\Delta - \Delta_3 - \Delta_4 + 2}{2}}
	{\frac{\Delta - \Delta_3 + \Delta_4 - d + 2}{2}}
	{\Delta - \frac{d}{2} + 1}{w}~.
\end{equation}
The normalization of this conformal partial wave is arbitrary, but we will adopt the general convention that 
\begin{equation}
	f_{\Delta, \ell}(w, \cos\theta) =
	w^{(\Delta - \Delta_1 - \Delta_2 - \ell)/2} ( \cos\theta )^\ell
	\left[ 1  + \mathcal{O}(w, \cos\theta^{-1}) \right]~.
	\label{eq:fnormalization}
\end{equation}
Similarly, the solution for an exchange operator of spin 1 is
\begin{equation}
\begin{aligned}
	f_{\Delta,1}(w, \cos\theta) &= w^{(\Delta - \Delta_1 - \Delta_2 - 1)/2} (1 - w)^{1 - \Delta_3} \\
	& \quad \times \bigg[
	\left( \cos\theta - \frac{\Delta_1 - \Delta_2}{\Delta - d + 1} \right)
	\Hypergeometric{\frac{\Delta - \Delta_3 - \Delta_4 + 1}{2}}
	{\frac{\Delta - \Delta_3 + \Delta_4 - d + 1}{2}}
	{\Delta - \frac{d}{2} + 1}{w}
	\\
	& \quad\qquad 
	+ 2 \frac{\left( \Delta - \frac{d}{2} \right) (\Delta_1 - \Delta_2)}
	{(\Delta - 1)(\Delta - d + 1)}
	\Hypergeometric{\frac{\Delta - \Delta_3 - \Delta_4 + 1}{2}}
	{\frac{\Delta - \Delta_3 + \Delta_4 - d + 1}{2}}
	{\Delta - \frac{d}{2}}{w} \bigg]~.
\end{aligned}
\end{equation}
These solutions motivate the following ansatz
\begin{equation}
	f_{\Delta,\ell}(w, \cos\theta) = w^{(\Delta - \ell - \Delta_1 - \Delta_2)/2} (1-w)^{1 - \Delta_3}
	\sum_{n = 0}^\infty  w^n h_{\Delta, \ell, n}(\cos\theta)~,
	\label{eq:f:distinctoperators}
\end{equation}
where the $h_{\Delta, \ell, n}(\cos\theta)$ are polynomials of degree $\ell$ in $\cos\theta$.
The Casimir equation \eqref{eq:Casimir:f} gives a recursive differential equation for them
\begin{equation}
\begin{aligned}
	&(1 - \cos\theta^2) h''_{\Delta, \ell, n}(\cos\theta)
	+ \left[ \Delta_2 - \Delta_1 - (\Delta - \ell + 2n) \cos\theta \right] h'_{\Delta, \ell, n}(\cos\theta)
	\\
	& \qquad
	+ \left[ \ell (\Delta - 1) - 2n \left( \Delta - \ell - \tfrac{d}{2} + n \right) \right] h_{\Delta, \ell, n}(\cos\theta)
	\\
	&= - \frac{1}{2} \left( \Delta - \ell - \Delta_3 - \Delta_4 + 2n \right)
	\left( \Delta - \ell - \Delta_3 + \Delta_4 - d + 2n \right) h_{\Delta, \ell, n-1}(\cos\theta)
\end{aligned}
\label{eq:h:recursion}
\end{equation}
as well as an ordinary differential equation for the first polynomial in the series,
\begin{equation}
\begin{aligned}
	(1 - \cos\theta^2) h''_{\Delta, \ell, 0}(\cos\theta)
	+ \left[ \Delta_2 - \Delta_1 - ( \Delta - \ell ) \cos\theta \right] h'_{\Delta, \ell, 0}(\cos\theta)
	& \\
	+ \ell (\Delta - 1) h_{\Delta, \ell, 0}(\cos\theta) &= 0~.
\end{aligned}
\end{equation}
This is the differential equation satisfied by the Jacobi polynomials $P_\ell^{(\alpha,\beta)}(\cos\theta)$. In the normalization \eqref{eq:fnormalization}, we find
\begin{equation}
	h_{\Delta, \ell, 0}(\cos\theta) = \frac{2^\ell \ell!}{(\Delta - 1)_\ell}
	P_\ell^{(\alpha,\beta)}(\cos\theta)
	\label{eq:Jacobipolynomials}
\end{equation}
with
\begin{equation}
	\alpha = \frac{\Delta - \ell + \Delta_1 - \Delta_2 - 2}{2}~, 
	\qquad
	\beta = \frac{\Delta - \ell + \Delta_2 - \Delta_1 - 2}{2}~.
\end{equation}
There is no obvious simple representation of the polynomials $h_{\Delta, \ell, n}$ for $n > 0$, but it is relatively simple to construct them using the recursion relation \eqref{eq:h:recursion}. They can for instance be written as a linear combination of Jacobi polynomials of the form $P_\ell^{(\alpha + n,\beta + m)}(\cos\theta)$ with $n, m \in \mathbb{N}$ and coefficients that do not depend on the scaling dimensions of the external operators (except for an overall multiplicative factor). For instance the second polynomial in the series expansion is
\begin{equation}
\begin{aligned}
	h_{\Delta, \ell, 1}(\cos\theta) &= -\frac{2^{\ell-2} \ell!
	\left( \Delta - \ell - \Delta_3 - \Delta_4 + 2 \right)
	\left( \Delta - \ell - \Delta_3 + \Delta_4 - d + 2 \right)}
	{(\Delta -  1)_\ell \left( \frac{d}{2} + \ell - 2 \right)}
	\\
	& \quad \times
	\bigg[ P_\ell^{(\alpha,\beta)}(\cos\theta) - \frac{\Delta - 1}{\Delta - \ell - d + 2}
	\left( P_\ell^{(\alpha + 1,\beta)}(\cos\theta) + P_\ell^{(\alpha,\beta + 1)}(\cos\theta) \right)
	\\
	& \quad \qquad
	+ \frac{\Delta (\Delta - 1)}{\left( \Delta - \frac{d}{2} + 1 \right) ( \Delta - \ell - d + 2)}
	P_\ell^{(\alpha + 1,\beta + 1)}(\cos\theta) \bigg]~.
\end{aligned}
\end{equation}
An interesting feature of this solution is that the leading order in $w$ given by a single Jacobi polynomial does not depend on the spacetime dimension $d$.
Moreover, since the Jacobi polynomials satisfy orthogonality properties, these conformal partial waves are particularly well-suited to write an OPE inversion formula, at least as long as $\Delta - \ell > |\Delta_1 - \Delta_2|$.%
\footnote{When $\Delta - \ell \leq |\Delta_1 - \Delta_2|$, i.e.~when $\alpha$ or $\beta \leq -1$, the integration kernel of the orthogonality relation
\begin{equation*}
	\int_{-1}^1 d\cos\theta \, (1 - \cos\theta)^\alpha (1 - \cos\theta)^\beta
\end{equation*}
becomes singular. In spacetime dimensions $d \geq 3$ the unitarity bound prevents this from happening. In $d = 2$, however, this situation is realized when Virasoro descendants of the identity operators are exchanged.
}
We leave this prospect for future study.

When all the external operators are identical, i.e.~when $\Delta_i = \Delta_\phi$ for $i = 1, \ldots 4$, the solution can be simplified further. The Jacobi polynomials with identical parameters become Gegenbauer polynomials, $P_\ell^{(\alpha,\alpha)}(\cos\theta) \propto \mathcal{C}_\ell^{(\alpha + 1/2)}(\cos\theta)$, and the solution at leading order in $w$ is
\begin{equation}
	f_{\Delta,\ell}(w, \cos\theta)
	= w^{(\Delta - \ell - 2 \Delta_\phi)/2}
	\frac{\ell!}{2^\ell \left( \frac{\Delta - \ell - 1}{2} \right)_\ell}
	\mathcal{C}_\ell^{(\Delta - \ell - 1)/2}(\cos\theta)
	\left[ 1 + \mathcal{O}(w) \right]~.
	\label{eq:f:smallzlimit}
\end{equation}
Moreover, in this case we found a general solution to the recursion relation \eqref{eq:h:recursion},
\begin{equation}
\begin{aligned}
	f_{\Delta,\ell}(w, \cos\theta)
	&= w^{(\Delta - \ell - 2 \Delta_\phi)/2} (1 - w)^{1 - \Delta_\phi}
	\sum_{n = 0}^\infty w^n
	\frac{\left( \frac{\Delta - \ell - d + 2}{2} \right)_n
	\left( \frac{\Delta - \ell - 2\Delta_\phi + 2}{2} \right)_n}
	{2^\ell \left( \frac{\Delta - \ell - 1}{2} \right)_\ell 
	\left( \Delta - \frac{d}{2} + 1 \right)_n}
	\\
	& \quad \times \!\!\!\!\!\! \sum_{j = 0}^{\min( \ell/2, n)} \!\!\!\!
	\frac{\ell!}{j! (n-j)!}
	\frac{\left( \Delta - \ell - d + 2 + n \right)_j
	\left( \frac{\Delta - \ell - 1}{2} \right)_j}
	{\left( \frac{d}{2} - 1 + \ell - j \right)_j
	\left( \frac{\Delta - \ell - d + 3}{2} \right)_j}
	\mathcal{C}_{\ell - 2j}^{(\Delta - \ell - 1 + 2j)/2}(\cos\theta),
\end{aligned}
\label{eq:f:doublesum}
\end{equation}
or equivalently in closed form, using the generalized hypergeometric function $_3F_2$,
\begin{equation}
\begin{aligned}
	f_{\Delta,\ell}(w, \cos\theta)
	&= w^{(\Delta - \ell - 2 \Delta_\phi)/2} (1 - w)^{1 - \Delta_\phi}
	\frac{\ell!}{2^\ell \left( \frac{\Delta - \ell - 1}{2} \right)_\ell}
	\sum_{j = 0}^{\ell/2} w^j \,
	\mathcal{C}_{\ell - 2j}^{(\Delta - \ell - 1 + 2j)/2}(\cos\theta)
	\\
	& \quad \times
	\frac{\left( \frac{\Delta - \ell - d + 2}{2} \right)_j
	\left( \frac{\Delta - \ell - 1}{2} \right)_j
	\left( \frac{\Delta - \ell - 2 \Delta_\phi + 2}{2} \right)_j
	\left( \Delta - \ell - d + 2 + j \right)_j}
	{j! \left( \frac{\Delta - \ell - d + 3}{2} \right)_j
	\left( \frac{d}{2} - 1 + \ell - j \right)_j
	\left( \Delta - \frac{d}{2} + 1 \right)_j}
	\\
	& \quad \times
	\GeneralizedHypergeometric{\Delta - \ell - d + 2 + 2j}
	{\frac{\Delta - \ell - d + 2 + 2j}{2}}{\frac{\Delta - \ell - 2\Delta_\phi + 2 + 2j}{2}}
	{\Delta - \ell - d + 2 + j}{\Delta - \frac{d}{2} + 1 + j}{w}~.
\end{aligned}
\label{eq:f:alternaterep}
\end{equation}
Remarkably, this solution can be written in an even simpler form in terms of a different set of Gegenbauer polynomials in which the twist $\Delta - \ell$ is replaced by $d-1$:%
\footnote{This is reminiscent of the symmetry $(\Delta, \ell) \to (\ell + d - 1, \Delta - d + 1)$ used in the Lorentzian OPE inversion formula of Ref.~\cite{Caron-Huot:2017vep}.}
\begin{equation}
\begin{aligned}
	f_{\Delta,\ell}(w, \cos\theta)
	&= w^{(\Delta - \ell - 2 \Delta_\phi)/2} (1 - w)^{1 - \Delta_\phi}
	\frac{\ell!}{2^\ell \left( \frac{d - 2}{2} \right)_\ell}
	\sum_{j = 0}^{\ell/2}
	\frac{\left( \frac{d - 2}{2} \right)_j \left( \frac{d - \Delta + \ell - 1}{2} \right)_j}
	{j! \left( \frac{3 - \Delta - \ell}{2} \right)_j}
	\\
	& \quad \times
	\GeneralizedHypergeometric{\frac{\Delta - \ell - d + 3 + 2j}{2}}
	{\frac{\Delta - \ell - d + 2}{2}}{\frac{\Delta - \ell - 2\Delta_\phi + 2}{2}}
	{\frac{\Delta - \ell - d + 3 - 2j}{2}}{\Delta - \frac{d}{2} + 1}{w}
	\mathcal{C}_{\ell - 2j}^{(d/2 - 1 + j)}(\cos\theta)~.
\end{aligned}
\label{eq:f:appendix}
\end{equation}
Note that this expression is consistent with the derivation of Appendix~\ref{sec:wightman3pt} where we found that the Wightman three-point function are given as a finite sum of hypergeometric $_2F_1$ functions, since we have the identity
\begin{equation}
\begin{aligned}
	& \GeneralizedHypergeometric{\frac{\Delta - \ell - d + 3 + 2j}{2}}
	{\frac{\Delta - \ell - d + 2}{2}}{\frac{\Delta - \ell - 2\Delta_\phi + 2}{2}}
	{\frac{\Delta - \ell - d + 3 - 2j}{2}}{\Delta - \frac{d}{2} + 1}{w}
	\\
	&= \sum_{k = 0}^{2j} \frac{(2j)!}{k! (2j - k)!}
	\frac{\left( \frac{\Delta - \ell - d + 2}{2} \right)_k
	\left( \frac{\Delta - \ell - 2\Delta_\phi + 2}{2} \right)_k}
	{\left( \frac{\Delta - \ell - d + 3 - 2j}{2} \right)_k
	\left( \Delta - \frac{d}{2} + 1 \right)_k}
	\Hypergeometric{\frac{\Delta - \ell - d + 2 + 2k}{2}}{\frac{\Delta - \ell - 2\Delta_\phi + 2 + 2k}{2}}
	{\Delta - \frac{d}{2} + 1 + k}{w}
	w^k.
\end{aligned}
\label{3F2to2F1}
\end{equation}

\subsection{Limit $q^2 \to 0$}
\label{sec:casimir:amplitude}

The limit $q^2 \to 0$ corresponds to $w \to -\infty$ (it requires $s > 0$). It cannot be taken directly in the Casimir equation \eqref{eq:Casimir}. More information is needed to resolve the dependence on $\cos\theta$ as $w \to -\infty$. One could for instance study the quartic Casimir equation. Instead, the limit can be obtained by direct multiplication of the three-point functions. This is what is done in section~\ref{sec:lsz:fourth} and in appendix~\ref{sec:lszdistinct}.

Nevertheless, in the case of identical external operators, the solution \eqref{eq:f:appendix}  in closed form allows to take the limit directly. Using
\begin{equation}
\begin{aligned}
	& \GeneralizedHypergeometric{\frac{\Delta - \ell - d + 3 + 2j}{2}}
	{\frac{\Delta - \ell - d + 2}{2}}{\frac{\Delta - \ell - 2\Delta_\phi + 2}{2}}
	{\frac{\Delta - \ell - d + 3 - 2j}{2}}{\Delta - \frac{d}{2} + 1}{w}
	\\
	& \qquad \xrightarrow{w \to - \infty} (- w )^{-(\Delta - \ell - d + 2)/2}
	(-1)^j
	\frac{\left( \frac{1}{2} \right)_j^2
	\Gamma\left( \frac{d}{2} - \Delta_\phi \right)
	\Gamma\left( \Delta - \frac{d}{2} + 1 \right)
	\Gamma\left( \frac{\Delta - \ell - d + 3 - 2j}{2} \right)}
	{\Gamma\left( \frac{\Delta + \ell}{2} \right)
	\Gamma\left( \frac{\Delta - \ell}{2} - \Delta_\phi + 1 \right)
	\Gamma\left( \frac{\Delta - \ell - d + 3 + 2j}{2} \right)},
\end{aligned}
\end{equation}
we obtain
\begin{equation}
	f_{\Delta,\ell}(w - i \epsilon, \cos\theta)
	\approx
	(-w)^{d/2 - 2\Delta_\phi}
	e^{- i \pi (\Delta - \ell - 2 \Delta_\phi)/2}
	\frac{\Gamma\left( \frac{d}{2} - \Delta_\phi \right)
	\Gamma\left( \Delta - \frac{d}{2} + 1 \right)}
	{\Gamma\left( \frac{\Delta + \ell}{2} \right)
	\Gamma\left( \frac{\Delta - \ell}{2} - \Delta_\phi + 1 \right)}
	g_{\Delta,\ell}(\cos\theta),
\end{equation}
where 
\begin{equation}
	g_{\Delta,\ell}(\cos\theta) = \frac{\ell!}{2^\ell \left( \frac{d - 2}{2} \right)_\ell}
	\sum_{n = 0}^{\ell/2}
	\frac{\left( \frac{1}{2} \right)_n^2
	\left( \frac{d - 2}{2} \right)_n}
	{n! \left( \frac{3 - \Delta - \ell}{2} \right)_n
	\left( \frac{\Delta - \ell - d + 3}{2} \right)_n}
	\mathcal{C}_{\ell - 2n}^{(d/2 - 1 + n)}(\cos\theta)~.
\label{eq:g:alt}
\end{equation}
When combined with the overall power $(-q^2)^{-\Delta_\phi}$ of the form factor, one recovers the expected scaling $(-q^2)^{\Delta_\phi - d/2}$ of the correlation function. Note that eqs.~\eqref{eq:g:alt} and \eqref{eq:g} are two different representations of the same polynomials.


\section{LSZ reduction with distinct external operators}
\label{sec:lszdistinct}

This appendix contains the results of a more general derivation of the LSZ reduction of section~\ref{sec:lsz} in the case of 4 distinct scalar operators $\phi_1$, $\phi_2$, $\phi_3$ and $\phi_4$ with scaling dimensions $\Delta_1$, $\Delta_2$, $\Delta_3$ and $\Delta_4$.

For the first operator, the derivation is identical to that of section~\ref{sec:lsz:first}.
We obtain
\begin{equation}
\begin{aligned}
	\lim_{p_1^0 \to |\vec{p}_1|} (p_1^2 - i \epsilon)^{d/2 - \Delta_1}
	& i \int d^dx_1 \, e^{i p_1 \cdot x_1}
	\langle 0 | T\{ \phi_1(x_1) \phi_1(x_2) \phi_3(x_3) \phi_4(x_4) \} | 0 \rangle
	\\
	&= \frac{1}{2 \sin\left[ \pi \left( \frac{d}{2} - \Delta_1 \right) \right]}
	\langle 0 | T\{ \phi_2(x_2) \phi_3(x_3) \phi_4(x_4) \} | \phi( \vec{p}_1) \rangle~.
\end{aligned}
\end{equation}
For the second operator, the early-time region still leads to an integral of the form of eq.~\eqref{eq:k0integral:p2}, but now instead of eq.~\eqref{eq:Wightman3pt:divergence}, the singularity of the matrix element around $(k - p_1)^2 = 0$ is given by
\begin{equation}
\begin{aligned}
	\langle \mathcal{O}^{\mu_1 \ldots \mu_\ell}(k) | \phi_2(0) | \phi_1(\vec{p}_1) \rangle
	&\approx
	\lambda_{12\mathcal{O}}
	\frac{(2\pi)^{d+1}
	2^{d - \Delta_1 - \Delta_2 - \Delta - \ell}
	\Gamma\left( \frac{d}{2} - \Delta_\phi \right)
	(\Delta - 1)_\ell}
	{\Gamma\left( \Delta_1 - \frac{d}{2} + 1 \right)
	\Gamma\left( \frac{\Delta_1 - \Delta_2 + \Delta + \ell}{2}  \right)
	\Gamma\left( \frac{\Delta_2 - \Delta_1 + \Delta + \ell}{2}  \right)}
	\\
	& \quad \times
	\left[ e^{i \pi (\Delta_1 + \Delta_2 - \Delta + \ell - 1)/2}
	\left( (k-p_1)^2 - i \epsilon \right)^{\Delta_2 - d/2}
	+ \text{c.c.} \right]
	\\
	& \quad \times
	(-k^2)^{(d - \Delta_1 - \Delta_2 - \Delta)/2}
	\mathcal{H}^{\mu_1 \ldots \mu_\ell}(k-p_1, p_1)~.
\end{aligned}
\end{equation}
This leads to the generalization of eq.~\eqref{eq:LSZ:second} for distinct operators, 
\begin{equation}
\begin{aligned}
	& \left( \prod_{i=1}^2\lim_{p_i^0 \to |\vec{p}_i|}
	(p_i^2 - i \epsilon)^{d/2 - \Delta_i}
	i \int d^dx_i \, e^{i p_i \cdot x_i} \right)
	\langle 0 | T\{ \phi_1(x_1) \phi_2(x_2) \phi_3(x_3) \phi_4(x_4) \} | 0 \rangle
	\\
	& \qquad = \sum_{\mathcal{O}} \lambda_{12\mathcal{O}}
	\frac{(2\pi)^{d/2-1}
	\Gamma\left( \frac{d}{2} - \Delta_1 \right) \Gamma\left( \frac{d}{2} - \Delta_2 \right)
	\Gamma(\Delta + \ell) \Gamma\left( \Delta - \frac{d}{2} + 1 \right)}
	{2^{\Delta_1 + \Delta_2 - \Delta + \ell - d/2}
	\Gamma\left( \frac{\Delta_1 - \Delta_2 + \Delta + \ell}{2}  \right)
	\Gamma\left( \frac{\Delta_2 - \Delta_1 + \Delta + \ell}{2}  \right)}
	\\
	& \quad\qquad\qquad \times
	\frac{e^{i \pi (\Delta_1 + \Delta_2 - \Delta + \ell)/2}}
	{[-(p_1 + p_2)^2 ]^{(\Delta_1 + \Delta_2 - \Delta)/2}}
	\\
	& \quad\qquad\qquad \times
	\tilde{\mathcal{H}}_{\mu_1 \ldots \mu_\ell}(p_2, p_1)
	\langle 0 | T\{ \phi_3(x_3) \phi_4(x_4) \} | \mathcal{O}^{\mu_1 \ldots \mu_\ell}(p_1 + p_2) \rangle~.
\end{aligned}
\end{equation}
To obtain the form factor, we combine the explicit expression
\begin{equation}
\begin{aligned}
	F(s,t,u) &= \sum_{\mathcal{O}} \lambda_{12\mathcal{O}}
	\frac{(2\pi)^{d/2-2}
	\left[ \prod\limits_{i=1}^3 \Gamma\left( \frac{d}{2} - \Delta_i \right) \right]
	\Gamma\left( \Delta_3 - \frac{d}{2} + 1 \right)
	\Gamma(\Delta + \ell) \Gamma\left( \Delta - \frac{d}{2} + 1 \right)}
	{2^{\Delta_1 + \Delta_2 - \Delta + \ell - d/2}
	\Gamma\left( \frac{\Delta_1 - \Delta_2 + \Delta + \ell}{2}  \right)
	\Gamma\left( \frac{\Delta_2 - \Delta_1 + \Delta + \ell}{2}  \right)}
	\\
	& \quad\qquad \times
	\frac{e^{i \pi (\Delta_1 + \Delta_2 - \Delta + \ell)/2}}
	{[-(p_1 + p_2)^2 ]^{(\Delta_1 + \Delta_2 - \Delta)/2}}
	\\
	& \quad\qquad \times
	\tilde{\mathcal{H}}_{\mu_1 \ldots \mu_\ell}(p_2, p_1)
	\langle \phi_3(\vec{p}_3) | \phi_4(0) | \mathcal{O}^{\mu_1 \ldots \mu_\ell}(p_1 + p_2) \rangle~.
\end{aligned}
\label{eq:distinctops:lsz:third}
\end{equation}
with the information from the Casimir differential equation of appendix~\ref{sec:casimir},
which establishes that the conformal blocks are proportional to the functions $f_{\Delta,\ell}(w,\cos\theta)$ of eq.~\eqref{eq:f:distinctoperators}. As explained there, we do not have a closed-form expression for these functions, but rather a recursion in powers of $w$, with the lowest order given by the Jacobi polynomial in eq.~\eqref{eq:Jacobipolynomials}.
To fix the proportionality constant, we compare it with the limit $w \to 0$ of eq.~\eqref{eq:distinctops:lsz:third}, and arrive at
\begin{equation}
\begin{aligned}
	F_{\Delta,\ell}( w, \cos\theta) &=
	\frac{(4\pi)^{3d/2}
	\left[ \prod\limits_{i=1}^3 \Gamma\left( \frac{d}{2} - \Delta_i \right) \right]
	\Gamma( \Delta + \ell)
	\left( \Delta - 1 \right)_\ell
	f_{\Delta,\ell}(w - i \epsilon,\cos\theta)}
	{2^{\Delta_\Sigma + 2\ell}
	\Gamma\left( \frac{\Delta_{12} + \Delta + \ell}{2}  \right)
	\Gamma\left( \frac{\Delta_{21} + \Delta + \ell}{2}  \right)
	\Gamma\left( \frac{\Delta_{43} + \Delta + \ell}{2}  \right)
	\Gamma\left( \frac{\Delta_3 + \Delta_4 - \Delta + \ell}{2} \right)}~,
\end{aligned}
\end{equation}
where we have introduced the notation $\Delta_\Sigma \equiv \Delta_1 + \Delta_2 + \Delta_3 + \Delta_4$ and $\Delta_{ab} \equiv \Delta_a - \Delta_b$.

Finally, the amplitude can be obtained from the limit $p_4^2 \to 0$ of eq.~\eqref{eq:distinctops:lsz:third}.
As explained in section~\ref{sec:lsz:fourth}, we take the limit with spacelike $p_4$, after which we obtain
\begin{equation}
\begin{aligned}
	A(s,t,u) &= i \, s^{(d - \Delta_\Sigma)/2} 
	(4\pi)^{3d/2 - 1} 2^{1 - \Delta_\Sigma}
	\left[ \prod_{i=1}^4 \Gamma\left( \tfrac{d}{2} - \Delta_i \right) \right]
	e^{i \pi (\Delta_1 + \Delta_2 - \Delta_3 - \Delta_4)/2}
	\\
	& \quad \times	
	\sum_{\mathcal{O}} \lambda_{12\mathcal{O}} \lambda_{34\mathcal{O}}
	\frac{\Gamma(\Delta + \ell)
	\Gamma\left( \Delta - \frac{d}{2} + 1 \right) (\Delta - 1)_\ell}
	{2^{2\ell}
	\Gamma\left( \frac{\Delta_{12} + \Delta + \ell}{2}  \right)
	\Gamma\left( \frac{\Delta_{21} + \Delta + \ell}{2}  \right)
	\Gamma\left( \frac{\Delta_{34} + \Delta + \ell}{2}  \right)
	\Gamma\left( \frac{\Delta_{43} + \Delta + \ell}{2} \right)}
	\\
	& \quad\qquad\quad \times
	\left[ 1 - e^{i \pi (\Delta_3 + \Delta_4  - \Delta + \ell)} \right]	
	g_{\Delta, \ell}(\cos\theta)
\end{aligned}
\end{equation}
where we have denoted
\begin{equation}
	g_{\Delta, \ell}(\cos\theta) = \mathcal{H}^{\mu_1 \ldots \mu_\ell *}(p_3, p_4)
	\tilde{\mathcal{H}}_{\mu_1 \ldots \mu_\ell}(p_2, p_1)~.
\end{equation}
These $g_{\Delta, \ell}(\cos\theta)$ are again polynomials of degree $\ell$ in $\cos\theta$, but we did not find simple representations like eq.~\eqref{eq:g} for them in the case of distinct operators.
Nevertheless, this amplitude still shows the properties of being trivial in generalized free field theory.


\section{Asymptotic behavior of the conformal partial waves}
\label{sec:convergence}

In this appendix we provide some formulas for the asymptotic behavior of the conformal partial waves $F_{\Delta, \ell}$ for the form factor at large scaling dimension and/or spin.

\subsection{Gegenbauer polynomials}

The Gegenbauer polynomials are a special case of the hypergeometric function,
\begin{equation}
	\mathcal{C}_\ell^\alpha(\cos\theta) = \frac{(2\alpha)_\ell}{\ell!}
	{}_2F_1\left( -\ell, 2\alpha + \ell; \alpha + \frac{1}{2}; \frac{1 - \cos\theta}{2} \right)~.
	\label{eq:Gegenbauer:hypergeometric}
\end{equation}
A convenient integral representation for this hypergeometric function is
\begin{equation}
	\mathcal{C}_\ell^\alpha(\cos\theta) = 
	\frac{\Gamma\left(\alpha + \ell + \frac{1}{2} \right) \Gamma\left(\alpha + \frac{1}{2} \right)}
	{\ell! \Gamma\left( \alpha \right)}
	\int_C \frac{dz}{2\pi i} \, \frac{z^{2\alpha + \ell - 1}}{(z-1)^{\alpha + \ell + 1/2}}
	\left[ 1 - z \frac{1-\cos\theta}{2} \right]^\ell~,
	\label{eq:Gegenbauer:integral}
\end{equation}
where the contour $C$ starts at $z=0$ above the cut  $(-\infty,1]$ and ends at $z = 0$ below the cut, going around the branch point at $z=1$.
This integral is convergent as long as $\alpha > \ell/2$. 
In this representation, the asymptotic behavior at large $\ell$ can be obtained from a saddle-point approximation:
there are two complex saddles as $z = 1 \pm i \sqrt{ \frac{1 + \cos\theta}{1 - \cos\theta}}$, 
provided that $\cos\theta \notin (-\infty, -1] \cup [1, \infty)$.
We obtain, for even $\ell$,
\begin{equation}
	\mathcal{C}_\ell^\alpha(\cos\theta) \approx
	\frac{(-1)^{\ell/2} \ell^{\alpha - 1}}{2^{\alpha - 1} \Gamma(\alpha)}
	\frac{\cos\left[ \left( \theta - \frac{\pi}{2} \right) (\ell + \alpha) \right]}
	{\left| \sin\theta \right|^\alpha}~.
\end{equation}
Instead, when $\cos\theta = 1$, one can directly use eq.~\eqref{eq:Gegenbauer:hypergeometric} to get
\begin{equation}
	\mathcal{C}_\ell^\alpha(1) = \frac{(2\alpha)_\ell}{\ell!}
	\approx \frac{\ell^{2\alpha - 1}}{\Gamma(2 \alpha)}~.
\end{equation}
When $\alpha > 0$ the asymptotic growth of the Gegenbauer polynomials is faster at $\theta = 0, \pi$ than at other scattering angles.

\subsection{Asymptotic behavior at fixed twist}

Using the previous result, we can write down an asymptotic expression for the conformal block $F_{\Delta, \ell}$ in the limit in which the twist $\tau = \Delta - \ell$ is fixed, while both $\Delta$ and $\ell$ are large. 
This is best understood in terms of the representation \eqref{eq:f:alternaterep} of the conformal block, 
for which it can be verified that the sum over $j$ is dominated by the term $j = 0$.
When $\cos\theta \notin (-\infty, -1] \cup [1, \infty)$, we have therefore
\begin{equation}
	f_{\Delta,\ell}(w, \cos\theta)
	\approx 2 \, \frac{w^{\tau/2 - \Delta_\phi} (1 - w)^{1 - \Delta_\phi}}
	{\left| 2 \sin\theta \right|^{(\tau - 1)/2}}
	\frac{(-1)^{\ell/2}}{2^\ell}
	\cos\left[ \left( \theta - \tfrac{\pi}{2} \right) \left( \ell + \tfrac{\tau - 1}{2} \right) \right]~.
\end{equation}
This implies
\begin{equation}
\begin{aligned}
	F_{\Delta, \ell}(w, \cos\theta)
	&\approx (4\pi)^{3d/2 - 1} \Gamma\left( \tfrac{d}{2} - \Delta_\phi \right)^3
	\frac{(16w - i \epsilon)^{\tau/2 - \Delta_\phi}  (1 - w)^{1 - \Delta_\phi}}
	{\left| 2 \sin\theta \right|^{(\tau - 1)/2}}
	\\
	& \quad\times
	\frac{(-1)^{\ell/2}2^\ell \ell^{1-\tau/2}}{\Gamma\left( \Delta_\phi - \frac{\tau}{2} \right)}
	\cos\left[ \left( \theta - \tfrac{\pi}{2} \right) \left( \ell + \tfrac{\tau - 1}{2} \right) \right]~.
\end{aligned}
\label{eq:F:asymptotics:appendix}
\end{equation}
On the other hand, in the forward limit $\cos\theta = 1$, 
\begin{equation}
	f_{\Delta,\ell}(w, 1)
	\approx (w - i \epsilon)^{\tau/2 -  \Delta_\phi} (1 - w)^{1 - \Delta_\phi}
	\frac{\Gamma\left( \frac{\tau - 1}{2} \right)}{\Gamma(\tau - 1)}
	\frac{\ell^{(\tau - 1)/2} }{2^\ell}
\end{equation}
and hence
\begin{equation}
	F_{\Delta, \ell}(w, 1) \approx
	\frac{(4\pi)^{(3d-1)/2} \Gamma\left( \tfrac{d}{2} - \Delta_\phi \right)^3}
	{2^\tau \Gamma\left( \frac{\tau}{2} \right)}
	(16w - i \epsilon)^{\tau/2 - \Delta_\phi} (1 - w)^{1 - \Delta_\phi}
	\frac{2^\ell \ell^{1/2}}{\Gamma\left( \Delta_\phi - \frac{\tau}{2} \right)}~.
\end{equation}

\subsection{Asymptotic behavior at large $\Delta$ and fixed spin}

A large-$\Delta$, fixed spin approximation of the conformal partial waves can be easily worked out using eq.~\eqref{3F2to2F1} and the following asymptotic expansion:
\begin{equation}
\Hypergeometric{\frac{\Delta}{2}+a}{\frac{\Delta}{2}+b}{\Delta+c}{w} 
\stackrel{\Delta \to \infty}{\approx}
(1-w)^{\frac{1}{2}\left(c-a-b-1/2\right)}\left(\frac{2}{1+\sqrt{1-w}}\right)^{\Delta+c-1}~.
\end{equation}
In the conformal partial wave, the dependence on $\Delta$ remarkably factorizes:
\begin{equation}
F_{\Delta,\ell}(w,\cos\theta) \stackrel{\Delta \to \infty}{\approx}
4^\Delta \Delta^{\frac{3}{2}-\Delta_\phi} \left(\frac{\sqrt{w}}{1+\sqrt{1-w}}\right)^\Delta
 \sin \left[\frac{\pi}{2}(2\Delta_\phi-\Delta+\ell)\right] \tilde{F}_{\ell}(w,\cos\theta) ~,
 \label{largeDeltaF}
\end{equation}
where $\tilde{F}_{\ell}(w,\cos\theta)$ is finite,
\begin{equation}
\begin{aligned}
\tilde{F}_{\ell}(w,\cos\theta) &= 
\pi^{\frac{3}{2}(d-1)} 2^{\frac{5}{2}(d-1)-\ell-3\Delta_\phi}
 \Gamma\left( \tfrac{d}{2} - \Delta_\phi \right)^3
(1-w)^{\frac{1}{4}+\frac{1}{2}(\ell-\Delta_\phi)}
w^{-\frac{1}{2}(\ell+2\Delta_\phi)}\\
	& \quad \times
 (1+\sqrt{1-w})^\frac{d}{2}
\frac{\ell!}{\left( \frac{d - 2}{2} \right)_\ell}
	\sum_{n = 0}^{\ell/2} \frac{\left( \frac{d - 2}{2} \right)_n}{n!} 
	\left(\frac{1}{\sqrt{1-w}}\right)^{2n}
	\mathcal{C}_{\ell - 2n}^{d/2 - 1 + n}(\cos\theta)~.
\end{aligned}
\end{equation}


\section{Form factor from Mellin representation}
\label{app:Mellin}

This appendix is devoted to proving eq.~\eqref{FFMellin}, which relates the form-factor to the Mellin transform of the four-point function.
Using the Mellin representation \eqref{eq:Mellindef}, the Fourier transform of the four-point function is given by
\be
(2\pi)^d \delta^d(\sum p_j ) G(p_1,\dots,p_4)=\int [d\gamma] M(\gamma_{ij}) 
\int \left( \prod_{j=1}^4 d^dx_j e^{i p_j\cdot x_j} \right)
\left( \prod_{i<j}^4 \frac{\Gamma(\g_{ij})}{(x_{ij}^2)^{\g_{ij}}} \right)~.
\ee
In order to remove the overall momentum conserving $\d-$function, we introduce 
\be
1 = \int d^dx_0 \,\delta^d\left (x_0 - \frac{1}{4}\sum x_j\right)
\ee
in the last integral and then change integration variables $x_j =x_0 +y_j$.
This leads to 
\be
G(p_1,\dots,p_4)=\int [d\gamma] M(\gamma_{ij}) 
\int \left( \prod_{j=1}^4 d^dy_j e^{i p_j\cdot y_j} \right)
\left( \prod_{i<j}^4 \frac{\Gamma(\g_{ij})}{(y_{ij}^2)^{\g_{ij}}} \right)
\delta^d\left (\frac{1}{4}\sum y_j\right)~.
\ee
Let us now focus on the  integrals over $y_j$. It is convenient to use the following representation of the $\d-$function,
\be
\delta^d\left (\frac{1}{4}\sum y_j\right) = \lim_{\epsilon \to 0} \left(\frac{4}{\epsilon \sqrt{\pi}}\right)^d
e^{-\frac{1}{\epsilon^2}\left(\sum y_j\right)^2}~,
\ee
and introduce Schwinger parameters $w_{ij}$ to write
\be
\frac{\Gamma(\g_{ij})}{(y_{ij}^2)^{\g_{ij}}} = \int_0^\infty \frac{dw_{ij}}{w_{ij}}w_{ij}^{\g_{ij}}
e^{-w_{ij} y_{ij}^2}
\ee
and make all $y_j$ integrals Gaussian.
In fact, we can write
\begin{align}
I\equiv&\int \left( \prod_{j=1}^4 d^dy_j e^{i p_j\cdot y_j} \right)
\left( \prod_{i<j}^4 \frac{\Gamma(\g_{ij})}{(y_{ij}^2)^{\g_{ij}}} \right)
\delta^d\left (\frac{1}{4}\sum y_j\right)\\
=&\lim_{\epsilon \to 0} \left(\frac{4}{\epsilon \sqrt{\pi}}\right)^d
\int_0^\infty \left( \prod_{i<j}^4 \frac{dw_{ij}}{w_{ij}}w_{ij}^{\g_{ij}} \right)
\int \left( \prod_{j=1}^4 d^dy_j e^{i p_j\cdot y_j} \right) e^{- \sum_{i,j=1}^4 y_i\cdot y_j\,Q_{ij} }
\end{align}
where $Q_{jj} = \frac{1}{\epsilon^2} +\sum_{k\neq j} w_{kj}$ and
$Q_{ij}=  \frac{1}{\epsilon^2}-w_{ij}$ for $i\neq j$.
Doing the gaussian integrals leads to
\be
I= 
\lim_{\epsilon \to 0} 
\int_0^\infty \left( \prod_{i<j}^4 \frac{dw_{ij}}{w_{ij}}w_{ij}^{\g_{ij}} \right)
\frac{4^d \pi^{3d/2}}{ ( \epsilon^2 \det(Q))^{\frac{d}{2}} }
e^{- \frac{1}{4} \sum_{i,j=1}^4 p_i\cdot p_j\,Q_{ij}^{-1} }~.
\ee
At this point, one can take the limit $\epsilon \to 0$ because both $Q_{ij}^{-1}$ and $\epsilon^2 \det(Q)$ are finite in this limit.  This simplifies the integrand, but not enough. Fortunately, we are interested in the limit $p_j^2 \to 0$. We shall take this limit for the first 3 momenta keeping $p_4^2>0$ fixed. For now we also keep $s<0$, $t<0$ and $u<0$. Notice that this is consistent with $s+t+u=- p_4^2$. 
To take this limit, we first rescale the Schwinger parameters
\be
w_{ij} \to w_{ij} \frac{ p_i^2\, p_j^2 }{p_4^2}
\ee
and then take the limit $p_j^2 \to 0$ for $j=1,2,3$ with fixed $w_{ij}$.
This gives  
\be
I\approx \pi^{3d/2}(p_4^2)^{\frac{d}{2}-\frac{1}{2}\sum_{j=1}^4\Delta_j} 
\left[ \prod_{j=1}^4 (p_j^2)^{\Delta_j -\frac{d}{2}} \right]
\int_0^\infty \left( \prod_{i<j}^4 \frac{dw_{ij}}{w_{ij}}w_{ij}^{\g_{ij}} \right)
\frac{e^{-H}}{( w_{14}w_{24} w_{34})^{\frac{d}{2}}}~,
\ee
where
\be
H=\frac{1}{4 w_{14}}+\frac{1}{ 4 w_{24}}+\frac{1}{ 4 w_{34}}
-\frac{s w_{12}}{4p_4^2 w_{14}w_{24}}
-\frac{t w_{13}}{4 p_4^2w_{14}w_{34}}
-\frac{u w_{23}}{4p_4^2 w_{24}w_{34}}>0~.
\ee
Now the integrals over $w_{12}$, $w_{13}$ and $w_{23}$ are trivial. After performing them,  the remaining integrals over  Schwinger parameters are also trivial.
The final result is
\be
I\approx \pi^{3d/2} 2^{3 d - \sum_{j=1}^4 \Delta_j}  
\frac{\Gamma(\g_{12})}{(-s)^{\gamma_{12}}}
\frac{\Gamma(\g_{13})}{(-t)^{\gamma_{13}}}
\frac{\Gamma(\g_{23})}{(-u)^{\gamma_{23}}}
\left[ \prod_{j=1}^3 (p_j^2)^{\Delta_j -\frac{d}{2}} \Gamma\left(\frac{d}{2}- \Delta_j \right) \right] ~.
\ee
We conclude that the form factor \eqref{def:FormFactor} is given by
\be
F(s,t,u)= C \int [d\gamma] M(\gamma_{ij}) 
\frac{\Gamma(\g_{12})}{(-s)^{\gamma_{12}}}
\frac{\Gamma(\g_{13})}{(-t)^{\gamma_{13}}}
\frac{\Gamma(\g_{23})}{(-u)^{\gamma_{23}}}~,
\ee
where
\be
C=\pi^{3d/2} 2^{3 d - \sum_{j=1}^4 \Delta_j}  
 \prod_{j=1}^3 \Gamma\left(\frac{d}{2}- \Delta_j \right) ~.
\ee


\section{On the Wick rotation in momentum space}
\label{app:wick}

In section~\ref{sec:lsz}, the form factor and the amplitude were derived from the Fourier transform of the Lorentzian time-ordered correlator. On the other hand, in section~\ref{sec:mellin} the same result followed from the analytic continuation of the Fourier transform of the Euclidean correlator. While a thorough discussion of the analyticity of conformal correlators in momentum space is beyond the scope of this work, the aim of this appendix is to make a few comments in this direction. We shall propose a strategy to relate the Fourier transform of a Euclidean correlator of primary scalars and the Fourier transform of the corresponding time-ordered correlator in Lorentzian signature. We begin by discussing the convergence of the Fourier transform of the Euclidean correlator. Next, we consider the partial Fourier transform where we only transition to momentum space in one direction, which we pick as the Eucidean time. We show that this partially Fourier transformed correlator can be Wick rotated. Along the way, new singularities are generated as a function of the position of the operators in the remaining directions. We do not attempt to establish the integrability of those singularities here. 

\subsection{Existence of the Fourier transform of the Euclidean correlator}

Let us Fourier transform three of the four operators in the connected Euclidean correlator:
\beq
G(p_1,p_2,p_3)= \int \prod_{i=1}^3 d^dx_i e^{i p_i \cdot x_i} \langle\phi(x_1)\phi(x_2)\phi(x_3)\phi(0)\rangle_\textup{conn}^E
\label{FTEuclidean}
\eeq
The convergence of the integral depends on the behavior of the integrand in several dangerous regions, corresponding to the singularities of the correlator as the insertions collide, and to the asymptotics when they scale to infinity. Let us start from the former. When a pair of points collide, the corresponding singularities are integrable if $\Delta_\phi<\frac{d}{2}$.\footnote{In fact, the connected correlator is even less singular.} To discuss the collision of three and four points, let us parametrize the correlator with the usual $u$ and $v$ cross-ratios:
\beq
\langle\phi(x_1)\phi(x_2)\phi(x_3)\phi(x_4)\rangle_\textup{conn}^E =
 \frac{\mathcal{G}_\textup{conn}(u,v)}{\left(x_{12}^2x_{34}^2\right)^{\Delta_\phi}}~, \qquad
 u=\frac{x_{12}^2x_{34}^2}{x_{13}^2x_{24}^2}~, \quad  v=\frac{x_{14}^2x_{23}^2}{x_{13}^2x_{24}^2}~.
 \label{corr_uv}
\eeq
Scale invariance implies that sending three operators close to each other at the same rate $r$ and fixed angle is the same as sending to infinity the fourth operator: the cross-ratios stay finite. We conclude that the correlator scales as $r^{-2\Delta_\phi}$, and is therefore integrable in this limit. Indeed, the measure is suppressed as $r^{2d-1}$, as it can be seen by translating one of the three points to the origin, and going to spherical coordinates in the $2d$-dimensional space parametrized by the position of other two insertions. Similarly, the cross-ratios are constant also when the four operators all scale to the same point, \emph{i.e.} $x_i=r n_i$ in eq.~\eqref{FTEuclidean}, with fixed $n_i$ and $r\to0$. The correlation function scales as $r^{-4\Delta_\phi}$ and the measure scales as $r^{3d-1}$, thus again ensuring integrability at small $r$. These considerations establish the absolute convergence of the Euclidean Fourier transform at short distances. Indeed, consider again the parametrization above, $x_i=r n_i$, with $n_1^2+n_2^2+n_3^2=1$, and perform first the integral over the unit $(3d-1)$-dimensional sphere. Such integral is absolutely convergent, because the dangerous regions where two or three of the $n_i$ coincide are integrable according to the previous discussion. Hence, integrating over the angles cannot enhance the singularity $r^{-4\Delta_\phi}$ as $r\to0$, which is fixed by dimensional analysis.

Let us now discuss the convergence of the integral at infinity. Here more care is needed, since in the cases of physical interest the Fourier transform is not absolutely convergent: if we naively cut-off the integration region and take a limit at the end, the result depends on the way this procedure is performed. 
For instance, consider the region where two points, say $x_1$ and $x_2$, scale to infinity at a finite angle. Then the $(34)$ OPE fixes the behavior of the correlator:
\beq
\mathcal{G}_\textup{conn}(u,v) \approx u^{\Delta_0/2} C_\ell^{d/2-1}\left(\frac{1-v}{2 \sqrt{u}}\right)~.
\eeq
Here, $(u,v)\approx (0,1)$ with the argument of the Gegenbauer polynomial fixed, and $\Delta_0$ is the scaling dimension of the lightest operator above the identity in the $\phi\times \phi$ OPE. If we set $x_1 = r n_1$ and $x_2=r n_2$, with $r\to \infty$ and $n_1^2+n_2^2=1$ parametrizing an $S^{2d-1}$, then $u \approx r^{-2}$, hence the correlator decays as $r^{-2\Delta_\phi-\Delta_0}$. If we now perform the integral in the sphere, we obtain the following large $r$ behavior:
\beq
\int^\infty\! dr\, r^{d-1/2-2\Delta_\phi-\Delta_0} \cos\bigg(r\sqrt{p_1^2+p_2^2}-\frac{2d-1}{4}\pi\bigg)~.
\eeq  
As long as $\sqrt{p_1^2+p_2^2}\neq 0$, the oscillating factor guarantees the convergence of the integral if the function decreases at infinity, \emph{i.e.} if $2\Delta_\phi+\Delta_0>d-1/2$.
However, this condition excludes various interesting examples, like the four-point function of $\sigma$ in the $3d$ Ising model. Moreover, it is not hard to see that different limiting procedures -- \emph{e.g.} integrating over two $S^{d-1}$ of radius $r$ in $x_1$ and $x_2$ -- would give a different condition for convergence.

Luckily, the Fourier transform of a power can always be defined by convolution with a test function. Here we shall use a Gaussian, to fix ideas. The following simple example is enough to capture any asymptotic region in the Fourier transform of the four-point function:
\beq
I=\int_0^{+\infty}\! dr\, e^{ipr} (r^2)^\beta e^{-2 r^2/\sigma^2}~.
\label{IGaussian}
\eeq
The convergence of the integral at $x=0$ is irrelevant for us, and clearly for any value of $\beta$ and $p$ the integral converges at infinity. The integral can be evaluated exactly, and the limit $\sigma\to\infty$ exists:
\beq
\lim_{\sigma\to\infty} I = (-i p)^{-1-2\beta}\, \Gamma(2\beta+1)~.
\label{IGaussianLim}
\eeq
With this prescription, any Euclidean four-point function of primary operators possesses a Fourier transform.

In order to perform the Wick rotation, the following observation is important. The contour of integration in eq.~\eqref{IGaussian} can be deformed, as long as $|\textup{Arg} \ r|<\pi/4$ when $r\to \infty$. Now, suppose that $p>0$, and the contour is deformed upwards at infinity, so that the phase $e^{ipr}$ becomes exponentially suppressed. Then, the Gaussian is not needed anymore and the limit in eq.~\eqref{IGaussianLim} can be taken before integration. In the next subsection, we shall perform various contour deformations of this sort. Even when we do not mention test functions, it is understood that, after a first small deformation of the contour, the $\sigma\to\infty$ limit has been taken inside the integral.

\subsection{The Wick rotation}

In position space, the Wick rotation amounts to the following analytic continuation of a correlation function in its Euclidean time variables:
\beq
\tau_i = \tau_{E,i} e^{\ii \theta}~, \qquad \theta:\ 0 \to \pi/2~.
\label{Wickpos}
\eeq
Here the $\tau_{E,i}$ are real, but not necessarily positive. If all the time coordinates are rotated together, the correlator is time-ordered for any $\theta$, meaning that if $\Im \tau_i>\Im \tau_j$ then $\phi(x_i)$ appears to the left of $\phi(x_j)$ in the correlator. Intuitively, the Wick rotation in momentum space amounts to the opposite rotation:
\beq
p_i^\tau = p_{E,i}^\tau e^{-\ii \theta}~, \qquad \theta:\ 0 \to \pi/2~,
\label{Wickmom}
\eeq
where $p^\tau$ denotes the Euclidean time component of the momentum, and the $p_{E,i}^\tau$ are real. Indeed, performing the two rotations \eqref{Wickpos} and \eqref{Wickmom} at the same time, the plane wave factor in the Fourier transform \eqref{FTEuclidean} remains oscillating. In the following, we would like to show that this is essentially correct, at least when considering the partial Fourier transform
\begin{equation}
G(\vec{x}_i,p_i^\tau)= \int \prod_{i=1}^3 d\tau_i\, e^{i p^\tau_i \tau_i} \langle\phi(x_1)\phi(x_2)\phi(x_3)\phi(0)\rangle_\textup{conn}^E~,
\label{partialFT}
\end{equation}
where we split the Euclidean coordinates of the insertions as $x^\mu=(\vec{x},\tau)$. Indeed,  $G(\vec{x}_i,p_i^\tau)$ admits an analytic continuation along the path \eqref{Wickmom} up to $\theta=\pi/2-\epsilon$ for any finite $\epsilon>0$, and the result is the Fourier transform of the time-ordered correlator, in the sense just described.\footnote{Note however the exception discussed at the end of this appendix:  if $p_{E,1}^\tau<0$, $|p_{E,1}^\tau|>p_{E,2}^\tau $, $|p_{E,1}^\tau|> p_{E,3}^\tau$   and $|p_{E,1}^\tau|<p_{E,2}^\tau+p_{E,3}^\tau$, we are not able to exhibit the analytic continuation along the path \eqref{Wickmom}. }

\begin{figure}
\centering
\begin{tikzpicture}[scale=1.2]
\draw [decorate,decoration={snake,amplitude=0.8pt}] (-2,1) -- (-2,3);
\filldraw [black] (-2,1) circle [radius=1pt] node[left] {$\tau_2+i|\vec{x}_{12}|$};
\draw [decorate,decoration={snake,amplitude=0.8pt}] (-2,-1) -- (-2,-3);
\filldraw [black] (-2,-1) circle [radius=1pt] node[left] {$\tau_2-i|\vec{x}_{12}|$};
\draw [decorate,decoration={snake,amplitude=0.8pt}] (2.5,1.5) -- (2.5,3);
\filldraw [black] (2.5,1.5) circle [radius=1pt] node[right] {$\tau_3+i|\vec{x}_{13}|$};
\draw [decorate,decoration={snake,amplitude=0.8pt}] (2.5,-0.5) -- (2.5,-3);
\filldraw [black] (2.5,-0.5) circle [radius=1pt] node[right] {$\tau_3-i|\vec{x}_{13}|$};
\draw [decorate,decoration={snake,amplitude=0.8pt}] (0.1,1.3) -- (0.1,3);
\filldraw [black] (0.1,1.3) circle [radius=1pt] node[right] {$i|\vec{x}_{1}|$};
\draw [decorate,decoration={snake,amplitude=0.8pt}] (0.1,-1.3) -- (0.1,-3);
\filldraw [black] (0.1,-1.3) circle [radius=1pt] node[right] {$-i|\vec{x}_{1}|$};
\draw[thick,->] (-4,0) -- (4,0);
\draw[thick,->] (0,-3) -- (0,3);
\draw [black] (4.2,2.8)-- (4,2.8) -- (4,3);
\node at (4.2,2.91) {$\tau_1$};
\end{tikzpicture}
\caption{Analytic structure of the correlator as a function of the Euclidean time of one of the insertions. The three pairs of branch points correspond the past and future lightcone OPEs of $\phi(x_1)$ with one of the three other insertions.}
\label{fig:analytic_structure}
\end{figure}
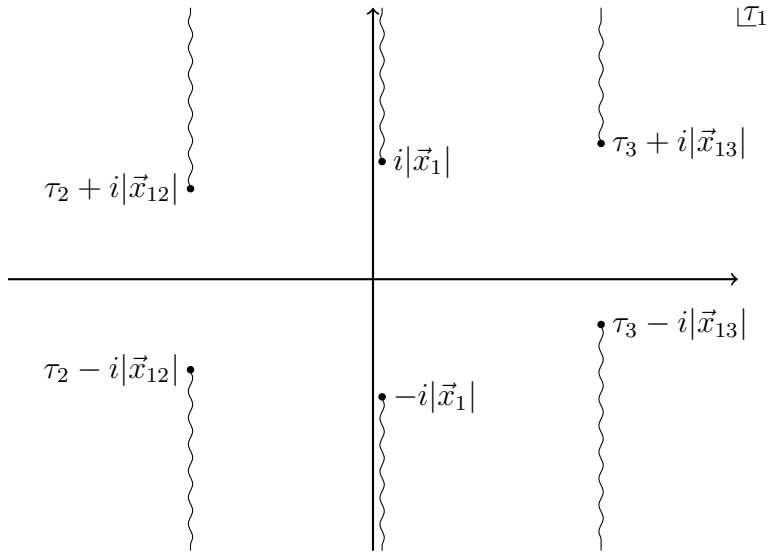

We need to first recall the analytic properties of the correlation function for complex times. When the $\vec{x}_i$ are all different, the correlator is analytic in each time $\tau_i$ if $\Re \tau_i\neq \Re \tau_j$. Such analyticity is exhibited by the multiple Laplace transform of the correlator. If, say $\Re\tau_1=\Re\tau_2$, the correlator is analytic in $\tau_1$ in the interval $|\Im \tau_1-\Im\tau_2|<|\vec{x}_{12}|$ \cite{Hartman:2015lfa}. The analytic structure is exemplified in figure~\ref{fig:analytic_structure}.

We shall perform the Wick rotation of eq.~\eqref{partialFT} in three steps. A first contour deformation will extend the analytic domain in the momenta from the real axis to part of the complex plane. Next we shall rotate the momenta according to eq.~\eqref{Wickmom}, and finally we will complete the deformation of the integration contours according to eq.~\eqref{Wickpos}. It will be necessary to consider separately the case where the Euclidean momenta $p_{E,i}^\tau$ have different signs.

\subsection*{$p_{E,i}^\tau$ all positive}

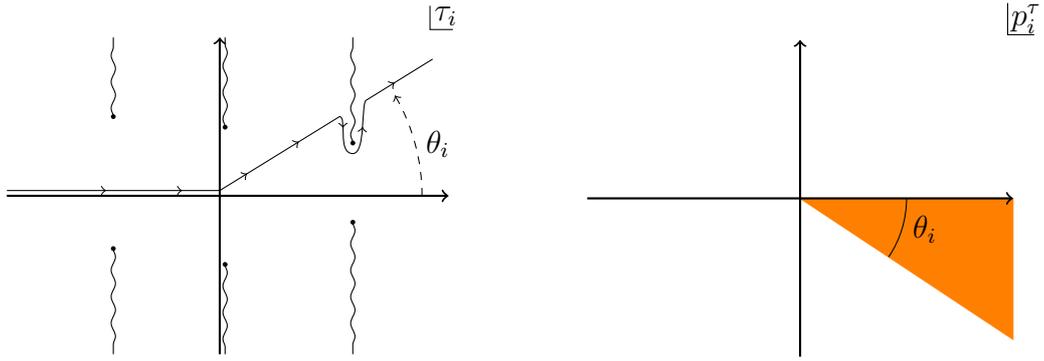
\begin{figure}
\begin{minipage}{0.5\textwidth}
\centering
\begin{tikzpicture}[scale=0.7]
\draw [decorate,decoration={snake,amplitude=0.8pt}] (-2,1.5) -- (-2,3);
\filldraw [black] (-2,1.5) circle [radius=1pt]; 
\draw [decorate,decoration={snake,amplitude=0.8pt}] (-2,-1) -- (-2,-3);
\filldraw [black] (-2,-1) circle [radius=1pt]; 
\draw [decorate,decoration={snake,amplitude=0.8pt}] (2.5,1) -- (2.5,3);
\filldraw [black] (2.5,1) circle [radius=1pt];
\draw [decorate,decoration={snake,amplitude=0.8pt}] (2.5,-0.5) -- (2.5,-3);
\filldraw [black] (2.5,-0.5) circle [radius=1pt];
\draw [decorate,decoration={snake,amplitude=0.8pt}] (0.1,1.3) -- (0.1,3);
\filldraw [black] (0.1,1.3) circle [radius=1pt];
\draw [decorate,decoration={snake,amplitude=0.8pt}] (0.1,-1.3) -- (0.1,-3);
\filldraw [black] (0.1,-1.3) circle [radius=1pt];
\draw[thick,->] (-4,0) -- (4.3,0);
\draw[thick,->] (0,-3) -- (0,3);
\draw [black] (4.38,3.15)-- (3.95,3.15) -- (3.95,3.6);
\node at (4.25,3.4) {$\tau_i$};
\draw [postaction = {decoration={markings, mark=between positions 0.1 and 0.95 step 0.8cm with {\arrow{>};}},decorate}] (0,0.1) -- (2.25,1.5) .. controls (2.4,1.5) and (2.20,0.8) .. (2.5,0.8) .. controls (2.75,0.8) and (2.65,1.75) .. (2.75,1.81)--(4,2.59);
\draw [postaction = {decoration={markings, mark=between positions 0.2 and 0.9 step 1cm with {\arrow{<};}},decorate}] (0,0.1) -- (-4,0.1);
\draw[dashed,->] (3.8,0)   arc  [start angle=0, end angle=30, radius=3.8] node [midway,right]  {$\theta_i$};
\end{tikzpicture}
\end{minipage}
\begin{minipage}{0.5\textwidth}
\centering
\begin{tikzpicture}[scale=0.7]
\filldraw [orange] (0,0) -- (4,-2.67) -- (4,0) -- cycle;
\draw[thick,->] (-4,0) -- (4,0);
\draw[thick,->] (0,-3) -- (0,3);
\draw [black] (4.4,3.1)-- (3.9,3.1) -- (3.9,3.7);
\node at (4.25,3.4) {$p^\tau_i$};
\draw (2,0)   arc  [start angle=0, end angle=-33.69, radius=2] node [midway,right]  {$\theta_i$};
\end{tikzpicture}
\end{minipage}
\caption{\emph{Left panel}. The contour deformation when all $p_{E,i}^\tau>0$. More precisely, the contour of the time component which is integrated first must avoid all the branch points as shown in the figure. After the first integration,  the integrand only has two pairs of branch points, and so on. \emph{Right panel}. The region of convergence of the Fourier transform after the contour deformation in the left panel. See the text for a precise characterization.}
\label{fig:rotation_plus}
\end{figure}

In this case, we first modify the contours of all the time variables as in the left panel of figure~\ref{fig:rotation_plus}. The angles $\theta_i$ for the three time variables do not need to all coincide: it is possible to rotate one coordinate at a time.\footnote{Depending on the values of $\theta_i$ and $\vec{x}_{ij}$, the contours may circumnavigate the upper cuts as in figure~\ref{fig:rotation_plus}, or the lower ones. All the following considerations are unaffected.} As we pointed out earlier, dropping the arcs when rotating the contours is allowed, thanks to the convolution with test functions at first, then thanks to the exponential suppression from the plane wave factors. After the rotation, we obtain a representation of $G(\vec{x}_i,p_i^\tau)$ with a larger region of analyticity. Indeed, the $\tau_i$-integrals converge as long as the plane wave factors do not grow exponentially in any of the asymptotic regions. This requires
\beq
\Im (p_1^\tau \tau_1+p_2^\tau \tau_2+p_3^\tau \tau_3)\geq 0~, \qquad \textup{large } \tau_i~.
\label{phase_condition}
\eeq
When only one time variable is large -- $\Re \tau_i \to \pm \infty$ -- the exponential is not growing when $p_i^\tau$ belongs to the wedge highlighted in the right panel in figure~\ref{fig:rotation_plus}. Because of the dip in the $\tau$-contour -- see again figure~\ref{fig:rotation_plus} -- we also need to worry about the region where multiple $\tau_i$ are of the same order and large. Exponential suppression is then guaranteed if all $p^\tau_i$'s belong to the same wedge, with width $\theta=\theta_\textup{min}$, the smallest angle of the three contours in the $\tau$ plane.

\begin{figure}
\begin{minipage}{0.5\textwidth}
\centering
\begin{tikzpicture}[scale=0.7]
\draw[thick,->] (-4,0) -- (4,0);
\draw[thick,->] (0,-3) -- (0,3);
\draw [black] (4.38,3.15)-- (3.95,3.15) -- (3.95,3.6);
\node at (4.25,3.4) {$\tau$};
\draw [postaction = {decoration={markings, mark=between positions 0.1 and 0.95 step 0.8cm with {\arrow{>};}},decorate}] (0,0) -- (3,3);
\draw [postaction = {decoration={markings, mark=between positions 0.2 and 0.9 step 1cm with {\arrow{<};}},decorate}] (0,0) -- (-4,-2);
\draw[dashed,->] (3,0)   arc  [start angle=0, end angle=44, radius=3] node [midway,right]  {$\theta_+$};
\draw[dashed,->] (-3,0)   arc  [start angle=180, end angle=206, radius=3] node [midway,left]  {$\theta_-$};
\end{tikzpicture}
\end{minipage}
\begin{minipage}{0.5\textwidth}
\centering
\begin{tikzpicture}[scale=0.7]
\filldraw [orange] (0,0) -- (3,-3) -- (4,-3) -- (4,-2) -- cycle;
\draw[thick,->] (-4,0) -- (4,0);
\draw[thick,->] (0,-3) -- (0,3);
\draw [black] (4.4,3.1)-- (3.9,3.1) -- (3.9,3.7);
\node at (4.25,3.4) {$p^\tau$};
\draw (1.7,0)   arc  [start angle=0, end angle=-26.5, radius=1.7] node [midway,right]  {$\theta_-$};
\draw (3.1,0)   arc  [start angle=0, end angle=-45, radius=3.1] node [midway,right]  {$\theta_+$};
\end{tikzpicture}
\end{minipage}
\caption{Schematic plot of the last step in the contour deformations when all $p_{E,i}^\tau>~0$, and of the associated region of convergence of the Fourier transform. The detailed procedure is completely analogous to the one illustrated in figure~\ref{fig:rotation_plus} and in the relative text.}
\label{fig:rotation_2_plus}
\end{figure}
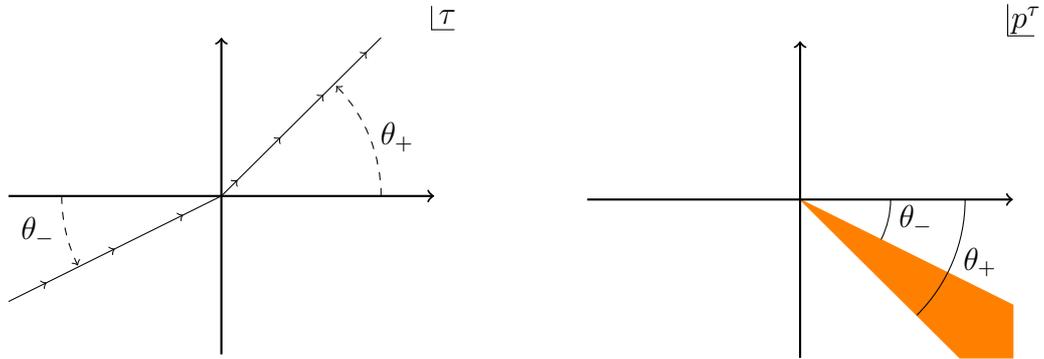

At this point, we can choose all $\theta_i$ to be equal, and rotate all $p^\tau_i$'s according to eq.~\eqref{Wickmom}. The last step in the procedure requires rotating the contours of integration in the half-plane $\Re \tau<0$. This restricts the region of analyticity in the $p^\tau$ plane, as shown schematically in figure~\ref{fig:rotation_2_plus}.
Finally, one can set $\theta_-=\theta_+=\theta$ for all the three variables, at the price of re-introducing Gaussian wave packets to dump the large $|\tau|$ tails.

It is important to notice that, after integrating each of the time variables, the integrand develops additional singularities with respect to the ones in the left panel of figure~\ref{fig:rotation_plus}. These correspond to the pinching of the contour of integration between two branch points. Let us say we perform the integral over $\tau_1$. At fixed $|\vec{x}_{1j}|$ the branch points corresponding to the past and future lightcones of the same operator never pinch the contour. On the contrary, when, say, $\Re \tau_2=\Re \tau_3$, the branch point corresponding to the future lightcone of $\phi(x_2)$ can collide with the one corresponding to the past lightcone of $\phi(x_3)$, or vice versa. This happens when
\beq
\tau_2\pm i |\vec{x_{12}}|=\tau_3 \mp i |\vec{x_{13}}|~.
\eeq
The two branch points then pinch the $\tau_1$-contour. This configuration is a double lightcone limit, where $x_1$ is lightlike separated from $x_2$ and $x_3$ (it intersects the future lightcone of one point and the past lightcone of the other). However, in the $\tau_2$ and $\tau_3$ complex planes, these new singularities lie always above or below the branch points associated to the ordinary lightcone singularities. Indeed, due to the triangular inequality,
\beq
|\Im \tau_2-\Im \tau_3|_\textup{double lightcone} \geq |\Im \tau_2-\Im \tau_3|_\textup{lightcone} 
= |\vec{x_{23}}|~.
\eeq
Hence, the new singularities do not affect our procedure for the contour deformation, because they lie on top of the cuts already present in figure \eqref{fig:analytic_structure}. On the other hand, if $|\vec{x_{ij}}|\to 0$ for $i,j=1,2,3$, all the branch points pinch the contour simultaneously. When performing the integral over the spacelike coordinates, one must be careful that the correlator remains integrable in this limit also after the contour deformation. Although it should be possible to use the OPE to prove this fact, at least as long as $\theta<\pi/2$, we leave this analysis to future work.

\subsection*{$p_{E,i}^\tau$ with mixed signs}

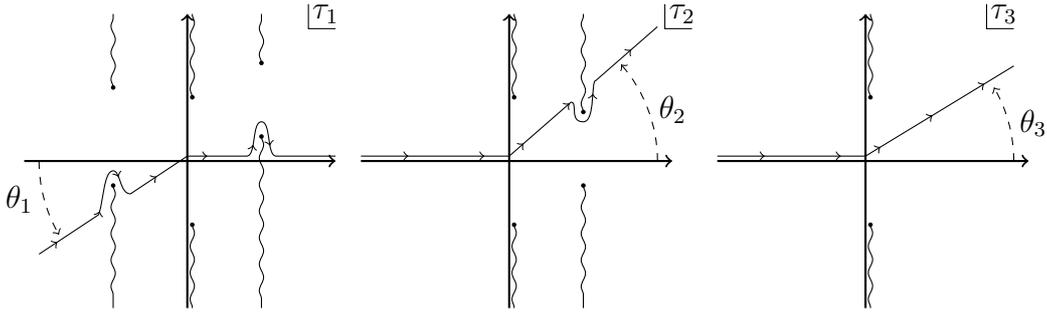
\begin{figure}
\centering
\begin{tikzpicture}[scale=0.65]
\draw [decorate,decoration={snake,amplitude=0.8pt}] (-1.5,1.5) -- (-1.5,3);
\filldraw [black] (-1.5,1.5) circle [radius=1pt]; 
\draw [decorate,decoration={snake,amplitude=0.8pt}] (-1.5,-0.5) -- (-1.5,-3);
\filldraw [black] (-1.5,-0.5) circle [radius=1pt]; 
\draw [decorate,decoration={snake,amplitude=0.8pt}] (1.5,2) -- (1.5,3);
\filldraw [black] (1.5,2) circle [radius=1pt];
\draw [decorate,decoration={snake,amplitude=0.8pt}] (1.5,0.5) -- (1.5,-3);
\filldraw [black] (1.5,0.5) circle [radius=1pt];
\draw [decorate,decoration={snake,amplitude=0.8pt}] (0.1,1.3) -- (0.1,3);
\filldraw [black] (0.1,1.3) circle [radius=1pt];
\draw [decorate,decoration={snake,amplitude=0.8pt}] (0.1,-1.3) -- (0.1,-3);
\filldraw [black] (0.1,-1.3) circle [radius=1pt];
\draw[thick,->] (-3.3,0) -- (3,0);
\draw[thick,->] (0,-3) -- (0,3);
\draw [black] (3,2.7)-- (2.45,2.7) -- (2.45,3.2);
\node at (2.8,3) {$\tau_1$};
\draw [postaction = {decoration={markings, mark=between positions 0.1 and 0.9 step 0.7cm with {\arrow{>};}},decorate}] (-3,-1.9) -- (-1.8,-1.1)  .. controls (-1.7,-0.9) and (-1.7,-0.2) .. (-1.5,-0.2) .. controls (-1.35,-0.2) and (-1.37,-0.67) .. (-1.15,-0.67)--(0,0.1);
\draw [postaction = {decoration={markings, mark=between positions 0.1 and 0.9 step 0.7cm with {\arrow{>};}},decorate}] (0,0.1) -- (1.2,0.1)  .. controls (1.4,0.1) and (1.3,0.8) .. (1.5,0.8) .. controls (1.7,0.8) and (1.6,0.1) .. (1.8,0.1)--(3,0.1);
\draw[dashed,->] (-3,0)   arc  [start angle=180, end angle=210, radius=3] node [midway,left]  {$\theta_1$};
\end{tikzpicture}
\begin{tikzpicture}[scale=0.65]
\draw [decorate,decoration={snake,amplitude=0.8pt}] (1.5,1) -- (1.5,3);
\filldraw [black] (1.5,1) circle [radius=1pt];
\draw [decorate,decoration={snake,amplitude=0.8pt}] (1.5,-0.5) -- (1.5,-3);
\filldraw [black] (1.5,-0.5) circle [radius=1pt];
\draw [decorate,decoration={snake,amplitude=0.8pt}] (0.1,1.3) -- (0.1,3);
\filldraw [black] (0.1,1.3) circle [radius=1pt];
\draw [decorate,decoration={snake,amplitude=0.8pt}] (0.1,-1.3) -- (0.1,-3);
\filldraw [black] (0.1,-1.3) circle [radius=1pt];
\draw[thick,->] (-3,0) -- (3.3,0);
\draw[thick,->] (0,-3) -- (0,3);
\draw [black] (3.7,2.7)-- (3.15,2.7) -- (3.15,3.2);
\node at (3.5,3) {$\tau_2$};
\draw [postaction = {decoration={markings, mark=between positions 0.08 and 0.98 step 0.8cm with {\arrow{>};}},decorate}] (0,0.1) -- (1.25,1.2) .. controls (1.4,1.2) and (1.20,0.8) .. (1.5,0.8) .. controls (1.75,0.8) and (1.65,1.65) .. (1.75,1.64)--(3,2.74);
\draw [postaction = {decoration={markings, mark=between positions 0.2 and 0.9 step 1cm with {\arrow{<};}},decorate}] (0,0.1) -- (-3,0.1);
\draw[dashed,->] (3,0)   arc  [start angle=0, end angle=41, radius=3] node [midway,right]  {$\theta_2$};
\end{tikzpicture}
\begin{tikzpicture}[scale=0.65]
\draw [decorate,decoration={snake,amplitude=0.8pt}] (0.1,1.3) -- (0.1,3);
\filldraw [black] (0.1,1.3) circle [radius=1pt];
\draw [decorate,decoration={snake,amplitude=0.8pt}] (0.1,-1.3) -- (0.1,-3);
\filldraw [black] (0.1,-1.3) circle [radius=1pt];
\draw[thick,->] (-3,0) -- (3.3,0);
\draw[thick,->] (0,-3) -- (0,3);
\draw [black] (3,2.7)-- (2.45,2.7) -- (2.45,3.2);
\node at (2.8,3) {$\tau_3$};
\draw [postaction = {decoration={markings, mark=between positions 0.1 and 0.95 step 0.8cm with {\arrow{>};}},decorate}] (0,0.1) -- (3,1.94);
\draw [postaction = {decoration={markings, mark=between positions 0.2 and 0.9 step 1cm with {\arrow{<};}},decorate}] (0,0.1) -- (-3,0.1);
\draw[dashed,->] (3,0)   arc  [start angle=0, end angle=30, radius=3] node [midway,right]  {$\theta_3$};
\end{tikzpicture}
\caption{Prescription to deform the $\tau$-contours when $|p_{E,1}^\tau|<p_{E,2}^\tau,\,p_{E,3}^\tau$. Recall that, according to the explanation in the main text, additional branch points corresponding to double lightcone limits lie on top of the cuts shown in the $\tau_2$ and $\tau_3$ planes.}
\label{fig:rotation_minus}
\end{figure}

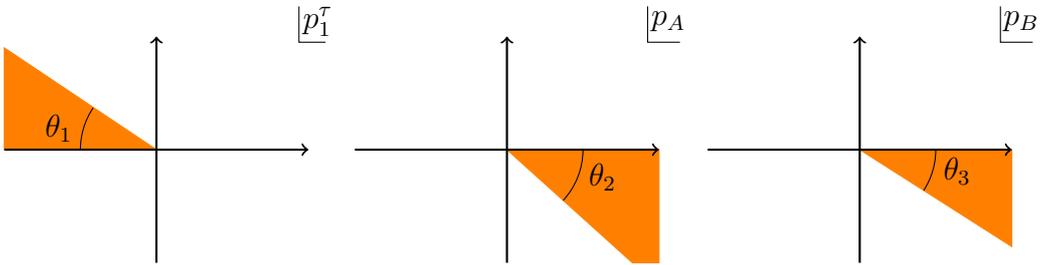
\begin{figure}
\centering
\begin{tikzpicture}[scale=0.5]
\filldraw [orange] (0,0) -- (-4,2.7) -- (-4,0) -- cycle;
\draw[thick,->] (-4,0) -- (4,0);
\draw[thick,->] (0,-3) -- (0,3);
\draw [black] (4.45,2.85)-- (3.75,2.85) -- (3.75,3.8);
\node at (4.25,3.4) {$p^\tau_1$};
\draw (-2,0)   arc  [start angle=180, end angle=146.3, radius=2] node [midway,left]  {$\theta_1$};
\end{tikzpicture}
\begin{tikzpicture}[scale=0.5]
\filldraw [orange] (0,0) -- (3.3,-3) -- (4,-3) -- (4,0) -- cycle;
\draw[thick,->] (-4,0) -- (4,0);
\draw[thick,->] (0,-3) -- (0,3);
\draw [black] (4.55,2.85)-- (3.7,2.85) -- (3.7,3.8);
\node at (4.25,3.4) {$ p_A$};
\draw (2,0)   arc  [start angle=0, end angle=-42.3, radius=2] node [midway,right]  {$\theta_2$};
\end{tikzpicture}
\begin{tikzpicture}[scale=0.5]
\filldraw [orange] (0,0) -- (4,-2.57) -- (4,0) -- cycle;
\draw[thick,->] (-4,0) -- (4,0);
\draw[thick,->] (0,-3) -- (0,3);
\draw [black] (4.55,2.85)-- (3.7,2.85) -- (3.7,3.8);
\node at (4.25,3.4) {$ p_B$};
\draw (2,0)   arc  [start angle=0, end angle=-32.75, radius=2] node [midway,right]  {$\theta_3$};
\end{tikzpicture}
\caption{Region of convergence of the Fourier transform after the contour deformations in figure~\ref{fig:rotation_minus}. In the second plot, $p_A$ stands for either of the following two combinations: $p_2^\tau$ or $p^\tau_1+p^\tau_2$. Similarly, in the third plot $p_B$ stands for $p_3^\tau$, $(p^\tau_1+p^\tau_3)$, $(p^\tau_2+p^\tau_3)$ or $(p^\tau_1+p^\tau_2+p^\tau_3)$. }
\label{fig:wedges}
\end{figure}

If we want to reach a Lorentzian configuration where, say, $p_1^\tau<0$ and the timelike components of the remaining momenta are positive, we need to start from $p_{E,1}^\tau<0,$ $p_{E,2}^\tau,\,p_{E,3}^\tau>0$. Now, the $\tau_1$-contour needs to be first deformed downward in the region $\Re \tau_1<0$. Let us also first assume that $|p_{E,1}^\tau|<p_{E,2}^\tau,\,p_{E,3}^\tau$. Then, the following prescription allows to exhibit the required analytic continuation. We first integrate in $\tau_1$, then in $\tau_2$ and finally in $\tau_3$.\footnote{Actually, only the fact that $\tau_1$ is integrated first is important.} The contours of integration are then shown in figure~\ref{fig:rotation_minus}.
Again, we should make sure that the phase factors are not growing in the asymptotic regions, which include, besides the limits $\Re\tau_i \to \pm\infty$ independently, also the following ones:
\begin{align}
&\tau_1\approx \tau_2\to \infty \quad \textup{along the path of } \tau_2~, \\
&\tau_1\approx \tau_3\to \infty \quad \textup{along the path of } \tau_3~, \\
&\tau_2\approx \tau_3\to \infty \quad \textup{along the path of } \tau_3~, \\
&\tau_1\approx \tau_2 \approx \tau_3 \to \infty \quad \textup{along the path of } \tau_3~.
\end{align}
We conclude that the region of convergence of the Fourier transform, after the contour deformation, is the domain in $\mathbb{C}_{p^\tau_1}\times \mathbb{C}_{p^\tau_2} \times \mathbb{C}_{p^\tau_3}$ formed by the intersection of the wedges depicted in figure~\ref{fig:wedges}.  Due to the ordering of the $p_{E,i}^\tau$, it is easy to convince oneself that this domain includes the path in eq.~\eqref{Wickmom}, once all $\theta_i$ are chosen equal. In the final step, again we straighten up all $\tau$-contours.

The argument generalizes to the other possible orderings among $|p_{E,1}^\tau|,\,p_{E,2}^\tau$ and $p_{E,3}^\tau$: we always integrate the time variables in increasing order of the associated momenta. There is, however, a caveat. If $|p_{E,1}^\tau|>p_{E,2}^\tau\,p_{E,3}^\tau$, but $|p_{E,1}^\tau|<p_{E,2}^\tau+p_{E,3}^\tau$, this procedure still does not allow to exhibit the Wick rotation along the path \eqref{Wickmom}. Indeed, when all $\tau_i$ become large along the path of $\tau_1$ the condition \eqref{phase_condition} is violated as soon as $\theta_1\neq 0$. Closing this loophole would likely require a more refined analysis, which lies outside the scope of this appendix. Notice in particular that the Fourier transform in the spatial coordinates should simplify the analytic structure of the result, thanks to the restored Lorentz invariance. For instance, when all the momenta are spacelike, the Fourier transform of the time-ordered correlator cannot depend on the sign of $p_1^0$, since the latter can be changed via a Lorentz transformation.  
\vspace{1 cm}

Let us conclude this discussion with the following simple example:
\beq
G(p_1,p_2)=\int\!d\tau_2\, e^{i p_2 \tau_2}\int\! d\tau_1\, e^{i p_1 \tau_1}
 \frac{1}{\left((\tau_1-\tau_2)^2+1\right)\left(\tau_1^2+1\right)\left(\tau_2^2+1\right)}~.
 \label{G3pointExample}
\eeq
In other words, we take the Fourier transform in the Euclidean variable $\tau$ of the three-point function of primaries with dimension $\Delta_\phi=2$, placed at unit distance from each other in the remaining coordinates. In this case, all the branch points corresponding to the lightcones are simple poles, and the Fourier transform is easily computed by closing the contour appropriately. After integrating in $\tau_1$ one finds
\beq
G(p_1,p_2)= \int\!d\tau_2\, e^{i p_2 \tau_2-|p_1|} \frac{\pi}{\tau_2 (1+\tau_2^2)}
\left( \frac{e^{i p_1 \tau_2}}{\tau_2+2 i\, \textup{sign} p_1}+ \frac{1}{\tau_2-2 i\, \textup{sign} p_1}   \right)~. 
\eeq
The integrand has now additional poles at $\tau_2=\pm 2i$ which, as explained above, correspond to the pinching of the $\tau_1$ contour between the (future lightcone) pole at $\tau_1=i$ and the (past lightcone) pole at $\tau_1=\tau_2-i$, or vice versa. The final result is
\beq
G(p_1,p_2) = 
\left\{
\begin{array}{lr}
 \pi^2 \left(e^{-|p_1|-|p_2|}-\frac{1}{3}e^{-2|p_1|-|p_2|}-\frac{1}{3}e^{-|p_1|-2|p_2|}\right)  & p_1p_2>0~, \\
 \pi^2 \left(e^{-|p_1|}-\frac{1}{3}e^{-2|p_1|+|p_2|}-\frac{1}{3}e^{-|p_1|-|p_2|}\right) & p_1p_2<0,\, p_1(p_1+p_2)>0~.
 \label{G3pointResult}
\end{array}
\right.
\eeq
The remaining case, where $p_1p_2<0$ and $p_1(p_1+p_2)<0$ can be deduced from the $p_1 \leftrightarrow p_2$ symmetry of eq.~\eqref{G3pointExample}. eq.~\eqref{G3pointResult} can be analytically continued along the path \eqref{Wickmom}, and the resulting function is piecewise analytic in the space spanned by complex $p_1,\,p_2$. The analytic continuation depends on the starting point of the path, according to the Euclidean Fourier transform eq.~\eqref{G3pointResult}. This all matches our general discussion, where the cases $p_1 p_2  \gtrless 0$ and $p_1+p_2 \gtrless 0$ had to be treated separately.


\bibliography{./bibliography}
\bibliographystyle{./JHEP}

\end{document}